\documentclass[12pt]{article}

\usepackage{epsfig,epsf}
\usepackage{amsmath}
\usepackage{amsthm}
\usepackage{amsfonts}
\usepackage{amssymb}

\usepackage{slashed}

\usepackage[active]{srcltx}

\renewcommand{\thefootnote}{\fnsymbol{footnote}}

\newcommand{\be}{\begin{equation}}
\newcommand{\ee}{\end{equation}}
\newcommand{\ba}{\begin{eqnarray}}
\newcommand{\ea}{\end{eqnarray}}
\newcommand{\baa}{\begin{eqnarray*}}
\newcommand{\btab}{\begin{tabular}}
\newcommand{\etab}{\end{tabular}}
\newcommand{\eaa}{\end{eqnarray*}}

\def \labeltest #1 {\label{#1}}

\newcommand \vev [1] {\langle{#1}\rangle}

\def\inbar{\,\vrule height1.5ex width.4pt depth0pt}
\def\IC{\relax\hbox{$\inbar\kern-.3em{\rm C}$}}
\def\IZ{\relax{\hbox{\cmss Z\kern-.4em Z}}}
\def\IR{{\hbox{{\rm I}\kern-.2em\hbox{\rm R}}}}

\def\IP{{\hbox{{\rm I}\kern-.2em\hbox{\rm P}}}}
\def\II{\hbox{{1}\kern-.25em\hbox{l}}}

\newcommand{\twist}{ t}

\numberwithin{equation}{section}

\begin{document}

\begin{titlepage}

\vskip1cm
\begin{center}
  {\large \bf
     Baryon Operators of Higher Twist in QCD and Nucleon Distribution Amplitudes
  \\}

\vspace{1cm}
{\sc V.M.~Braun\,$^{a}$},
{\sc A.N.~Manashov\,$^{a,b}$}
and {\sc J. Rohrwild\,$^{a}$}
\\[0.5cm]
\vspace*{0.1cm} $^{a}${\it
   Institut f\"ur Theoretische Physik, Universit\"at
   Regensburg, \\ D-93040 Regensburg, Germany}  \\[5mm]
\vspace*{0.1cm} $^{b}${\it Department of Theoretical Physics,  St.-Petersburg State
University\\
199034, St.-Petersburg, Russia}\\[10mm]

%{\em submitted to Nuclear Physics B\\Version of \today}\\[1cm]

\def\thefootnote{\arabic{footnote}}
\setcounter{footnote} 0

\vskip0.8cm
{\bf Abstract:\\[10pt]} \parbox[t]{\textwidth}{
We develop a general theoretical framework for the description
of higher--twist baryon operators which makes maximal use
of the conformal symmetry of the QCD Lagrangian.
The conformal operator basis is constructed for all twists.
The complete analysis of the one-loop renormalization of twist-4 operators is given.
The evolution equation for three-quark operators of the same chirality
turns out to be completely integrable. The spectrum of anomalous
dimensions coincides in this case with the energy spectrum of the twist-4 subsector
of the $SU(2,2)$ Heisenberg spin chain.
The results are applied to give a general classification and
calculate the scale dependence of subleading twist-4 nucleon distribution 
amplitudes that are relevant for hard exclusive reactions involving a helicity flip.
In particular we find an all-order expression (in conformal spin) for the contributions
of geometric twist-3 operators to the (light-cone) twist-4 nucleon
distribution amplitudes, which are usually referred to
as Wandzura--Wilczek terms.
}
\vskip1cm

\end{center}

\end{titlepage}

{\small \tableofcontents}

\newpage

%%%%%%%%%%%%%%%%%%%%%%%%%%%%%%%%%%%%%%%%%%%%%%%%%%%%%%%%%%%%%%%%%%%%%%%%%%%%%%%%%%%%%%%%%%%%%%%%%%%
%
\section{Introduction}
%
%%%%%%%%%%%%%%%%%%%%%%%%%%%%%%%%%%%%%%%%%%%%%%%%%%%%%%%%%%%%%%%%%%%%%%%%%%%%%%%%%%%%%%%%%%%%%%%%%%%

Higher-twist effects in hard processes in QCD generically correspond to  corrections to
physical observables  that are suppressed by powers of the hard scale. They are important
in order to achieve high accuracy, and interesting because higher-twist corrections are
sensitive to fine details of the hadron structure. A theoretical description of higher
twist effects within QCD factorization involves contributions of a large number of local operators
which are much more numerous compared to the leading twist so that the choice of a proper
operator basis is important. This choice is not unique, as exemplified by the two
existing classical approaches to the twist-4 effects in deep-inelastic lepton hadron
scattering~\cite{Jaffe:1982pm,Ellis:1982cd}. The ``transverse'' basis of
Ref.~\cite{Ellis:1982cd} leads to simpler coefficient functions whereas the
``longitudinal'' basis of Ref.~\cite{Jaffe:1982pm} (see also \cite{Shuryak:1981kj})
allows for a parton-model-like interpretation \cite{Jaffe:1983hp}.

The renormalization of higher-twist operators corresponds to the scale dependence of the
physical observables.   For twist three, the corresponding study  is essentially
completed.  The anomalous dimension matrix for baryon operators was first calculated  in
\cite{Peskin:1979mn},  for chiral-even quark-antiquark-gluon and three-gluon operators in
\cite{Bukhvostov:1984as}, and for chiral-odd in \cite{Koike:1994st}.
The structure of the spectrum of twist-three anomalous dimensions is well understood
\cite{BDM,BDKM,Belitsky:1999qh,Belitsky:1999ru,Belitsky:1999bf,DKM-00,Braun:2000av,Braun:2001qx} 
and in some cases explicit WKB-type
expansions are available that allow to calculate anomalous dimensions  to arbitrary
accuracy: The size of the mixing matrix plays the role of the expansion parameter.
Beyond twist three much less is known. Up to now, anomalous dimensions have only been
calculated for a few operators of lowest dimension (e.g. \cite{Morozov:1983qr,Pivovarov:1991nk}).  In
addition, the structure of  the most singular parts of the mixing kernels for small values
of the  Bjorken  variable that are relevant for the contribution of two-pomeron  cuts in
high-energy scattering processes was considered in \cite{Levin:1992mu,Bartels:1993it}.

The modern approach for the calculation of leading-order anomalous dimensions of
higher-twist operators  makes maximal use of the conformal symmetry of the renormalization
group equations.   Historically, the importance of  conformal symmetry in the present  context
was first understood for the leading twist pion distribution amplitude and it was
instrumental  for the proof of QCD factorization for the pion form factor
\cite{ER80,LB80,Brodsky:1980ny}.  A general formalism  was  developed in
\cite{Bukhvostov:1985rn} for the special class of so-called quasipartonic operators
that are built of ``plus'' components of quark and gluon fields. For each  twist, the set
of quasipartonic operators is closed under renormalization and the renormalization group
(RG) equation can be written in a Hamiltonian form that involves two-particle kernels
given in terms of  two-particle Casimir operators of the collinear subgroup
$SL(2,{\mathbb{R}})$  of the conformal group. In this formulation symmetries of the RG
equations  become explicit.  Moreover, for a few important cases the corresponding
three-particle  quantum-mechanical problem turns out to be completely integrable  and in
fact  equivalent to a specific Heisenberg spin chain \cite{BDM}. An almost complete
understanding  achieved at present of the renormalization of twist-three operators  is due
to all  these formal developments, see \cite{Braun:2003rp,Belitsky:2004cz} for a review
and further  references.

The goal of this paper is to generalize some of the above techniques to the situation
where not all  contributing operators are quasipartonic, as it proves to be the case
starting with  twist four.
Apart of the needs of practical applications to QCD phenomenology, our work is fuelled by
the recent  study \cite{Beisert,BFKS04} where it was shown that
diagonal part of
one-loop QCD  RG equations (for arbitrary twist) can be written in a Hamiltonian form in
terms of  quadratic Casimir operators of the full conformal group $SO(4,2)$ instead of its
collinear subgroup. Moreover,
all kernels can be obtained from the known  kernels for the collinear $SL(2,\mathbb{R})$ subgroup
~\cite{Bukhvostov:1985rn}.
Although much of the formalism appears to be general, in this paper we concentrate on the
simplest  example of non-quasipartonic twist-four baryon operators that contain two
``plus''  and one ``minus'' quark field, schematically
$$q_+ q_- q_+\,,$$
and their mixing with (quasipartonic) four-particle operators involving a gluon field,  of the type
$$q_+ q_+ q_+ F_{+\perp}.$$
Our main results can be summarized as follows.

{}First, we construct a complete conformal operator basis for arbitrary twist,
with "good" transformation properties. We then specialize
to the case of twist-4 baryonic operators, calculate all one-loop evolution
kernels including the mixing with four-particle operators involving a gluon
field, and check that the kernels are $SL(2)$ invariant, as expected.
The operators involving three quark fields with the same chirality do
not mix with the operators involving both chiral and antichiral quarks,
so that these two cases can be considered separately.
The evolution equation for three-quark operators of the same chirality
turns out to be completely integrable. The spectrum of anomalous
dimensions coincides in this case with the energy spectrum of the twist-4 subsector of the $SU(2,2)$
Heisenberg spin chain, confirming the prediction
of \cite{BFKS04}.
For both cases, we present a detailed study of the spectra of the anomalous
dimensions.
Finally, these results are applied to give a general classification and
calculate the scale dependence of subleading twist-4 nucleon distribution
amplitudes that are relevant for hard exclusive reactions involving a helicity flip.
In particular we introduce novel four-particle distribution amplitudes
involving a gluon field, and derive explicit expressions for the expansion
of all distribution amplitudes in contributions of multiplicatively
renormalizable operators in first three orders of the conformal expansion.
As a byproduct of our analysis, we
give an %all-order 
expression %(in conformal spin) 
for the contributions
of geometric twist-3 operators to the (light-cone) twist-4 nucleon
distribution amplitudes, which are usually referred to
as Wandzura--Wilczek terms.

The presentation is organized as follows. We begin in Sect.~2 with
a short exposition of the spinor formalism that is used throughout
our work. This formalism is standard in the studies of SUSY theories but
is used rarely by the QCD community so we felt that a short summary
is necessary. Next, conformal transformation properties of the fields
are considered in some detail. A complete basis of one-particle
light-ray operators is constructed for chiral quark and self-dual
gluon fields in QCD, cf. Eq.~(\ref{basis}), which is one of our main results.
In Sect.~3 we specialize to the particular case of baryonic
twist-4 operators which are the main subject of the rest of the paper.
Renormalization group equations for the light-ray baryonic operators
are derived in Sect.~4. We discuss general properties of these equations,
give a summary of the relevant conformal invariant evolution kernels,
introduce a convenient scalar product on the space of the solutions and,
finally, give explicit expressions of the Hamiltonians
for all cases of interest. Solutions of the renormalization group
equations for twist-4 operators are considered in Sect.~5.
{}For three-quark operators of the same chirality the problem
turns out to be completely integrable. We find the
corresponding conserved charge and discuss the relation of
this result to the approach of \cite{Beisert,BFKS04}.
A simple analytic expression is found for the lowest anomalous dimension
in the spectrum of chiral quark twist-4 operators with odd number $N=2k+1$
of covariant derivatives. For other cases the spectra are studied numerically.
The results are presented in the Figures and for the first few $N$
also in table form. It turns out that
differences between twist-4 and twist-3 operators mostly
affect a few lowest eigenstates (for a given $N$);
the upper part of the spectrum is universal:
the anomalous dimensions appear to be almost independent on twist and chirality.
{}Explicit expressions for the nucleon distribution amplitudes taking
into account first three orders in conformal spin and Wandzura-Wilczek
corrections are given in Sect.~6. The final Sect.~7 is reserved
for summary and conclusions.

%%%%%%%%%%%%%%%%%%%%%%%%%%%%%%%%%%%%%%%%%%%%%%%%%%%%%%%%%%%%%%%%%%%%%%%%%%%%%%%%%%%%%%%%%%%%%%%%%%%
%
\section{Spinors and Conformal Symmetry}
%
%%%%%%%%%%%%%%%%%%%%%%%%%%%%%%%%%%%%%%%%%%%%%%%%%%%%%%%%%%%%%%%%%%%%%%%%%%%%%%%%%%%%%%%%%%%%%%%%%%%
For applications it is important to have an operator basis with good transformation
properties with respect to the collinear $SL(2,\mathbb{R})$ subgroup of the conformal group.
It is well known that analysis of  tensor properties  of operators
is greatly simplified in the spinor representation. Although this formalism is standard,
a number of  different prescriptions exist in the literature for raising and lowering
indices, normalization etc. In order to make our presentation self-contained
we choose to begin with a summary of the definitions and basic relations  of the spinor algebra, 
and also introduce some general notation that is used throughout the paper.
Our conventions are similar but not identical to the ones accepted in Ref.~\cite{Sohnius}.

%%%%%%%%%%%%%%%%%%%%%%%%%%%%%%%%%%%%%%%%%%%%%%%%%%%%%%%%%
%%%%%%%%%%%%%%%%%%%%%%%%%%%%%%%
\subsection{Spinor formalism}
%%%%%%%%%%%%%%%%%%%%%%%%%%%%%%%%%%%%%%%%%%%%%%%%%%%%%%%%%
%%%%%%%%%%%%%%%%%%%%%%%%%%%%%%%
The Lorentz group $SO(3,1)$ is locally isomorphic to the group of complex unimodular $2\times 2$
matrices, $SL(2,\mathbb{C})$.  To make this explicit, each covariant four-vector $x_\mu$ can be
mapped to a hermitian matrix $x$
\begin{align}\label{xd}
x=\begin{pmatrix}
x_0+x_3 & x_1-ix_2\\
x_1+ix_2 & x_0-x_3
\end{pmatrix} \equiv x_{\mu}\sigma^\mu\,
\end{align}
where $\sigma^\mu=(\II,\vec{\sigma})$ and $\vec{\sigma}$ are the usual Pauli matrices.
A Lorentz transformation $x'_\mu={\Lambda_{\mu}}^{\nu} x_\nu$ corresponds to
a rotation  $x'=MxM^\dagger$, where $M\in SL(2,\mathbb{C})$, and the homomorphism
$\Lambda\to M$ defines a two-dimensional (spinor) representation of Lorentz group,
$u'=Mu$. The correspondence between $\Lambda$ and $M$ is not unique and in general
one might consider four representations defined by the homomorphisms
$\Lambda\to M, M^*,M^{-1,T}$ and $M^{-1\dagger}$.
The vectors from the corresponding representation spaces~--~spinors~--~
are usually denoted as $u_\alpha,\bar u_{\dot\alpha}, u^{\alpha}$ and
$\bar u^{\dot\alpha}$, respectively, i.e.
$u'_{\alpha}={M_{\alpha}}^{\beta}\, u_{\beta}$,
$\bar u'_{\dot\alpha}={M^*_{\dot\alpha}}^{\dot\beta}\, \bar u_{\dot\beta}$
 etc.
The representations  $M$ and $M^{-1,T}$ (also $M^*$ and $M^{-1,\dagger}$) are
equivalent since  $\sigma_2 M=M^{-1,T}\sigma_2$.
The intertwining operator $\sigma_2$ is proportional to the Levi-Civita
tensor $\epsilon$. We define
\begin{align}\label{eps-n}
\epsilon_{12}=\epsilon^{12}=-\epsilon_{\dot1\dot2}=-\epsilon^{\dot1\dot2}=1
\end{align}
and  accept the following rule for raising and lowering of spinor indices (cf.~\cite{Sohnius})
\begin{align}\label{raise}
u^\alpha=\epsilon^{\alpha\beta}u_\beta\,,&& u_\alpha=u^\beta\epsilon_{\beta\alpha}\,,&&
\bar u^{\dot\alpha}=\bar u_{\dot\beta}\epsilon^{\dot\beta\dot\alpha}\,,&&
\bar u_{\dot\alpha}=\epsilon_{\dot\alpha\dot\beta}\bar u^{\dot\beta}\,,
\end{align}
which is consistent with~(\ref{eps-n}). Note  that
${\epsilon_\alpha}^\beta=-{\epsilon^\beta}_\alpha=\delta^\beta_\alpha$ and
${\epsilon^{\dot\alpha}}_{\dot\beta}=-{\epsilon_{\dot\beta}}^{\dot\alpha}=
\delta_{\dot\beta}^{\dot\alpha}$.

When it is not displayed explicitly it is implied that  undotted indices
are contracted ``up--down'',
$
(u v)\overset{\text{def}}{=}u^\alpha v_\alpha = - u_\alpha v^\alpha
$
and dotted ones ``down--up'',
$(\bar u \bar v)\overset{\text{def}}{=}\bar u_{\dot\alpha} \bar v^{\dot\alpha}
=-\bar u^{\dot\alpha} \bar v_{\dot\alpha}$

Next, we define $(u_\alpha)^*=\bar u_{\dot\alpha}$ and $(u^\alpha)^*=\bar u^{\dot\alpha}$
that is, again,   consistent with (\ref{eps-n}) and results in $(uv)^*=(\bar v \bar u)$.
The Fierz transformation for Weyl spinors reads
\begin{align}\label{}
(u_1u_2)(v_1v_2)=(u_1 v_1)(u_2 v_2)-(u_1 v_2)(u_2 v_1)\,
\end{align}
which is a consequence of the identity
\begin{align}\label{e-e}
\epsilon_{ab}\epsilon_{cd}=\epsilon_{ac}\epsilon_{bd}-\epsilon_{ad}\epsilon_{bc}
\end{align}

In addition to  $\sigma^\mu_{\alpha\dot\beta} = (\II,\vec{\sigma})$
it is convenient to introduce
$(\bar\sigma^\mu)^{\dot\alpha \beta}= (\II,-\vec{\sigma})$ so that
$(\bar\sigma^\mu)^{\dot\alpha \beta}=(\sigma^\mu)^{\beta\dot\alpha}$,
and define  $\bar x=x_\mu\bar\sigma^\mu$, cf. ~(\ref{xd}).
One easily finds that
$$
a_\mu=\frac12{(a\bar\sigma_\mu)_\alpha}^\alpha =\frac12{(\bar
a\sigma_\mu)^{\dot\alpha}}_{\dot\alpha}\,,
\qquad a_\mu b^\mu=\frac12 a_{\alpha\dot\alpha}\bar b^{\dot\alpha\alpha}\,.
$$

For completeness, we give below some useful identities involving $\sigma_\mu$ matrices:
\begin{align}\label{id-1}
&\sigma^\mu_{\alpha\dot\alpha}\,(\bar
\sigma^\nu)^{\dot\alpha\alpha}=2g^{\mu\nu}\,,
&&\sigma^\mu_{\alpha\dot\alpha}\,\bar
\sigma_\mu^{\dot\beta\beta}=2\delta_\alpha^\beta\,\delta_{\dot\alpha}^{\dot\beta}\,,
%\notag\\
%& \sigma^\mu_{\alpha\dot\alpha}\,(\sigma_\mu)_{\beta\dot\beta}=-
%2\epsilon_{\alpha\beta}\epsilon_{\dot\alpha\dot\beta}\,,
%&& \bar\sigma_\mu^{\dot\alpha\alpha}\,(\bar\sigma^\mu)^{\dot\beta\beta}=-
%2\epsilon^{\alpha\beta}\epsilon^{\dot\alpha\dot\beta}
\end{align}
\begin{align}\label{id-2}
{(\sigma^\mu\bar\sigma^\nu+\sigma^\nu\bar\sigma^\mu)_{\alpha}}^{\beta}=&
2g^{\mu\nu}{\delta_{\alpha}}^\beta\,,&&
{(\bar\sigma^\mu\sigma^\nu+\bar\sigma^\nu\sigma^\mu)^{\dot\alpha}}_{\dot\beta}=
2g^{\mu\nu}{\delta^{\dot\alpha}}_{\dot\beta}\,.
\end{align}
Generators of the Lorentz group read
\begin{align}
{(\sigma^{\mu\nu})_{\alpha}}^{\beta}=
\frac{i}2{\left[\sigma^{\mu}\bar\sigma^\nu-\sigma^{\nu}\bar\sigma^\mu\right
]_{\alpha}}^\beta\,, &&
{(\bar\sigma^{\mu\nu})^{\dot\alpha}}_{\dot\beta}=
\frac{i}2{\left[\bar\sigma^{\mu}\sigma^\nu-\bar\sigma^{\nu}\sigma^\mu\right
]^{\dot\alpha}}_{\dot\beta}\,,
\end{align}
or, in  the explicit form
\begin{align}
\sigma^{0i}=-i\sigma^{i}\,,&&
\sigma^{ik}=i\epsilon^{ikj}\sigma^j\,,&&
\bar\sigma^{0i}=i\sigma^{i}\,,&&
\bar \sigma^{ik}=i\epsilon^{ikj}\sigma^j\,,
\end{align}
They satisfy the self-duality relations
\begin{align}\label{sigma-dual}
\sigma^{\mu\nu}=\frac{i}{2}\varepsilon^{\mu\nu\rho \omega}\sigma_{\rho\omega}\,,&&
\bar \sigma^{\mu\nu}=-\frac{i}2\varepsilon^{\mu\nu\rho \omega}\bar\sigma_{\rho\omega}\,.
\end{align}
where $\epsilon_{0123}=1$.

A four-dimensional Dirac bispinor is written as
\begin{align}\label{}
q=\begin{pmatrix}\psi_\alpha\\ \bar\chi^{\dot\beta}\end{pmatrix}\,,
&& \bar q=(\chi^\beta,\bar\psi_{\dot\alpha})
\end{align}
and the  $\gamma_\mu$ matrices take  the form
\begin{align}\label{}
\gamma^{\mu}=\begin{pmatrix}0&[\sigma^\mu]_{\alpha\dot\beta}\\
                            [\bar\sigma^{\mu}]^{\dot\alpha\beta}&0 \end{pmatrix}\,,&&
\slashed{a}=\begin{pmatrix}0&a_{\alpha\dot\beta}\\
                            \bar a^{\dot\alpha\beta}&0 \end{pmatrix}\,.
\end{align}
For the common $\sigma^{\mu\nu}=\frac{i}2[\gamma^\mu,\gamma^\nu]$,
$\gamma_5=i\gamma^0\gamma^1\gamma^2\gamma^3$ and the charge conjugation matrix
$C=i\gamma^ 2\gamma^0$
one finds
\begin{align}
\sigma^{\mu\nu}=\begin{pmatrix}{[\sigma^{\mu\nu}]_{\alpha}}^{\beta}&0\\
                           0& {[\bar\sigma^{\mu\nu}]^{\dot\alpha}}_{\dot\beta} \end{pmatrix}\,,
&&
\gamma_5=\begin{pmatrix}-\delta_{\alpha}^\beta&0\\
                           0&\delta^{\dot\alpha}_{\dot\beta}  \end{pmatrix}\,,
                           &&
C=\begin{pmatrix}-\epsilon_{\alpha\beta}&0\\
                           0&-\epsilon^{\dot\alpha\dot\beta}  \end{pmatrix}\,.
\end{align}

Irreducible representations of the Lorentz group are labeled by two spins $(s,\bar s)$.
%$({\sf j},\bar {\sf j})$.
The representation space is spanned by tensors
$T_{\alpha_1\ldots\alpha_{2{\sf s}},\dot\beta_1\ldots\dot\beta_{2\bar {\sf s}}}$
which are symmetric  in dotted and undotted indices separately. In particularly, the Weyl
spinors $\psi$ (chiral) and $\bar\chi$ (antichiral) belong to the representations
$(1/2,0)$ and $(0,1/2)$, respectively, whereas the Dirac spinor transforms
as $(1/2,0)\oplus (0,1/2)$.

The gluon strength tensor $F_{\mu\nu}$ transforms as $(1,0)\oplus (0,1)$
and can be decomposed as
\begin{align}\label{}
F_{\alpha\beta,\dot\alpha\dot\beta}=\sigma^\mu_{\alpha\dot\alpha}\sigma^\nu_{\beta\dot\beta}
F_{\mu\nu}=
2\left(\epsilon_{\dot\alpha\dot\beta} f_{\alpha\beta}-
\epsilon_{\alpha\beta} \bar f_{\dot\alpha\dot\beta}
\right)
%\epsilon_{\alpha\beta} \bar f_{\dot\alpha\dot\beta}\right)\,
%\frac{i}2\left(\epsilon_{\dot\alpha\dot\beta} f_{\alpha\beta}+
%\epsilon_{\alpha\beta} \bar f_{\dot\alpha\dot\beta}\right)\,,
\end{align}
where $f_{\alpha\beta}$ and $\bar f_{\dot\alpha\dot\beta}$ are chiral and antichiral
symmetric tensors, $f^*=\bar f$, which belong to  the representations
$(1,0)$ and $(0,1)$, respectively. One obtains
\begin{align}\label{}
f_{\alpha\beta}=&\frac{i}{4}\sigma_{\alpha\beta}^{\mu\nu} F_{\mu\nu}\,,&&
\bar f_{\dot\alpha\dot\beta}=-\frac{i}4\bar \sigma_{\dot\alpha\dot\beta}^{\mu\nu} F_{\mu\nu}\,,.
\end{align}
or in terms of the gauge field $A_{\dot\alpha \alpha}$:
\begin{align}\label{}
f_{\alpha\beta}=\frac14\left({D_{\alpha}}^{\dot\alpha} \bar A_{\dot\alpha\beta}+
 {D_{\beta}}^{\dot\alpha} \bar A_{\dot\alpha\alpha}\right)\,, &&
\bar f_{\dot\alpha\dot\beta}=\frac14\left({\bar D_{\dot \alpha}}^{\phantom{\beta}\alpha}
  A_{\alpha\dot\beta}+
{\bar D_{\dot \beta}}^{\phantom{\beta}\alpha}  A_{\alpha\dot\alpha} \right)\,,
\end{align}
where the covariant derivative is defined as $D_\mu=\partial_\mu-ig A_\mu$.
The expressions for $F^{\mu\nu}$ and the dual strength tensor
$\widetilde F^{\mu\nu}=\dfrac12 \epsilon^{\mu\nu\rho\sigma} F_{\rho\sigma}$ are
\begin{align}\label{}
F^{\mu\nu}=\frac{i}2\left( \sigma^{\mu\nu}_{\alpha\beta} f^{\alpha\beta}-
\bar\sigma^{\mu\nu}_{\dot\alpha\dot\beta} \bar f^{\dot\alpha\dot\beta}\right)\,,&&
\widetilde F^{\mu\nu}=\frac12\left( \sigma^{\mu\nu}_{\alpha\beta} f^{\alpha\beta}+
\bar\sigma^{\mu\nu}_{\dot\alpha\dot\beta} \bar f^{\dot\alpha\dot\beta}\right)\,.
\end{align}

The Dirac equation for the quark fields reads
\begin{align}\label{Dirac}
\bar D^{\dot\alpha\alpha} \psi_\alpha(x)=0\,,&& D_{\alpha\dot\alpha}\bar\chi^{\dot\alpha}(x)=0\,
\end{align}
where the covariant derivative is defined as $D_\mu=\partial_\mu-ig A_\mu$.
The equation of motion (EOM) for the fields $f,\bar f$ becomes
\begin{align}\label{}
{\bar D_{\dot\beta}}^{\phantom{\beta}\alpha}
f_{\alpha\beta}^a=%{i}
g\left(\bar\psi_{\dot\beta} T^a\psi_\beta+
\chi_\beta T^a \bar\chi_{\dot\beta}\right)\,,&&
%{D_{\beta}^{\dot\alpha}}
%{\bar f^{\dot\alpha}}_{\phantom{\alpha}\dot\beta}=-%{i}
%g\left(\bar\psi_{\dot\beta}\psi_\beta+
%\chi_\beta\bar\chi_{\dot\beta}\right)\,.
{D_{\beta}}^{\dot\alpha}
\bar f_{\dot\alpha\dot\beta}^a=%-{i}
g\left(\bar\psi_{\dot\beta} T^a\psi_\beta+
\chi_\beta T^a\bar\chi_{\dot\beta}\right)\,.
\end{align}
The class of the operators which are proportional to the equation of motion is closed
under renormalization (for a more precise statement see e.g. Ref.~\cite{Collins}). On-shell
matrix elements of such operators vanish and one can consider two operators which
difference is an EOM operator as being equivalent.

The equation
\begin{align}\label{}
T_{\alpha_1\ldots\alpha_n,\dot\beta_1\ldots\dot\beta_n}=
\sigma^{\mu_1}_{\alpha_1\dot\beta_1}\ldots
\sigma^{\mu_n}_{\alpha_n\dot\beta_n} T_{\mu_1\ldots\mu_n}
\end{align}
establishes the relation between generic tensors in the usual vector and spinor representations.
The symmetrization over spinor indices is most conveniently achieved
contracting the open indices with an auxiliary spinor $\xi$.
We define
\begin{align}\label{}
T_\xi=\,
\xi^\alpha_1\ldots\xi^{\alpha_{n}}\,
T_{\alpha_1\ldots\alpha_{n},\dot\beta_1\ldots\dot\beta_{\bar n}}\,
\bar\xi^{\dot\beta_1}\ldots\bar\xi^{\dot\beta_{\bar n}}\,.
\end{align}
In particular
\begin{align}\label{xi-1}
 \psi_\xi = (\xi \psi) = \xi^\alpha \psi_\alpha && f_\xi = \xi^\alpha \xi^\beta f_{\alpha\beta}\,,
\nonumber\\
 \bar \chi_\xi = (\bar\chi \bar \xi) = \bar\chi_{\dot\alpha} \bar\xi^{\dot\alpha} &&
\bar f_\xi = \bar f_{\dot\alpha \dot\beta}\bar \xi^{\dot\alpha}\bar \xi^{\dot\beta}\,,
\end{align}
etc.

It is obvious that a symmetric tensor
$T_{\alpha_1\ldots\alpha_{n},\dot\beta_1\ldots\dot\beta_{\bar n}}$
can  unambiguously  be restored from the convolution $T_\xi$ by applying multiple
derivatives over $\xi$.  We define
\begin{align}\label{}
\partial_\beta \xi^\alpha=
\dfrac{\partial}{\partial\xi^\beta}\xi^\alpha={\epsilon_\beta}^\alpha=\delta^\alpha_\beta\,,&&
\bar\partial^{\dot\beta}\bar\xi_{\dot\alpha}=
\dfrac{\partial}{\partial\bar\xi_{\dot\beta}}\bar\xi_{\dot\alpha }=
{\epsilon^{\dot\beta}}_{\dot\alpha}=\delta^{\dot\beta}_{\dot\alpha}\,,
\end{align}
so that
\begin{align}\label{T-sym}
T_{\alpha_1\ldots\alpha_{n},\dot\beta_1\ldots\dot\beta_{\bar n}}=\frac{(-1)^{\bar
n}}{n!\,\bar n!}
\frac{\partial}{\partial\xi^{\alpha_1}}\ldots\frac{\partial}{\partial\xi^{\alpha_n}}
\frac{\partial}{\partial\bar\xi^{\dot\beta_1}}\ldots
\frac{\partial}{\partial\bar\xi^{\dot\beta_{\bar n}}} T_\xi\,.
\end{align}
Note that the rule for raising and lowering of indices for derivatives over spinor variables
is different from that for the spinors themselves, cf. Eq.~(\ref{raise}):
\begin{align}
\dfrac{\partial}{\partial\xi^\beta}
=\epsilon_{\beta\alpha}\dfrac{\partial}{\partial\xi_\alpha}\,,
%= \epsilon_{\beta\alpha} \partial^\alpha\,,
&&
%\bar\partial^{\dot\beta}=
\dfrac{\partial}{\partial\bar\xi_{\dot\beta}}
=\epsilon^{\dot\beta\dot\alpha}\dfrac{\partial}{\partial\bar\xi^{\dot\alpha}}\,.
%=\epsilon^{\dot\beta\dot\alpha}\bar\partial_{\dot\alpha}
\end{align}
%

%%%%%%%%%%%%%%%%%%%%%%%%%%%%%%%%%%%%%%%%%%%%%%%%%%%%%%%%%%%%%%%%%%%%%%%%%%%%%%%%%%%%%
\subsection{Conformal symmetry}
%%%%%%%%%%%%%%%%%%%%%%%%%%%%%%%%%%%%%%%%%%%%%%%%%%%%%%%%%%%%%%%%%%%%%%%%%%%%%%%%%%%%%
 It is known that the QCD enjoys conformal symmetry at the classical  level.
 Although this symmetry is broken in the full quantum theory, it leads to strong
constraints on the form of (one-loop) operator counterterms and will be quite useful in the
subsequent  analysis.
The action on the generators of the conformal group on the fundamental fields in 
the spinor representation, $\Phi=(\Phi_\xi,\bar \Phi_\xi)$ with $\Phi_\xi=\{\psi_\xi,\chi_\xi, f_\xi\}$ and
$\bar \Phi_{\xi}=\{\bar\psi_\xi,\bar \chi_\xi, \bar f_\xi\}$, takes the form \cite{MS69}
\begin{subequations}\label{conf}
\begin{align}\label{}
i[{\mathbf P}_{\alpha\dot\alpha},\Phi(x)]=&\partial_{\alpha\dot\alpha} \Phi(x)
\,\equiv\, iP_{\alpha\dot\alpha}\Phi(x)\,,
\\
i[{\mathbf D},\Phi(x)]=&\frac12\left(x_{\alpha\dot\alpha}\partial^{\alpha\dot\alpha}+2\twist+
\xi^\alpha\frac{\partial}{\partial\xi^\alpha}
+\bar\xi_{\dot\alpha}\frac{\partial}{\partial\bar\xi_{\dot\alpha}}\right)\Phi(x)
\,\equiv\, iD\,\Phi(x) \,,
\\
i[{\mathbf M}_{\alpha\beta},\Phi(x)]=&\frac14\left(x_{\alpha\dot\gamma} {\partial_{\beta}}^{\dot\gamma}+
x_{\beta\dot\gamma} {\partial_{\alpha}}^{\dot\gamma}
-2\xi_{\alpha}\frac{\partial}{\partial{\xi^\beta}}-2\xi_{\beta}\frac{\partial}{\partial{\xi^\alpha}}
\right)\Phi(x)
\,\equiv\, iM_{\alpha\beta} \Phi(x)\,,
\\
i[ \bar{\mathbf  M}_{\dot\alpha\dot\beta},\Phi(x)]=&\frac14\left(x_{\gamma\dot\alpha}
{\partial^\gamma}_{\dot\beta}
+
x_{\gamma\dot\beta}{\partial^\gamma}_{\dot\alpha}
-
2\bar\xi_{\dot\alpha}\frac{\partial}{\partial{\bar\xi^{\dot\beta}}}-
2\bar\xi_{\dot\beta}\frac{\partial}{\partial{\bar\xi^{\dot\alpha}}}
\right)\Phi(x)
\,\equiv\, i\bar M_{\dot\alpha\dot\beta} \Phi(x)\,,
\\
i[{\mathbf K}_{\alpha\dot\alpha},\Phi(x)]=&
\left(x_{\alpha\dot\gamma} x_{\gamma\dot\alpha}\,\partial^{\gamma\dot\gamma}+
2\twist x_{\alpha\dot\alpha}+2\xi_\alpha {\bar x_{\dot\alpha}}^{\phantom{\alpha}\beta}
\frac{\partial}{\partial\xi^\beta}
+
2\bar\xi_{\dot\alpha} x_{\alpha\dot\beta}\frac{\partial}{\partial\bar\xi_{\dot\beta}}\right)\Phi(x)
\,\equiv\, iK_{\alpha\dot\alpha} \Phi(x)
\,,
\end{align}
\end{subequations}
where $\partial_{\alpha\dot\alpha}=\sigma^\mu_{\alpha\dot\alpha}\partial_\mu$ and $\twist=1$ is the
{\em geometric}\, twist \cite{Gross:1971wn}:
for the field with canonical scaling dimension $\ell^{\rm can}$
and Lorentz spin $(s,\bar s)$ it is defined as  $\twist = \ell^{\rm can}-s-\bar s$.
Note that we use boldface letters for the generators acting on quantum
fields to distinguish them from the
corresponding differential operators acting on the field coordinates. % cf. \cite{Braun:2003rp}.
The transformations of the gauge field
$A_\xi=A_{\alpha\dot\alpha}\xi^{\alpha}\bar\xi^{\dot\alpha}$ are given by the same
expressions with $\twist=0$.

In the applications of QCD to high-energy scattering the separation of transverse and longitudinal
degrees of freedom proves to be essential.
It is conveniently achieved by the introduction of two independent light-like vectors
\begin{align}
  n_{\alpha\dot\alpha}=\lambda_{\alpha}\bar\lambda_{\dot\alpha}\,, \qquad n^2=0\,,
\nonumber  \\
  \tilde n_{\alpha\dot\alpha}=\mu_\alpha\bar\mu_{\dot\alpha}\,, \qquad \tilde n^2 =0\,,
\label{n-ntilde}
\end{align}
which we choose to be normalized to
\begin{align}
(\mu\lambda)=-(\lambda\mu) = 1\,, \qquad (n\cdot\tilde n)=1/2 \,.
\end{align}
Without loss of generality one can take
\begin{align}
  \lambda^{\alpha}=(1,0)\,, \qquad  \lambda_\alpha=(0,1)\,,
\nonumber \\
 \mu^\alpha=(0,1)\,,\qquad  \mu_\alpha=(-1,0) \,.
\label{n-choice}
\end{align}
Then, for example
\begin{align}
   \partial^{2\dot 2} = 2(n\cdot\partial)\,, &&    \partial^{1\dot 1} = 2(\tilde n\cdot\partial)\,
\label{partial1122}
\end{align}
are the derivatives in the two chosen light-like directions whereas  the remaining two,
$\partial^{1\dot 2}$ and $\partial^{2\dot 1}$, are the derivatives in the transverse plane.

{}Fast moving hadrons can be viewed as a collection of partons that move in the same direction, 
say $\tilde n_\mu$. Whenever this picture applies, quantum fields ``living'' on the light ray
\begin{align}
   \Phi(x) \to \Phi(z n)
\end{align}
play a special role.
Such light-ray fields can be viewed as generating functions  for local
operators that arise through the (formal) Taylor expansion
\begin{align}
   \Phi(z) \equiv \Phi(z n) = \sum_k \frac{z^k}{k!}(n\partial)^k\Phi(0) =
   \sum_k \frac{z^k}{2^kk!}(\partial^{2\dot 2})^k\Phi(0) \,.
\label{light-ray-oper}
\end{align}
Note that all local operators on the r.h.s. of (\ref{light-ray-oper})
have the same collinear twist as the
field $\Phi$ itself since each $\partial^{2\dot 2}$ derivative adds one unit of dimension
and spin projection, simultaneously.
We will use a shorthand notation $\Phi(z)$ for $ \Phi(nz)$ in what follows.

With the restriction to light-ray operators the four-dimensional conformal transformations
are reduced to the collinear subgroup $SL(2,\mathbb{R})$ corresponding to
projective   (M\"obius)  transformations of the line~$x=zn$:
$$
z\to\frac{az+b}{cz+d}\,,\qquad ab-cd=1\,,
$$
where  $a,b,c,d$ are  real numbers.
The generators of the collinear subgroup, $S_\pm, S_0$ can be chosen as
\begin{align}\label{}
S_+=\frac{i}2(\mu\, K\,\bar\mu)\,, &&S_-=-\frac{i}2(\lambda\, P\,\bar\lambda)\,,&&
S_0=\frac{i}2\Big(D-\mu^\alpha\lambda^\beta M_{\alpha\beta}-
\bar M_{\dot\alpha\dot\beta}\bar\mu^{\dot\alpha}\bar\lambda^{\dot\beta}\Big)\,,
\end{align}
or, using the convention  in Eq.~(\ref{n-choice}),
\begin{align}\label{}
S_+=\frac{i}2K_{2\dot 2}=\frac{i}2K^{1\dot 1}\,, &&S_-=-\frac{i}2P^{2\dot2}=-\frac{i}2P_{1\dot 1}\,,&&
S_0=\frac{i}2\Big(D -M_{21}-\bar M_{\dot 1\dot 2}\Big)\,.
\end{align}
The explicit expressions are
\begin{subequations}
\begin{align}\label{generators}
S_+=&%\frac{i}2K_{2\dot 2}=%\frac{i}2K^{1\dot 1}=
\frac12 x_{2\dot\gamma}x_{\gamma\dot 2}\partial^{\gamma\dot\gamma}+
x_{2\dot 2}\left(t+\xi^\beta\frac{\partial}{\partial\xi^\beta}+
\bar\xi^{\dot\beta}\,\frac{\partial}{\partial\bar\xi^{\dot\beta}}\right)-
 x_{\beta\dot 2}\,\xi^\beta\frac{\partial}{\partial\xi^2}-x_{2\dot\beta}\,\bar\xi^{\dot\beta}
\frac{\partial}{\partial\bar\xi^{\dot 2}}\,,\\
S_-=&%-\frac{i}2P^{2\dot2}=%-\frac{i}2P_{1\dot 1}=
-\frac12\partial^{2\dot2}\,,\\
S_0=&%\frac{i}2\Big(D -M_{21}-\bar M_{\dot 1\dot 2}\Big)=
\frac12\left(x_{2\dot 2}\partial^{2\dot 2}+\frac12\left(x_{2\dot 1}\partial^{2\dot1}+x_{1\dot
2}\partial^{1\dot2}\right)+t
+\xi^1\dfrac{\partial}{\partial\xi^1}+\bar\xi^{\dot
1}\dfrac{\partial}{\partial\bar\xi^{\dot 1}}
\right)\,
\,.
\end{align}
\end{subequations}
They obey the standard commutation relations
\begin{align}\label{algebra}
[S_+,S_-]=2S_0\,, && [S_0,S_\pm]=\pm S_\pm\,.
\end{align}
In addition, there exist two operators  that commute with all $SL(2,\mathbb{R})$ generators:
% one defines two operators~--~
%the collinear twist operator $E$ and the helicity operator $H$~--~which commute with all
%$SL(2,\mathbb{R})$ generators, by
%
\begin{align}\label{E}
E=&{i}\Big( D +M_{21}+\bar M_{\dot 1\dot 2}\Big)
=
x_{1\dot 1}\partial^{1\dot 1}+\frac12\left(x_{2\dot 1}\partial^{2\dot1}+
x_{1\dot2}\partial^{1\dot2}+2t\right)
+\xi^2\dfrac{\partial}{\partial\xi^2}+\bar\xi^{\dot
2}\dfrac{\partial}{\partial\bar\xi^{\dot 2}}\,,
\\
\label{H}
H=&i(\bar M_{\dot 1\dot2}-M_{21})
=\frac12\left(x_{2\dot 1}\partial^{2\dot1}-x_{1\dot 2}\partial^{1\dot2}+
\xi^1\dfrac{\partial}{\partial\xi^1}-\xi^2\dfrac{\partial}{\partial\xi^2}
-\bar\xi^{\dot 1}\dfrac{\partial}{\partial\bar\xi^{\dot 1}}+
\bar\xi^{\dot 2}\dfrac{\partial}{\partial\bar\xi^{\dot 2}}
\right).
\end{align}
$E$ is usually called the collinear twist operator:  collinear twist $E$ counts the dimension of
the field minus spin projection, as opposed to the geometric twist $t$ which is  dimension minus spin.
In a slight abuse of language we will refer to $H$ as the helicity operator; the name can be justified
by observing that for ``good'' components of the fields (see below) the eigenvalue of $H$ coincides
with helicity of the corresponding one-particle state.

A light-ray operator with definite collinear twist $E$ transforms according to the
irreducible representation of the $SL(2,\mathbb{R})$ group with  the conformal spin
\begin{align}
    j = \ell^{\rm can} -E/2\,.
\end{align}
In particular the $SL(2)$ generators acquire their canonical form
\begin{align}\label{diff-form}
S_+=z^2\partial_z+2jz\,, && S_0=z\partial_z+j\,, && S_-=-\partial_z\,,
\end{align}
i.e. first order differential operators
acting on functions of the light-cone coordinate $z$. The finite form of the group transformations is
\begin{align}\label{Tg}
[T^j(g^{-1})\Phi](z)=\frac1{(cz+d)^{2j}}\Phi\left(\frac{az+b}{cz+d}\right)\,,\qquad
g=\begin{pmatrix}a&b\\c&d\end{pmatrix}.
\end{align}

{}For example, a chiral field $\psi$ should be decomposed as
\begin{align}
 \psi(z)=\lambda \,\psi_-(z)-\mu \,\psi_+(z)\,,
\end{align}
where
\begin{align}\label{}
\psi_+(z)=\lambda^\alpha\psi_\alpha(z)\equiv \psi_1(z)\,, && [E\psi_+](z) =
\phantom{2}\psi_+(z)\,,&&
[H\psi_+](z)=\frac12  \psi_+(z)\,,
\nonumber\\
\psi_-(z)=\mu^\alpha\psi_\alpha(z)\equiv\psi_2(z)\,,&& [E\psi_-](z) = 2 \psi_-(z)
\,,&&[H\psi_-](z)=-\frac12  \psi_-(z)\,.
\end{align}
Note that $iM_{21}$ appearing in (\ref{E}), (\ref{H})
counts the difference in the number of ``first'' and ``second'' spinor indices,
which is nothing but the
Lorentz spin projection on the light-ray direction.
In particular
 $\psi_+$ and $\psi_-$ correspond to spin projections $+1/2$ and $-1/2$, respectively.
Using explicit expressions in Eq.~(\ref{conf}) it is easy to check that  the fields
$\psi_+$ and $\psi_-$ indeed transform according to Eq.~(\ref{diff-form})
with the conformal spin $j=1$ and $j=1/2$, respectively.

Similarly, for the anti-chiral field $\bar\chi$ we define the ``plus'' and ``minus'' projections as
\begin{align}\label{}
\bar\chi_+\,=\,\bar\chi_{\dot\alpha}\bar\lambda^{\dot\alpha}\,\,,&&
\bar\chi_-\,=\,\bar\chi_{\dot\alpha}\bar\mu^{\dot\alpha}
\end{align}
and for the self-dual vector field $f_{\alpha\beta}$
\begin{align}
f_{++}(z)=&\lambda^\alpha\lambda^\beta \,f_{\alpha\beta}(z)\,,&
f_{+-}(z)=&\lambda^\alpha\mu^\beta \,f_{\alpha\beta}(z)\,,&
f_{--}(z)=&\mu^\alpha\mu^\beta \,f_{\alpha\beta}(z)\,.
\end{align}
The projections for the conjugate fields are defined as $\bar\psi_\pm=(\psi_\pm)^*$ etc.
The $SL(2,\mathbb{R})$ quantum numbers of the fundamental fields --- conformal spin,
(collinear) twist and helicity--- are collected in  Table~\ref{es}.
\begin{table}[t]
\begin{center}
\begin{tabular}{|c|c|c|c|c|c|c|c|}
\hline  \hline
&$\psi_+$&$\psi_-$ & $\bar\chi_+$&$\bar\chi_-$& $f_{++}$& $f_{+-}$& $f_{--}$\\
\hline \hline
$j$&$1$& $1/2$ & $1$& $1/2$ & $3/2$& $1$ &  $1/2$   \\
\hline
 $E$&$1$& $2$ & $1$& $2$ & $1$& $2$ &  $3$ \\
\hline
 $H$&$1/2$& $-1/2$ & $-1/2$& $1/2$ & $1$& $0$ &  $-1$ \\
 \hline     \hline
\end{tabular}
\end{center}
\caption{The $SL(2,\mathbb{R})$ spin and twist for the fundamental fields}
\label{es}
\end{table}

The plus components of the fields,
$\Phi_+(z)=\Phi_\lambda(zn)=\{\psi_+(z),\chi_+(z),f_{++}(z)\}$,
and their anti-chiral
counterparts --- "good" components in conventional terminology --- have the lowest twist.
The product of the plus fields taken at the different points on  the light-ray
$$
\Phi_+(z_1)\bar\Phi_+(z_2)\ldots\Phi_+(z_N)
$$
serves as a generating function\footnote{For the moment we ignore the color structure and
all issues related to gauge invariance.} for  the so-called quasipartonic
operators~\cite{Bukhvostov:1984as}. An operator constructed from $N$ "plus" fields has
collinear twist $E$ equal to $N$ which is  the lowest possible twist for $N-$particle
operators. The set of $N-$particle quasipartonic operators is
closed under renormalization at the one-loop level. The renormalization group equation can be reinterpreted as
a Schr{\"o}dinger equation where the scale $\mu$ plays the role of time. The
corresponding Hamiltonian  contains pairwise interactions only and can be written in
terms of the two-particle Casimir operators of the collinear conformal
group~\cite{Bukhvostov:1984as}.

The light-ray $N-$particle operators containing minus components of the fields, \hfill\break
$\{\psi_-,\bar\chi_-,f_{+-},\ldots\}$, have twist larger than N and provide one with
examples of operators
that are not quasipartonic. Renormalization of non-quasipartonic operators in QCD
has never been studied
systematically, to the best of our knowledge. On this way, there are two complications.

{}First, the number of fields (``particles'') is not conserved. 
To one-loop accuracy, the mixing matrix
of operators with a given twist $E$ has a block-triangular structure as the operators with less
fields can mix with ones containing more fields but not vice versa. Operators with the maximum
possible number of fields $N=E$ are quasipartonic.

Second, operators involving  minus field components can
mix with operators of the same twist  containing minus, $\partial^{1\dot 1}$, or
transverse, $\partial^{1\dot 2},\partial^{2\dot 1}$, derivatives.
These operators, therefore, also must be included.
The problem is that transverse derivatives generally do not have good
transformation properties with respect to the $SL(2,\mathbb{R})$ group. In concrete applications
it may be possible to get rid of such operators using EOM
and exploiting the specific structure of the matrix elements of interest,  e.g. if
there is no transverse momentum transfer between the initial and the final state.
Two well known examples are
the twist-four contributions to the deep-inelastic scattering (DIS) \cite{Jaffe:1982pm}
and to  meson distributions amplitudes \cite{Braun:1989iv,Ball:1998sk}.
The main problem as far as the operator renormalization
is concerned is that after this reduction conformal symmetry becomes obscured.
This procedure  is also not universal and probably cannot be applied beyond twist four.

In this work we suggest a different, general approach based on the
construction of a complete conformal operator basis for all twists. In this basis, the
$SL(2,{\mathbb R})$ symmetry of the renormalization group  equations is manifest.
To begin with, we will explain our construction on the example of a free chiral field $\psi$,
the extension to the other fields is straightforward.

Let us examine the action of the $SL(2,\mathbb{R})$ generators in Eq.~(\ref{diff-form}) on
the light-ray operator with a transverse or ``minus'' derivative, $[\partial^{1\dot 2}\psi_\pm](z)$,
$[\partial^{2\dot 1}\psi_\pm](z)$ and $[\partial^{1\dot 1}\psi_\pm](z)$. It is easy to see
that $S_0$ and
$S_-$ retain their form, and complications only arise in the case of $S_+$ which is related to
special conformal transformations:
\begin{align}\label{K-psi}
i[K_{\alpha\dot\alpha}\psi_\beta](x)=
\left(x_{\alpha\dot\gamma} x_{\gamma\dot\alpha}\,\partial^{\gamma\dot\gamma}+
4 x_{\alpha\dot\alpha}\right)\psi_\beta(x)-2x_{\beta\dot\alpha}\psi_\alpha(x)\,,
\end{align}
cf (\ref{conf}). In particular
\begin{align}\label{Kpm}
i[K_{2\dot2}\psi_-](x)=&
\left(x_{2\dot\gamma} x_{\gamma\dot2}\,\partial^{\gamma\dot\gamma}+
2 x_{2\dot2}\right)\psi_-(x)\,,
\nonumber\\
i[K_{2\dot2}\psi_+](x)=&
\left(x_{2\dot\gamma} x_{\gamma\dot2}\,\partial^{\gamma\dot\gamma}+
4 x_{2\dot2}\right)\psi_+(x)-2x_{1\dot2}\psi_-(x)\,.
\end{align}
The action of the ``spin-up'' generator $S_+=iK_{2\dot 2}/2$ on the light-ray operator with
a transverse derivative follows readily from Eq.~(\ref{Kpm}) observing that, e.g.
$$
\Big[{\mathbf K}_{2\dot 2},[\partial^{2\dot 1}\psi_\pm]\Big](z)\equiv
\Big(\partial^{2\dot 1}\Big[K_{2\dot 2}\psi_\pm]\Big](x)\Big)_{x=zn}
$$
and taking into account that $\partial^{\alpha\dot\alpha}
x_{\beta\dot\beta}=2\delta^\alpha_\beta \delta^{\dot\alpha}_{\dot\beta}$ and $x_{2\dot2}=z$.
One obtains
\begin{align}
 S_+[\partial^{2\dot 1}\psi_+](z)&=(z^2\partial_z+3z)[\partial^{2\dot 1}\psi_+](z)\,,
\nonumber\\
S_+[\partial^{1\dot 1}\psi_+](z)&=(z^2\partial_z+2z)[\partial^{1\dot 1}\psi_+](z)\,,
\nonumber\\
S_+[\partial^{1\dot 2}\psi_+](z)&=(z^2\partial_z+3z)[\partial^{1\dot 2}\psi_+](z)-2\psi_-(z)\,
\end{align}
and
\begin{align}
S_+[\partial^{2\dot 1}\psi_-](z)&=(z^2\partial_z+2z)[\partial^{2\dot 1}\psi_-](z)\,,
\nonumber\\
S_+[\partial^{1\dot 1}\psi_-](z)&=(z^2\partial_z+z)[\partial^{1\dot 1}\psi_-](z)\,,
\nonumber\\
 S_+[\partial^{1\dot 2}\psi_-](z)&=(z^2\partial_z+2z)[\partial^{1\dot 2}\psi_-](z)\,.
\end{align}
We see that the generator $S_+$ take the standard from~(\ref{diff-form}) for all cases
except for $[\partial^{1\dot 2}\psi_+](z)$.  Fortunately, this derivative can be eliminated
with the help of EOM (\ref{Dirac}):
\begin{align}
  \partial^{1\dot  2}\psi_+=-\partial^{2\dot 2}\psi_-\,,  &&
 \partial^{2\dot  1}\psi_-=-\partial^{1\dot 1}\psi_+\,.
\label{spinorEOM}
\end{align}
The first equation in (\ref{spinorEOM}) allows to replace all occurrences of
$\partial^{1\dot  2}\psi_+(z)$
by $-\partial^{2\dot  2}\psi_-(z)= -\partial_z\psi_-(z)$. The second one, in principle,
 can be used in either direction since $\partial^{2\dot  1}\psi_-$ and
$\partial^{1\dot 1}\psi_+$ both have ``good'' transformation properties. It turns out,
however, that eliminating $\partial^{2\dot  1}\psi_-$ in favor of $\partial^{1\dot
1}\psi_+$ is advantageous since it leads to a simpler complete operator basis in a
general situation and we adopt this option for what follows.
The remaining four independent operators $\partial^{1\dot 1}\psi_+$,  $\partial^{1\dot
1}\psi_-$, $\partial^{2\dot 1}\psi_+$ and $\partial^{1\dot 2}\psi_-$ transform according
to the irreducible representations of the collinear conformal group with spin $j=1$,
$j=1/2$, $j=3/2$ and $j=1$, respectively. Note that the ``minus'' derivative does not
change the conformal spin of the light-ray operator, whereas  a ``good'' transverse
derivative increases the spin by 1/2.

The above construction can be generalized for an arbitrary number of derivatives.
It is easy to verify that the following fields
%\begin{subequations}
\begin{align}\label{b-q}
\psi_{+}^{(j,m)}(z)%\equiv &\psi^{l}_{j,m}(z)
=&
[(\partial^{2\dot 1})^{2j-2} (\partial^{1\dot 1})^{2m} \psi_+](z)\,,
\nonumber \\
%q^{l-}_{j,m}(z)=&[(\partial^{2\dot 1})^{2j-1} (\partial^{1\dot 1})^{2m} q_-](z)\,,\\
\psi_{-}^{(j,m)}(z)%\equiv& \psi^{r}_{j,m}(z)
=&
[(\partial^{1\dot 2})^{2j-1} (\partial^{1\dot 1})^{2m} \psi_-](z)\,
\end{align}
%\end{subequations}
transform according to the spin-$j$ representation of the $SL(2,\mathbb{R})$ group,
Eq.~(\ref{diff-form}).
All other combinations of derivatives can be reduced to this basis with the help of EOM.
In particular, all pairs $\partial^{1\dot 2}\partial^{2\dot 1}$ can
be replaced by $\partial^{2\dot 2}\partial^{1\dot 1}$ which is a consequence of (\ref{spinorEOM}).

Next,  we consider which modifications have to be done in the (interacting) gauge theory.
In the first place  we have to replace ordinary derivatives by the covariant ones.
This is achieved  by modifying the definition of the light-ray operator
(\ref{light-ray-oper})
to include  the factor
\begin{align}
   \Phi(z) \to [0,z]\Phi(z)\,,
\end{align}
 where
\begin{align}
   [0,z] = \mbox{Pexp}\left[- \frac{1}{2}ig z \int_0^1 du\, A^{2\dot 2} (u z)\right]
\end{align}
is the light-like Wilson line in the appropriate (fundamental or adjoint)
representation of the color group.
In this way the Taylor expansion goes over covariant derivatives:
\begin{align}
  [0,z]\Phi(z) = \sum_k \frac{z^k}{2^kk!}(D^{2\dot 2})^k\Phi(0) \,.
\end{align}
In what follows the $[0,z]$--factors are not shown for brevity, but they are always implied.
Note that dropping  the gauge links  can be viewed as  going over to the Fock-Schwinger gauge
$x_{\alpha \dot \alpha} A^{\alpha \dot \alpha} (x) =0$, $A^{\alpha\dot\alpha}(0)=0$,
or, alternatively, the light-cone gauge $A^{2\dot 2} =0$.

In addition,  we have to replace ordinary derivatives by covariant ones in Eqs.~(\ref{b-q}).
All  relations which we have used to reduce an arbitrary combination of derivatives to
this particular form
hold true  up to commutator terms $[D^{\alpha\dot\alpha}, D^{\beta\dot\beta}]$ which
can be expressed in terms of gluon field strength. Such terms
contain two or more fundamental light-ray fields and do not  affect the proof of
completeness in the  one-particle sector which we are considering at present.

We still have to check, however, that the replacement $\partial\to D$ does not spoil transformation
properties of the basis fields~(\ref{b-q}).
The special conformal transformation for the gauge field $A$ takes the form
\begin{align}
i[K_{\alpha\dot\alpha} A_{\beta\dot\beta}](x)=
x_{\alpha\dot\gamma}x_{\gamma\dot\alpha}\partial^{\gamma\dot\gamma} A_{\beta\dot\beta}(x)+
2\left(x_{\alpha\dot\alpha}
A_{\beta\dot\beta}(x)-x_{\beta\dot\beta}A_{\alpha\dot\alpha}(x)\right)-
2\epsilon_{\alpha\beta}\epsilon_{\dot\alpha\dot\beta} (x^{\gamma\dot\gamma} A_{\gamma\dot\gamma}(x))\,,
\end{align}
and for the components of interest  becomes
%
%\begin{subequations}
\begin{align}\label{}
i[K_{2\dot 2}A^{1\dot1}](x)=&x_{2\dot\gamma}
x_{\gamma\dot 2}\,\partial^{\gamma\dot\gamma} A^{1\dot1}(x)\,,
\nonumber\\
i[K_{2\dot 2}A^{2\dot1}](x)=&\left(x_{2\dot\gamma}
x_{\gamma\dot 2}\,\partial^{\gamma\dot\gamma}
+2x_{2\dot 2} \right)A^{2\dot 1}(x)-2x^{2\dot 1} A_{2\dot 2}(x)\,,
\nonumber\\
i[K_{2\dot 2}A^{1\dot2}](x)=&\left(x_{2\dot\gamma}
x_{\gamma\dot 2}\,\partial^{\gamma\dot\gamma}
+2x_{2\dot 2} \right)A^{1\dot 2}(x)-2x^{1\dot 2} A_{2\dot 2}(x)\,.
\end{align}
%\end{subequations}
%
Using these expressions it is easy to check (by induction in $m$ and $j$) that, e.g. for
$\psi_+^{j,m}(x)=\bigl(D^{2\dot1}\bigr)^{2j-2} \bigl(D^{1\dot1}\bigr)^{2m}\psi_+(x)$,
the transformation is
\begin{align}\label{}
i\left[K_{2\dot 2},\psi_+^{j,m}\right](x)=
 \left(x_{2\dot\gamma}
x_{\gamma\dot 2}\,\partial^{\gamma\dot\gamma}
+2jx_{2\dot 2} \right) \psi_+^{j,m}(x)+x_{1\dot 2} {\mathcal G}\,,
\end{align}
where ${\mathcal G}$ is a light-ray operator containing some combination of the chiral field $\psi$,
gauge field $A$ and derivatives.
This inhomogeneous term vanishes on the light-cone, $x=zn$, so that one ends up with
\begin{align}
S_+\,\psi_+^{j,m}(z)=(z^2\partial_z+2j z)\psi_+^{j,m}(z)\,,
\end{align}
as required.

The complete basis of one-particle light-ray operators for chiral quark and self-dual gluon
fields in QCD contains seven fields:
%
%\begin{subequations}
\begin{align}
\label{basis}
\psi^{(j,m)}_+(z) &= \bigl(D^{2\dot1}\bigr)^{2j-2} \bigl(D^{1\dot1}\bigr)^{2m} \psi_1(z)\,,
\nonumber\\
\psi^{(j,m)}_-(z)&=  \bigl(D^{1\dot2}\bigr)^{2j-1} \bigl(D^{1\dot1}\bigr)^{2m}\psi_2(z)\,,
\nonumber\\
\bar\chi_{+}^{(j,m)}(z)&=
\bigl(D^{1\dot 2}\bigr)^{2j-2} \bigl(D^{1\dot 1}\bigr)^{2m} \bar\chi_{\dot1}(z)\,,
\nonumber\\
\bar\chi_{-}^{(j,m)}(z)&=
\bigl(D^{2\dot 1}\bigr)^{2j-1} \bigl(D^{1\dot 1}\bigr)^{2m}\bar\chi_{\dot2}(z)\,,
\nonumber\\
f_{++}^{(j,m)}(z)&=
\bigl(D^{2\dot1}\bigr)^{2j-3} \bigl(D^{1\dot1}\bigr)^{2m} f_{11}(z)\,,
\nonumber\\
f_{--}^{(j,m)}(z)&=
 \bigl(D^{1\dot2}\bigr)^{2j-1} \bigl(D^{1\dot1}\bigr)^{2m}f_{22}(z)\,,
\nonumber\\
f_{+-}^{(1,m)}(z)&= \bigl(D^{1\dot1}\bigr)^{2m} f_{12}(z)\,.
\end{align}
%\end{subequations}
%
The field carrying the superscript $j$ transforms according to the representation
$T^j$ of the $SL(2,{\mathbb R})$ group, see Eq.~(\ref{Tg}).
Note that ordering of the covariant derivatives in (\ref{basis}) does not affect
the transformation properties. The twist $E$ and helicity $H$ take the following values:
%
%\begin{subequations}
\begin{align}\label{}
E\,\psi^{(j,m)}_\pm=&\bigl(2j+4m\mp1\bigr)\psi^{(j,m)}_\pm\,,&
E\,\bar\chi^{(j,m)}_\pm=&\bigl(2j+4m\mp1\bigr)\bar\chi^{(j,m)}_\pm\,,
\nonumber\\
E\,f_{\pm\pm}^{(j,m)}=&\bigl(2j+4m\mp2\bigr)f_{\pm\pm}^{(j,m)}\,, &
E\,f_{+-}^{(1,m)}=&\bigl(2+4m\bigr)f_{+-}^{(1,m)}\,,
\\
H\,\psi^{(j,m)}_\pm=&\pm\left(2j-1\mp\frac12\right)\psi^{(j,m)}_\pm\,,&
H\,\bar\chi^{(j,m)}_\pm=&\mp\left(2j-1\mp\frac12\right)\psi^{(j,m)}_\pm\,,
\notag\\
H\,f^{(j,m)}_{\pm,\pm}=&\pm\bigl(2j-1\mp 1\bigr)\,f^{(j,m)}_{\pm,\pm}\,,&
H\,f_{+-}^{(1,m)}=&0\,,
\end{align}
%\end{subequations}
%
The basis fields in the antichiral sector can be defined  as $\bar\psi_+^{(s,m)}=(\psi_+^{(s,m)})^*$ and
similarly for all other cases.

The proof that the fields in (\ref{basis}) form a complete basis in the one-particle sector essentially
follows the above discussion of a chiral field.  To this end one can consider the
derivatives as commuting ones and assume that the fundamental fields satisfy ``free'' EOM.
One has to demonstrate  that all possible combinations of derivatives acting on the self-dual strength
tensor can be reduced to the combinations appearing in  (\ref{basis}).
This can be achieved by inspection.  The first step, as above, is to get rid of all pairs
$D^{1\dot 2} D^{2\dot 1}$ replacing them by
$D^{2\dot 2} D^{1\dot 1}\to \partial_z D^{1\dot 1}$,
and then check that all remaining combinations can be rewritten in the desired form, e.g.
\begin{align}
\bigl(D^{1\dot2}\bigr)^{2k}\bigl(D^{1\dot1}\bigr)^{2m} f_{12}(z)
 &\to
-\partial_z \bigl(D^{1\dot2}\bigr)^{2k-1}\bigl(D^{1\dot1}\bigr)^{2m}f_{22}(z)+O(f^2)
\nonumber\\
\bigl(D^{2\dot1}\bigr)^{2k}\bigl(D^{1\dot1}\bigr)^{2m}
f_{12}(z)
&\to
 -\bigl(D^{2\dot1}\bigr)^{2k-1}\bigl(D^{1\dot1}\bigr)^{2m+1}f_{22}(z)+O(f^2)\,,
\end{align}
etc.

Finally, taking a color-singlet product of  the basis fields defined in Eq.~(\ref{basis}), \hfill\break
$ \Phi^{j,m} = \{\psi_\pm^{j,m},\ldots, f_{+-}^{(j=1,m)}\}$, and their antichiral counterparts,
$\bar \Phi^{j,m}$,  at different light-ray positions $z_1,\ldots,z_N$, one obtains a complete
basis of gauge-invariant $N$-particle operators
\begin{align}
  \mathcal{O}(z_1,\ldots,z_N) &= \Phi^{j_1,m_1}(z_1)\ldots \Phi^{j_N,m_N}(z_N)
\label{Qnonlocal}
\end{align}
that transform according to the representation $T^{j_1}\otimes\ldots\otimes T^{j_N}$  of the
collinear conformal  group $SL(2,\mathbb{R})$ and serve as generating functions
for towers of the local operators of  twist $E = E_1+\ldots+E_N$.
If $E>N$ then these operators get mixed under renormalization with the operators
of the same twist $E$ and the number of fields ranging from $N$ to $E$.  Hence the
mixing matrix has a block-triangular form.
The anomalous dimensions are determined by the diagonal blocks only,
the off-diagonal blocks are, however,  important for the construction of multiplicatively
renormalizable operators.
The premium and main rationale for using the conformal
basis (\ref{basis}) is that the  $SL(2,\mathbb{R})$ symmetry imposes severe
constraints on the form of the kernels and also allows one to apply many of the
technical tools that were developed earlier for quasipartonic operators.
The explicit construction of this basis presents one of the main results of this paper.

Since the maximum light-cone spin projection coincides, obviously, with the Lorentz
spin, quasipartonic operators have definite geometric twist $T=E=N$. On the other hand,
non-quasipartonic operators contain both $T=E$ contributions and those with a lower
twist, $T<E$. Operators with different values of $T$ do not mix.
Thus, introducing operators with different geometrical twist
would bring the mixing matrix in the block-diagonal form at the cost, however, that
the  $SL(2,\mathbb{R})$ symmetry of the kernels is  lost.
A better strategy is to separate the (highest) geometric twist of interest by imposing
the appropriate symmetry conditions on the solutions of the
renormalization group equation for the operators in (\ref{Qnonlocal}) and
maintain  the $SL(2,\mathbb{R})$ covariance.

{}For illustration, let us consider a simple example: renormalization of twist-3
operators that one encounters in the study of chiral odd  pion distribution amplitudes
\cite{Braun:1989iv}. The complete set includes in this case three $E=3$ light-ray operators
\begin{eqnarray}
   {\mathcal O}_1(z_1,z_2) &=& \chi_+(z_1)\psi_-(z_2)\,,
\nonumber\\
   {\mathcal O}_2(z_1,z_2) &=&  \chi_-(z_1)\psi_+(z_2)\,,
\nonumber\\
{\mathcal O}_3(z_1,z_2,z_3)&=&\chi_+(z_1)\bar f_{++}(z_2) \psi_+(z_3)\,
\label{example:1}
\end{eqnarray}
that transform according to the representations $T^{j=1}\otimes T^{j=1/2}$,
$T^{j=1/2}\otimes T^{j=1}$ and $T^{j=1}\otimes T^{j=3/2}\otimes T^{j=1}$ of the collinear
conformal  group $SL(2,\mathbb{R})$, respectively. The renormalization group equation can
be written, schematically, as
\begin{align}
\left\{\mu \frac{\partial}{\partial \mu}+\beta(g)\frac{\partial }{\partial g} \right\}
\begin{pmatrix}
{\mathcal O}_1\\
{\mathcal O}_2\\
{\mathcal O}_3
\end{pmatrix} &= -\frac{\alpha_s}{2\pi}
\begin{pmatrix}
{\mathcal H}_{1,1}& {\mathcal H}_{1,2} & {\mathcal H}_{1,3}\\
{\mathcal H}_{2,1}& {\mathcal H}_{2,2} & {\mathcal H}_{2,3}\\
       0 &  0 & {\mathcal H}_{3,3}
\end{pmatrix}
\begin{pmatrix}
{\mathcal O}_1\\
{\mathcal O}_2\\
{\mathcal O}_3
\end{pmatrix}\,,
\label{example:2}
\end{align}
where the kernels ${\mathcal H}_{2,3}$ have simple $SL(2,\mathbb{R})$ transformation properties.
The operator ${\mathcal O}_3$ is  quasipartonic with $T=E=3$, the other two are non-quasipartonic
and do not have definite geometric twist.

{}For example, consider a generic local operator obtained by the Taylor expansion
of the light-ray operator  ${\mathcal O}_1$
$$
\lambda^{\alpha}\lambda^{\alpha_1}\ldots\lambda^{\alpha_k}\,\mu^{\beta}\,
\bar\lambda^{\dot\alpha_{1}}\ldots\bar\lambda^{\dot\alpha_{\bar k }}
\chi_{\alpha} D_{\alpha_1\dot\alpha_1}\ldots D_{\alpha_k\dot\alpha_k} \psi_\beta
$$
It is symmetric in all dotted indices, but does not have definite symmetry in the undotted: $\beta$
can be either symmetrized or antisymmetrized with (either) one of the $\alpha, \alpha_1,\ldots\alpha_k$.
These two possibilities correspond to picking up contributions of different geometric twist
$T=2$ and $T=3$, respectively. Going over to local operators is in fact not necessary as the
separation of contributions of different symmetry can be achieved in the nonlocal form:
\begin{align}
  {\mathcal O}_{1,2}^{T=2} &= \lambda^\alpha \frac{\partial}{\partial \mu^\alpha}{\mathcal O}_{1,2} =
\chi_+(z_1)\psi_+(z_2)\,,
\nonumber\\
 {\mathcal O}_{1,2}^{T=3}&=\epsilon^{\alpha\beta}\frac{\partial}{\partial\mu^\beta}
\frac{\partial}{\partial\lambda^{\alpha}} {\mathcal O}_{1,2}\,.
\label{example:3}
\end{align}
It is convenient to introduce
 \begin{align}
  {\mathcal O}_\pm (z_1,z_2) =  {\mathcal O}_1 (z_1,z_2) \pm  {\mathcal O}_2 (z_1,z_2)\,.
\end{align}
The operator ${\mathcal O}_-$ can be written as
\begin{align}
  {\mathcal O}_- (z_1,z_2) =  (\lambda\mu)\,
 \chi^\alpha(z_1)\psi_\alpha(z_2)
\end{align}
and is pure twist $T=3$, whereas for ${\mathcal O}_+$  inverting the
relations in (\ref{example:3}) one obtains after some algebra
\begin{align}\label{red2}
{\mathcal O}_+(z_1,z_2)=\mu^\alpha\frac{\partial}{\partial\lambda^{\alpha}}\int_0^1{d\tau}\,
 {\mathcal O}_+^{T=2}(\tau z_1,\tau z_2)-
(\mu\lambda) \int_0^1d\tau\,\tau\,{\mathcal O}_+^{T=3}(\tau z_1,\tau z_2)\,.
\end{align}	
Using ${\mathcal O}_-$ and ${\mathcal O}_+^{T=3}$ as basis fields instead of
${\mathcal O}_1$ and ${\mathcal O}_2$ one can avoid the contamination by twist-two
operators altogether. The problem is that the renormalization group equation will in this
form involve linear combinations of the kernels (\ref{example:2}),  of the type
${\mathcal H}_{11}\pm   {\mathcal H}_{12}$ etc.,
with different $SL(2,{\mathbb R}) $ transformation properties.

In the present case the problem can be simplified drastically using the operator
identities \cite{Braun:1989iv} that
%\begin{align}
% {\mathcal O}_+(n,-n)=&\mu^\alpha\frac{\partial}{\partial\lambda^{\alpha}}\int_0^1{d\tau}
% O^{t=2}(\tau n ,-\tau n)+\text{total derivatives}+
%\nonumber\\ &
%+(\mu\lambda)2ig \int_0^1d\tau\,\tau\,\int_{-\tau}^\tau ds\,s\, \chi_+(\tau n)\bar f_{++}(s
%n)
%\psi_+(-\tau n)\,,
%\nonumber\\
% {\mathcal O}_-(n,-n) = & (\lambda\mu)\, \Big(
%                                     \chi^\alpha(0)\psi_\alpha(0) +\text{total derivatives}
%\notag
%\\
%&
%-g \int_0^1d\tau\,\int_{-\tau}^\tau ds\, \chi_+(\tau n)\bar f_{++}(s n)\psi_+(-\tau n)
%+\text{total derivatives}\Big)\,.
%\end{align}
allow to rewrite both two-particle operators ${\mathcal Q}_+^{T=3}$ and  ${\mathrm Q}_3^{T=3}$
in terms of ${\mathcal O}_3$ (up to a local term $\chi^\alpha(0)\psi_\alpha(0)$ \cite{Braun:1989iv})
so that they do not need to be considered separately.
In this way the matrix renormalization group equation (\ref{example:2}) is reduced to the
single term ${\mathcal H}_{33}$.  Unfortunately, a similar reduction
to the quasipartonic sector does not hold in the general situation.

%%%%%%%%%%%%%%%%%%%%%%%%%%%%%%%%%%%%%%%%%%%%%%%%%%%%%%%%%%%%%%%%%%%%%%%%%%%%%%%%%%%%%%%%%%%%%%%%%%%%
%%%%%%%%%%%%%%%%%%%%%%%%%%%%%%%%%%%%%%%%%%%%%%%%%%%%%%%%%%%%%%%%%%%%%%%%%%%%%%%%%%%%%%%%%%%%%%%%%%%%
%
\section{Complete Operator Basis for twist-4}\label{sect3}
%
%%%%%%%%%%%%%%%%%%%%%%%%%%%%%%%%%%%%%%%%%%%%%%%%%%%%%%%%%%%%%%%%%%%%%%%%%%%%%%%%%%%%%%%%%%%%%%%%%%%%
%%%%%%%%%%%%%%%%%%%%%%%%%%%%%%%%%%%%%%%%%%%%%%%%%%%%%%%%%%%%%%%%%%%%%%%%%%%%%%%%%%%%%%%%%%%%%%%%%%%%

After this general discussion we proceed to the systematic study of the renormalization of QCD baryon
operators of twist-4. From the theory side, this is the simplest example where non-quasipartonic
operators enter nontrivially and cannot be excluded by using EOM; a generalization of the
approach of Ref.~\cite{Bukhvostov:1985rn} to such situations is our primary goal.
The main application of twist-4 baryon operators to QCD phenomenology has been to the studies
of hard exclusive reactions involving  helicity flip, for example the Pauli electromagnetic
form factor of the nucleon $F_2(Q^2)$ using pQCD factorization \cite{Belitsky:2002kj} or
light-cone sum rules \cite{Braun:2001tj}. The necessary nonperturbatuve input is given
in this case by the three nucleon (proton) distribution amplitudes $\Phi_4,\Psi_4$ and $\Xi_4$
defined in Ref.~\cite{Braun:2000kw} as matrix elements of twist-4 three-quark operators.
They can be represented in spinor notation as follows:
%
%\begin{subequations}\label{definitions}
\begin{align}
\vev{0|\epsilon^{ijk}\psi_+^{u,i}(z_1)\bar\chi_+^{u,j}(z_2)\psi^{d,k}_-(z_3)|P}=&
-\frac14(\mu\lambda)\,
m_N N^{\uparrow}_+\int \mathcal{D}x \, e^{-i(pn)\,\sum {x_iz_i}}\Phi_4(x)\,,
%\phantom{-}\,\frac12(pn)
%m_N N^{\downarrow}_-\int \mathcal{D}x \, e^{-i(pn)\,\sum {x_iz_i}}\Phi_4(x)\,,
\nonumber\\
\vev{0|\epsilon^{ijk}\,\bar\chi_+^{u,i}(z_1)\, \psi_-^{u,j}(z_2)\, \psi_+^{d,k}(z_3) \,|P}=&
-\frac14(\mu\lambda)\,
m_N N^{\uparrow}_+\int \mathcal{D}x \, e^{-i(pn)\,\sum {x_iz_i}}\Psi_4(x)\,,
\nonumber\\
\vev{0|\epsilon^{ijk}\psi_-^{u,i}(z_1)\psi_+^{u,j}(z_2)\psi^{d,k}_+(z_3)|P}=&
 - \frac14(\mu\lambda)\,m_N
N^{\downarrow}_+\int \mathcal{D}x \, e^{-i(pn)\,\sum {x_iz_i}}\Xi_4(x)\,.
\label{definitions}
\end{align}
%\end{subequations}
%
{}For comparison, the leading twist-3 distribution amplitude can be defined as
\begin{align}\label{def:Phi3}
 \vev{0|\epsilon^{ijk}\psi_+^{u,i}(z_1)\bar\chi_+^{u,j}(z_2)\psi^{d,k}_+(z_3)|P}=&\phantom{-}
\frac12 (p n)\, N^{\downarrow}_+\,\int \mathcal{D}x \, e^{-i(pn)\,\sum {x_iz_i}}\Phi_3(x)\,,
\end{align}
and only involves ``plus'' quark fields.
In addition, we introduce three independent twist-4 four-particle distribution amplitudes
involving a gluon field:
%
%\begin{subequations}\label{Gdefinitions}
\begin{align}
\vev{0|ig\epsilon^{ijk}\psi_+^{u,i}(z_1)\bar\chi_+^{u,j}(z_2)[\bar f_{++}(z_4)\psi^{d}_+(z_3)]^k|P}=&
\frac14
m_N (pn)^2 N^{\uparrow}_+\int \mathcal{D}x \, e^{-i(pn)\,\sum {x_iz_i}}\,\Phi^g_4(x)\,,
\nonumber\\
\vev{0|ig\epsilon^{ijk}\,\bar\chi_+^{u,i}(z_1)\, [\bar f_{++}(z_4)\psi_+^{u}(z_2)]^j\,
\psi_+^{d,k}(z_3) \,|P}=&
\frac14
m_N (pn)^2 N^{\uparrow}_+\int \mathcal{D}x \, e^{-i(pn)\,\sum {x_iz_i}}\,\Psi^g_4(x)\,,
\nonumber\\
\vev{0|ig\epsilon^{ijk} [\bar f_{++}(z_4) \psi^{u}_+(z_1)]^i\, \psi^{u,j}_+(z_2)\,
\psi^{d,k}_+(z_3)|P}=
&
\frac14
m_N (pn)^2 N^{\downarrow}_+\int \mathcal{D}x \, e^{-i(pn)\,\sum {x_iz_i}}\,\Xi^g_4(x)\,.
\label{Gdefinitions}
\end{align}
%\end{subequations}
%
Here the first and the second superscripts of $\psi^{a,i}$ ($\bar\chi^{a,i}$) are the flavor,
$a=u,d$, and color, $i=1,2,3$ indices of the quark field, respectively;
 $m_N$ and $p_\mu$ are the nucleon mass and momentum,
and $N^{\uparrow(\downarrow)}= (1/2)(1\pm\gamma_5)\,N(p)$ is the antichiral (chiral) part
of the Dirac spinor. %
%\footnote{Alternatively $
%\vev{0|\epsilon^{ijk}\psi_+^{u,i}(z_1)\bar\chi_-^{u,j}(z_2)\bar\chi^{d,k}_+(z_3)|P^\uparrow}=
%\phantom{-}\frac14(\bar\mu\bar\lambda)m_N
%N^{\uparrow}_+\int \mathcal{D}x \, e^{-i(pn)\,\sum {x_iz_i}}\Psi_4(x)\,.
%$}
 In order to keep the auxiliary spinors $\lambda,\mu$ dimensionless
we choose
$$
n=m_N^{-1}\lambda\otimes\bar\lambda,\qquad
\tilde n =m_N\,\mu\otimes\bar\mu\,.
$$
The distribution amplitudes depend on the
set of parton momentum fractions $x = \{x_1,\ldots,x_n \}$ and the integration measure is defined as
\begin{align}
\label{Dx}
 \int {\mathcal D}x = \int_0^1 dx_1\ldots dx_n\, \delta(1-\sum x_k)\,.
\end{align}
We hope that using the same notation $\int {\mathcal D}x$  for the three-particle and the four-particle
integration measure will not create confusion.
We  keep the factors $(\mu\lambda)=1$ on r.h.s. of Eq.~(\ref{definitions}) to maintain
the balance between the spinors $\mu$ and $\lambda$ on the both sides.

The scale dependence of the distribution amplitudes is driven by the renormalization of
the nonlocal light-ray
operators on the l.h.s. of Eq.~(\ref{definitions}). Our first task is to construct the
complete operator basis.
Note that  $\Xi_4$ involves chiral quarks only whereas
$\Phi_4$ and $\Psi_4$  involve both chiral and antichiral fields. Since chirality is
conserved in QCD perturbation theory,
there is no mixing and the two cases (pure chirality and mixed chirality operators) can be
considered separately.

%%%%%%%%%%%%%%%%%%%%%%%%%%%%%%%%%%%%%%%%%%%%%%%%%%%%%%%%%%%%%%%%%%%%%%%%%%%%%%%%%%%%%%%%%%%%%%%
\subsection{ Chiral operators}
%%%%%%%%%%%%%%%%%%%%%%%%%%%%%%%%%%%%%%%%%%%%%%%%%%%%%%%%%%%%%%%%%%%%%%%%%%%%%%%%%%%%%%%%%%%%%%%
The distribution amplitude $\Xi_4$ is related to the matrix element of the operators with
collinear twist $E=4$ and helicity $H=1/2$
\begin{align}\label{Qoperators}
Q_1(z_1,z_2,z_3)=&\epsilon^{ijk} \psi^{a,i}_-(z_1)\, \psi^{b,j}_+(z_2)\, \psi^{c,k}_+(z_3)\,,
\nonumber\\
Q_2(z_1,z_2,z_3)=&\epsilon^{ijk} \psi^{a,i}_+(z_1)\, \psi^{b,j}_-(z_2)\, \psi^{c,k}_+(z_3)\,,
\nonumber\\
Q_3(z_1,z_2,z_3)=&\epsilon^{ijk} \psi^{a,i}_+(z_1)\, \psi^{b,j}_+(z_2)\, \psi^{c,k}_-(z_3)\,.
\end{align}
For the discussion of the renormalization it is convenient to consider
the three quark fields of different flavor  in which case the
three light-ray operators in (\ref{Qoperators}) are independent.  We do not assign
flavor indices to $Q_i$ assuming that the flavors are always ordered as in the above
expressions.

The three-quark operators in (\ref{Qoperators}) mix with quasipartonic operators
containing an additional gluon field  $F_{+,\mu\bar\lambda}=-(\mu\lambda) \bar f_{++}$
\begin{align}\label{Goperators}
{G}_1(z_1,z_2,z_3,z_4)=&
        ig\epsilon^{ijk}(\mu\lambda)\,
 [\bar f_{++}(z_4) \psi^{a}_+(z_1)]^i\, \psi^{b,j}_+(z_2)\, \psi^{c,k}_+(z_3)\,,
\notag\\
{G}_2(z_1,z_2,z_3,z_4)=&
           ig\epsilon^{ijk}(\mu\lambda)\,
 \psi^{a,i}_+(z_1)\, [\bar f_{++}(z_4)\psi^{b}_+(z_2)]^{j}\,
\psi^{c,k}_+(z_3)\,,
\notag\\
{G}_3(z_1,z_2,z_3,z_4)=&
        ig\epsilon^{ijk}(\mu\lambda)\,
 \psi^{a,i}_+(z_1)\, \psi^{b,j}_+(z_2)\,[\bar f_{++}(z_4) \psi^{c}_+(z_3)]^k\,,
\end{align}
which are, however, not all independent because of the identity
\begin{align}\label{3gidentity}
{G}_1(z_1,z_2,z_3,z_4)+{G}_2(z_1,z_2,z_3,z_4)+{G}_3(z_1,z_2,z_3,z_4)=0\,.
\end{align}
Recall that $\psi_-$, $\psi_+$  and $\bar f_{++}$ have  conformal spins
$j=1/2$, $j=1$ and $j=3/2$, respectively. Hence $Q_1$ transforms
according to the representation $T^{(1/2)}\otimes T^{(1)}\otimes T^{(1)}$,
of the $SL(2,\mathbb{R})$ group (and similar for $Q_2,Q_3$), whereas  $G_i$
belong to the $T^{(1)}\otimes T^{(1)}\otimes T^{(1)}\otimes T^{(3/2)}$ representation..

{}For the matrix elements of the operators $Q_i$ and $G_i$  between the vacuum and proton state
there are more relations that follow from identity of the two $u$-quarks in the proton and
the requirement that the nucleon has isospin $1/2$. Let
\begin{align}
q_i(z_1,z_2,z_3)&=\vev{0|Q_i(z_1,z_2,z_3)|P}\,,
\nonumber\\
g_i(z_1,z_2,z_3,z_4)&=\vev{0|G_i(z_1,z_2,z_3,z_4)|P}\,,
\end{align}
where we put $a=b=u$ and $c=d$.  One finds
\begin{align}\label{q-symmetry}
q_2(z_1,z_2,z_3)&=q_1(z_2,z_1,z_3)\,,
\nonumber\\
q_3(z_2,z_3,z_1)&=-q_1(z_1,z_2,z_3)-q_1(z_1,z_3,z_2)\,
\end{align}
and similarly
\begin{align}\label{ggr-1}
g_2(z_1,z_2,z_3,z_4)&=g_1(z_2,z_1,z_3,z_4)\,,
\nonumber\\
g_3(z_2,z_3,z_1,z_4)&=-g_1(z_1,z_2,z_3,z_4)-g_1(z_1,z_3,z_2,z_4)\,.
\end{align}
Taking into account the identity~(\ref{3gidentity}) it follows that the remaining
independent function $g_1$ satisfies the
symmetry relation
\begin{align}\label{ggr-2}
g_1(z_1,z_2,z_3,z_4)-g_1(z_3,z_2,z_1,z_4)=
g_1(z_3,z_1,z_2,z_4)-g_1(z_2,z_1,z_3,z_4)\,.
\end{align}
Going over to the momentum fraction representation, we define a
new  twist-4 four-particle proton distribution amplitude $\Xi_4^g$ in Eq.~(\ref{Gdefinitions}).

%%%%%%%%%%%%%%%%%%%%%%%%%%%%%%%%%%%%%%%%%%%%%%%%%%%%%%%%%%%%%%%%%%%%%%%%%%%%%%%%%%%%%%%%%%%%%%%%%%%%
%
\subsection{Mixed chirality operators}
%
%%%%%%%%%%%%%%%%%%%%%%%%%%%%%%%%%%%%%%%%%%%%%%%%%%%%%%%%%%%%%%%%%%%%%%%%%%%%%%%%%%%%%%%%%%%%%%%%%%%%
The distribution amplitudes $\Phi_4$ and $\Psi_4$  are given by  the matrix elements of
the operators with
collinear twist $E=4$ and helicity $H=-1/2$.
There exist  three independent three-quark operators with these quantum numbers:
\begin{align}\label{qbx}
{\mathcal{Q}}_1(z_1,z_2,z_3)=&
\epsilon^{ijk}\,\psi_-^{a,i}(z_1)\, \psi_+^{b,j}(z_2)\,\bar\chi_+^{c,k}(z_3)\,,\notag\\	
\mathcal{Q}_2(z_1,z_2,z_3)=&
\epsilon^{ijk}\,\psi_+^{a,i}(z_1)\,\psi_-^{b,j}(z_2)\,\bar\chi_+^{c,k}(z_3)\,,\notag\\
\mathcal{Q}_3(z_1,z_2,z_3)=& \frac12
%-\epsilon^{ijk}\,\psi_+^{i,a}(z_1)\,\psi_+^{j,b}(z_2)\,[D_{\mu\bar\lambda}\bar\chi_+^{k,c}](z_3)=
\epsilon^{ijk}\,\psi_+^{a,i}(z_1)\,\psi_+^{b,j}(z_2)\,
[\bar\chi_+^{3/2}]^{c,k}(z_3)\,,
\end{align}
%
%({\it This $Q_3$=$2Q_3^{my\, notes}$}.)
where $\bar \chi_+^{3/2}\equiv \bar \chi_+^{(3/2,0)} = -(\mu D\bar\lambda)\bar\chi_+
\equiv  - D_{\mu\dot\lambda} \bar\chi_+$, cf. Eq.~(\ref{basis}).
In addition there are three operators containing an extra gluon field
\begin{align}\label{qbg}
\mathcal{G}_1(z_1,z_2,z_3,z_4)=&
ig\epsilon^{ijk}\,(\mu\lambda)\,
[\bar f_{++}(z_4)\psi_+^a(z_1)]^i\,  \psi_+^{b,j}(z_2)\,\bar\chi_+^{c,k}(z_3)\,,
\notag\\
\mathcal{G}_2(z_1,z_2,z_3,z_4)=&
ig\epsilon^{ijk}\,(\mu\lambda)\,
\psi_+^{a,i}(z_1)\,[\bar f_{++}(z_4)\psi_+^{b}(z_2)]^j\,\bar\chi_+^{c,k}(z_3)\,,
\notag\\
\mathcal{G}_3(z_1,z_2,z_3,z_4)=&
ig\epsilon^{ijk}\,(\mu\lambda)\,
\psi_+^{a,i}(z_1)\,  \psi_+^{b,j}(z_2)\,[\bar f_{++}(z_4)\bar\chi_+^{c}(z_3)]^k\,,
\end{align}
which, again, are subjected to the constraint
\begin{align}\label{3g}
\mathcal{G}_1(z_1,z_2,z_3,z_4)+\mathcal{G}_2(z_1,z_2,z_3,z_4)+\mathcal{G}_3(z_1,z_2,z_3,z_4)=0\,.
\end{align}
%
%It follows that 
Thus there are two independent distribution amplitudes in this case, $\Phi_4^g$
and $\Psi_4^g$,
which can be  defined as in Eq.~(\ref{Gdefinitions}).

The operators $\mathcal{Q}_1(z_2,z_3,z_1)$ and $\mathcal{Q}_2(z_1,z_3,z_2)$ with the
choice of flavors $b=d$ and $a=c=u$ enter
directly the definition of $\Psi_4$ and $\Phi_4$ in (\ref{definitions}).
For practical applications it can be convenient to introduce an
additional distribution amplitude related to the operator $\mathcal{Q}_3$
\begin{align}\label{def:D4}
\frac12\vev{0|\epsilon^{ijk}\psi_+^{u,i}(z_1)[\bar\chi_+^{3/2}]^{u,j}(z_2)\psi^{d,k}_+(z_3)|P}=&
\frac{i}4(\mu\lambda) (pn)
m_N N^{\uparrow}_+\int \mathcal{D}x \, e^{-i(pn)\,\sum {x_iz_i}}\,D_4(x)\,,
\end{align}
The amplitude $D_4$ is not independent and can be reduced to the combination of other amplitudes
using EOM.  
For the matrix elements in coordinate space
\begin{align}\label{}
\varphi_k(z_1,z_2,z_3)=\vev{0|\mathcal{Q}_k(z_1,z_2,z_3)|P}\,,&&
\varphi^g_k(z_1,z_2,z_3,z_4)=\vev{0|\mathcal{G}_k(z_1,z_2,z_3,z_4)|P}\,,
\end{align}
this relation reads
\begin{multline}\label{123gg}
\varphi_3(z_1,z_2,z_3)= \frac{\partial}{\partial{z_1}}\varphi_1(z_1,z_2,z_3)+
\frac{\partial}{\partial{z_2}}\varphi_2(z_1,z_2,z_3)\\
-\frac12\int_0^1d\tau \,\Big(z_{13}\,\varphi^g_1(z_1,z_2,z_3,z_{13}^\tau)+z_{23}\,
\varphi^g_2(z_1,z_2,z_3,z_{23}^\tau)
\Big),
\end{multline}
where we use the notation
\begin{align}
 z_{ik}=z_i-z_k\,, \qquad \bar \tau = 1-\tau  \,, \qquad z_{ik}^\tau=z_i \bar \tau+z_k \tau.
\label{notate}
\end{align}
Going to the momentum fractions one derives
\begin{align}\label{DPPrel}
D_4(x_1,x_2,x_3)=&x_3\Phi_4(x_1,x_2,x_3)+x_1\Psi_4(x_2,x_1,x_3)\notag\\
&+\frac12\biggl(\int^{x_2}_{0}\frac{dx}{x}\Phi_4^g(x_1,x_2-x,x_3,x)
-\int^{x_3}_{0}\frac{dx}{x}\Phi_4^g(x_1,x_2,x_3-x,x)
\notag\\
&+\int_{0}^{x_1}\frac{dx}{x}\Psi_4^g(x_2,x_1-x,x_3,x)-
\int_{0}^{x_3}\frac{dx}{x}\Psi_4^g(x_2,x_1,x_3-x,x)
\biggl)\,.
\end{align}
%
%which is a consequence of Eq.~(\ref{123gg}).

%%%%%%%%%%%%%%%%%%%%%%%%%%%%%%%%%%%%%%%%%%%%%%%%%%%%%%%%%%%%%%%%%%%%%%%%%%%%%%%%%%%%%%%%%%%%%%%%%%%%
%
\section{Renormalization Group Equations}
%
%%%%%%%%%%%%%%%%%%%%%%%%%%%%%%%%%%%%%%%%%%%%%%%%%%%%%%%%%%%%%%%%%%%%%%%%%%%%%%%%%%%%%%%%%%%%%%%%%%%%
\subsection{General properties}

The matrix of anomalous dimensions for a set of local operators $O_i$ is defined as
\begin{align}
\gamma_{ik}=  Z^{-1}_{ij}\mu\frac{\partial}{\partial\mu} Z_{jk} \,,&&[O]_{i}=Z^{-1}_{ik} O^B_{k},
\end{align}
where $[O]_{i}$, $O^B_{i}$ are  renormalized and bare operators, respectively.
The renormalized operators satisfy the matrix RG equation
\begin{align}\label{rg-0}
\left(\mu\frac{\partial }{\partial\mu}+
      \beta(g)\frac{\partial }{\partial g} +\gamma_{ik}(g)\right)\,[O]_{k}=0\,,
\end{align}
which, equivalently, can be cast in the form of an integro-differential equation for the
generating functions (light-ray operators):
\begin{align}\label{rg-1}
\left(\mu\frac{\partial }{\partial\mu}+
      \beta(g)\frac{\partial }{\partial g} +\gamma (g)\right)\,[O](z_1,\ldots,z_N)=0\,.
\end{align}
Here $\gamma$ is an integral operator, which we write as
\begin{align}
\gamma=\frac{\alpha_s}{2\pi}\, {\mathbb H}\,,
\end{align}
where $\alpha_s=g^2/4\pi$. In what follows we will refer to the evolution
kernel ${\mathbb H}$ as the Hamiltonian.
It is determined by  one-loop counterterms to the nonlocal operator.

{} For the both cases of pure chirality and mixed chirality operators we will be dealing with
three light-ray three-quark operators, (\ref{Qoperators}) or (\ref{qbx}), and three
operators with an additional gluon field, (\ref{Goperators}) or (\ref{qbg}), respectively.
Thus  ${\mathbb H}$ is in both cases a $6\times 6$ matrix which can be written as
\begin{align}\label{hamiltonian}
{\mathbb H}=\begin{pmatrix}{\mathbb H}_{q} & {\mathbb H}_{qg}\\ 0&{\mathbb H}_g\end{pmatrix}.
\end{align}
Each block, ${\mathbb H}_q, {\mathbb H}_g, {\mathbb H}_{qg}$, is a $3\times 3$ matrix
where the entries are integral operators.
The diagonal blocks, ${\mathbb H}_q$ and ${\mathbb H}_g$,
are given by the sum of two-particle operators, whereas  the off-diagonal
block  ${\mathbb H}_{qg}$  describes $3\to 2$ transitions.
The explicit expressions will be given below.

 One of the three light-ray operators involving a gluon field,
e.g. $G_{3}(z_1,z_2,z_3,z_4)$ (\ref{Goperators}) and $\mathcal{G}_{3}(z_1,z_2,z_3,z_4)$
(\ref{qbg}), can be excluded from
%$\mathfrak{O}$
consideration with the help of the operator identity~(\ref{3gidentity}).
Let
\begin{align}\label{vector-operator}
{\mathbb O}^{\rm chiral}(\vec{z})=\begin{pmatrix}
 Q_1(z_1,\ldots,z_3)\\Q_2(z_1,\ldots,z_3)\\Q_3(z_1,\ldots,z_3)\\G_1(z_1,\ldots,z_4)\\G_2(z_1,\ldots,z_4)
                   \end{pmatrix}
\end{align}
be the vector of the remaining five independent chiral operators, and similar for mixed chirality.
This vector satisfies the RG equation with the modified Hamiltonian
\begin{align}\label{mod-Hamiltonian}
\left(\mu\frac{\partial }{\partial\mu}+
      \beta(g)\frac{\partial }{\partial g}  +\frac{\alpha_{s}}{2\pi}\,
        \widetilde{\mathbb{H}}\right)\mathbb{O}(\vec{z}) = 0\,.
\end{align}
where $\widetilde{\mathbb{H}}$ is a $5\times 5$ matrix such that
\begin{align}
 [\widetilde{\mathbb{H}}_q]_{ik}&=[\mathbb{H}_q]_{ik}\,, & i,k=1,2,3
\nonumber\\
 [\widetilde{\mathbb{H}}_g]_{ik}&=[\mathbb{H}_g]_{ik}-[\mathbb{H}_g]_{i3}\,, & i,k=1,2
\nonumber\\
[\widetilde{\mathbb{H}}_{qg}]_{ik}&=[\mathbb{H}_{qg}]_{ik}-[\mathbb{H}_{qg}]_{i3}\,, & i=1,2,3,~~ k=1,2
\end{align}
The nonlocal operator $\mathbb{O}(\vec{z})$ can be expanded over a complete  basis of local operators
\begin{align}
\mathbb{O}(\vec{z})=\sum_{N,q}\Psi_{N,q}(\vec{z})\, \mathbb{O}_{N,q}\,,
\label{def:coef-fun}
\end{align}
where operator $\mathbb{O}_{N,q}$ has  canonical dimension $N+9/2$ and $q$
enumerates independent local operators of the same dimension.
 If  a local operator $\mathbb{O}_{N,q}$ satisfies the RG equation
$(\mu\partial_\mu+\beta(g) \partial_g + \gamma_{N,q})\,\mathbb{O}_{N,q}=0$ then the corresponding
``coefficient'' function $ \Psi_{N,q}(\vec{z})$ is the eigenvector of the Hamiltonian
$\widetilde{\mathbb{H}}$
\begin{align}\label{def:Schr-eq}
[\widetilde{\mathbb{H}}\Psi_{N,q}](\vec{z})=E_{N,q}\Psi_{N,q}(\vec{z})\,,
\end{align}
and $\gamma_{N,q}=(\alpha_s/2\pi)E_{N,q}$ so that
\begin{align}\label{scaledep}
 \mathbb{O}_{N,q}(\mu_2) =  \left(\frac{\alpha_s(\mu_2)}{\alpha_s(\mu_1)}\right)^{E_{N,q}/\beta_0}
 \mathbb{O}_{N,q}(\mu_1)\,,
\end{align}
where $\beta_0 = 11/3 N_c - 2/3 n_f$.

The coefficient functions of local
operators $\Psi_{N,q}(\vec{z})$ are homogeneous polynomials of three variables
\begin{align}\label{3expand}
\Psi_{N,q}^{(i)}(\vec{z})=\sum_{\substack{{k_1,\ldots, k_{3}}\\ k_1+k_2+k_3=N}}
 \psi^{(i)N,q}_{k_1\, k_2\, k_3}\,z_1^{k_1}\,z_2^{k_2}\, z_3^{k_3},
\end{align}
for the quark components, $i=1,2,3$, and four variables
\begin{align}\label{4expand}
\Psi_{N,q}^{(i)}(\vec{z})=\sum_{\substack{{k_1,\ldots, k_{4}}\\ k_1+\ldots+k_4=N-2}}
 \psi^{(i)N,q}_{k_1\, k_2\, k_3\,k_4}\,z_1^{k_1}\,z_2^{k_2}\, z_3^{k_3}\,z_{4}^{k_4}
\end{align}
for the quark-gluon components, $i=4,5$.

%The matrix elements of the nonlocal operator  $\mathbb{O}(\vec{z})$ between the vacuum and
%the nucleon state define nucleon distribution amplitudes, cf.  (\ref{definitions}), (\ref{Gdefinitions}):
%
%\begin{align}\label{}
%\vev{0|\mathbb{O}^{(i)}(\vec{z})|P}\sim \int \mathcal{D}x\, e^{i(pn)\sum z_k x_k} \Phi^{(i)}(\vec{x})\,.
%\end{align}
%
%The moments of the distribution amplitudes
%
%\begin{align}\label{}
%\phi_{\vec{k}}^{(i)}=\int \mathcal{D}x\, x_1^{k_1}\,\ldots\, x_n^{k_n}\,\Phi^{(i)}(\vec{x})
%\end{align}
%
%can be written in terms of the coefficients appearing in (\ref{3expand}), (\ref{4expand}) as
%
%\begin{align}\label{}
%\phi_{\vec{k}}^{(i)}=\sum_{q}  \psi^{(i)N,q}_{\vec{k}}
%\langle\langle\mathcal{Q}_{N,q}\rangle\rangle \,,
%\end{align}
%
%where $\vev{0|\mathcal{Q}_{N,q}|P}\sim \langle\langle\mathcal{Q}_{N,q}\rangle\rangle$ are reduced matrix elements of the
%multiplicatively renormalizable local operators.

%%%%%%%%%%%%%%%%%%%%%%%%%%%%
%%%%%%%%%%%%%%%%%%%%%%%%%%%%
\subsection{Conformally invariant evolution kernels}
%%%%%%%%%%%%%%%%%%%%%%%%%%%%
%%%%%%%%%%%%%%%%%%%%%%%%%%%%

The structure of the Hamiltonian is severely constrained by conformal symmetry.
It follows from the group theory that a nontrivial operator mapping the
representation $T^{j_1}\otimes T^{j_2}$ to $T^{i_1}\otimes T^{i_2}$
only exists if the difference $i_1+i_2-j_1-j_2$ is an integer.
If $i_1+i_2=j_1+j_2$  a conformally invariant operator $K$ can be written in the form
(see Ref.~\cite{BDKM05} for details)
\begin{align}\label{K}
[K_{j_1j_2}^{i_1i_2}\varphi](z_1,z_2)=
\int_0^1d\alpha\int_0^1 d\beta \,\bar\alpha^{i_1+j_1-2}\, \alpha^{i_2-j_2}\,
\bar\beta^{i_2+j_2-2} \,\beta^{i_1-j_1}\,\kappa\left(\frac{\alpha\beta}{\bar\alpha\bar\beta}\right)\,
\varphi(z_{12}^\alpha,z_{21}^\beta)\,,
\end{align}
where the notation follows Eq.~(\ref{notate}); the function $\kappa(x)$ is arbitrary.

The two-particle one-loop kernels fall in two groups. The kernels in the first group involve a single
integration, $\kappa(x)\sim \delta(x)$, 
and their form is completely fixed (up to prefactor) by the conformal spins
of the fields.  For the
case that the conformal spins are conserved, $i_i=j_1$ and $i_2=j_2$, one important example is%
\footnote{Here and below the subscripts ${\mathcal H}_{ik}$, $i,k=1,2,3,4$,  indicate that
the kernel acts on the coordinates
of the $i$-th and $k$-th partons (particles). }
\begin{multline}\label{Hv}
[\mathcal{H}_{12}^v\varphi](z_1,z_2)=
\int_0^1 \frac{d\alpha}{\alpha}
\Big\{\bar\alpha^{2j_1-1}[\varphi(z_1,z_2)-\varphi(z^\alpha_{12},z_2)]  +
\bar\alpha^{2j_2-1}[\varphi(z_1,z_2)-\varphi(z_1,z^\alpha_{21})]\Big\}\,.
\end{multline}
This structure is specific for gauge theories and arises from the diagrams involving the
gluon field  in the light-like Wilson lines (in covariant gauges).

Another kernel of this type is the ``exchange'' kernel, $\mathcal{H}^{e}$. It maps
$T^{j_1}\otimes T^{j_2}\to T^{j_2}\otimes T^{j_1}$ and is fixed by
conformal symmetry up to a prefactor~\cite{BDKM05}.  Assuming  that $j_1>j_2$:
\begin{align}\label{He}
[\mathcal{H}_{12}^e\varphi](z_1,z_2)=\int_0^1
{d\alpha}\,\bar\alpha^{2j_2-1}\,\alpha^{2(j_1-j_2)-1}\,\varphi(z_{12}^\alpha,z_2)\,.
\end{align}
The two-particle kernels~(\ref{Hv}),~(\ref{He}) depend implicitly on the conformal spins of the
fields which they act on; this dependence will always be implied.

The kernels  (\ref{Hv}) and (\ref{He}) are both known from the studies of leading twist operators.
In addition, we will need
\begin{align}\label{Hd}
[\mathcal{H}^d_{12}\varphi](z_1,z_2)=
%                      \int_0^1 d\alpha \int_0^{\bar \alpha} d\beta \,
% \bar\alpha^{2j_1-1}\,\alpha^{2j_2-1}\,\delta\left(1-\frac{\alpha\beta}{\bar\alpha\bar\beta}\right)\,
\varphi(z_{12}^\alpha,z_{21}^\beta)=
          \int_0^1 d\alpha\, \bar\alpha^{2j_1-1}\,\alpha^{2j_2-1}
\,\varphi(z_{12}^\alpha,z_{12}^\alpha)\,,
\end{align}
which corresponds to the ``diagonal'' mapping without the spin exchange:
 $T^{j_1}\otimes T^{j_2}\to T^{j_1}\otimes T^{j_2}$ with $\kappa(x)=\delta(1-x)$.

The kernels in the second group retain both integrations as in~(\ref{K}) and usually
involve  a theta--function $\kappa(x)\sim \theta(1-x)$ or $\kappa(x) \sim \theta(x-1)$ which
restricts the region of integration to $\alpha+\beta\leq 1$ and  $1\leq\alpha+\beta$,
respectively. We define two more  kernels $\mathcal{H}^\pm$ by
\begin{align}\label{Hpm}
[\mathcal{H}^+_{12}
\varphi](z_1,z_2)=&\int_0^1d\alpha\int_0^{\bar\alpha}d\beta\,\bar\alpha^{2j_1-2}\,
\bar\beta^{2j_2-2}\,\varphi(z_{12}^\alpha,z_{21}^\beta)\,,
\\
[\mathcal{H}^-_{12}
\varphi](z_1,z_2)=&\int_0^1d\alpha\int_{\bar\alpha}^1d\beta\,\bar\alpha^{2j_1-2}\,
\bar\beta^{2j_2-2}\,\varphi(z_{12}^\alpha,z_{21}^\beta)\,.
\end{align}

The off-diagonal block ${\mathbb H}_{qg}$ maps
the $SL(2,\mathbb{R})$  representations with a different number of fields
$T^{1}\otimes T^{1}\otimes T^{3/2} \to T^{j_1}\otimes T^{j_2}$, where
either $j_1=1/2$, $j_2=1$ or $j_1=1/2$, $j_2=1$ so that in the both cases $j_1+j_2=3/2$.
The general form of the corresponding kernel consistent with conformal symmetry is
\begin{align}\label{}
[R f](z_1,z_2)=z_{12}^2\int_0^1d\alpha\int_0^1 d\beta\int_0^1 d\gamma
\bar\beta^{2j_2-1}\beta^{2j_1-1}
r\left(\frac{\alpha\gamma}{\bar\alpha\bar\gamma}, \frac{\gamma\bar\beta}{\beta\bar\gamma}
\right)\,f(z_{12}^\alpha,z_{21}^\gamma,z_{21}^\beta)\,.
\end{align}
where $r(x,y)$ is an arbitrary function.
In one-loop diagrams the following kernels appear:
\begin{align}
[\mathcal{V}_{12(3)}^{(1)}\,f](z_1,z_2)=&
z_{12}^2\int_0^1d\alpha\int_{\bar\alpha}^1d\beta
\,\frac{\bar\alpha\bar\beta}{\alpha}\, f(z_{12}^\alpha,z_2,z_{21}^\beta)\,,
\label{V1}\\
[\mathcal{V}_{12(3)}^{(2)}\,f](z_1,z_2)=&z_{12}^2\int_0^1d\alpha\int_0^{\bar\alpha}d\beta
\,\beta\, f(z_{12}^\alpha,z_2,z_{21}^\beta)\,,
\label{V2}\\
[\mathcal{V}_{12(3)}^{(3)}\,f](z_1,z_2)=&z_{12}^2\int_0^1d\beta\int_{\beta}^1d\gamma\,
\frac{\beta\bar\gamma}{\gamma}\left(\frac{\bar\gamma}{\gamma}-2\frac{\bar\beta}{\beta}\right)\,
f(z_{1},z_{21}^\gamma,z_{21}^\beta)\,,
\label{V3}\\
[\mathcal{V}_{12(3)}^{(4)}\,f](z_1,z_2)=&z_{12}^2\int_0^1d\beta\,\bar\beta\left\{
f(z_{1},z_{2},z_{21}^\beta)+\frac{\bar\beta}{\beta}\int_0^\beta d\gamma
f(z_{1},z_{21}^\gamma,z_{21}^\beta)\right\},
\label{V4}\\
[\mathcal{V}^{(a)}_{12(3)}f](z_1,z_2)=&z_{12}^2
               \int_0^1 d\alpha \int_{\bar\alpha}^1 d\beta\int_0^{\bar\alpha} d\gamma
               \,\bar\beta\,f(z_{12}^\alpha,z_{21}^\gamma,z_{21}^\beta)\,,
\label{Va}
\\
[\mathcal{V}^{(b)}_{12(3)}f](z_1,z_2)=&z_{12}^2
		\int_0^1d\alpha \int_0^{\bar\alpha} d\beta \int_0^{\bar\alpha}d\gamma
  		\,\bar\beta\,f(z_{12}^\alpha,z_{21}^\gamma,z_{21}^\beta)\,,
\label{Vb}
\\
[\mathcal{V}^{(c)}_{12(3)}f](z_1,z_2)=&z_{12}^2
		\int_0^1d\alpha \int_0^{\bar\alpha} d\beta \int_0^{\beta}d\gamma
		\,\bar\beta\,f(z_{12}^\alpha,z_{21}^\gamma,z_{21}^\beta)\,,
\label{Vc}
\end{align}
which correspond to the choices
\begin{align}\label{}
r^{(1)}(x,y)&=\delta(x)\,\theta(x/y-1), &
r^{(2)}(x,y)&=\delta(y)\,\theta(1-x/y),\\
r^{(3)}(x,y)&=\delta(x)\,\theta(y-1)\,\left(\frac1{y}-2\right), &
r^{(4)}(x,y)&=\theta(1-y)\delta(x/y)\,(1+\delta(y)),\\
r^{(a)}(x,y)&=\,\theta(1-x)\theta(x/y-1),&
r^{(b)}(x,y)&=\,\theta(1-x)\theta(1-x/y),\\
r^{(c)}(x,y)&=\,\theta(1-y)\theta(1-x/y)\,.
\end{align}
%

%%%%%%%%%%%%%%%%%%%%%%%%%%%%%%%%%%%%%%%%%%%%%%%%%%%%%%%%%%%%%%%%%%%%%%%%%%%%%%%%%%%%%
\subsection{The scalar product}
%%%%%%%%%%%%%%%%%%%%%%%%%%%%%%%%%%%%%%%%%%%%%%%%%%%%%%%%%%%%%%%%%%%%%%%%%%%%%%%%%%%%%
We will be looking for solutions of the Schr{\"o}dinger equation (\ref{def:Schr-eq})
that are polynomials in light-cone variables $z_k$. The scalar product on this space
can be constructed as follows. First of all, we allow $z_k$ to take complex values.
It is convenient at this point to go over from the $SL(2,R)$ group to $SU(1,1)$ which has
the same algebra. In particular, the $SU(1,1)$ generators have the same form.
The  $SU(1,1)$ invariant scalar product is defined as \cite{Vilenkin}
\begin{align}\label{scp-1}
\vev{f_1,f_2}_j=\int_{|z|<1} D_jz \,\overline{f_1(z)}\,f_2(z)\,,\qquad\,
D_jz=\frac{2j-1}{\pi}(1-|z|^2)^{2j-2} d^2z\,,
\end{align}
where $j$ is the conformal spin, $j\geq1/2$, and functions $f_k(z)$ are polynomials in $z$;
$\overline{f(z)}=(f(z))^\ast$ stands for complex conjugation.
For the special case $j=1/2$ the $SU(1,1)$ scalar product takes the form
\begin{align}\label{}
\vev{f_1,f_2}_{j=1/2}=\frac1{2\pi}\int_0^{2\pi} d\varphi \, \overline{f_1(e^{i\varphi})}\,
f_2(e^{i\varphi})\,.
\end{align}
One easily finds $||z^n||^2\equiv \vev{z^n,z^n}=\dfrac{\Gamma(2j)n!}{\Gamma(2j+n)}$.  For 
polynomials of several variables $f(z_1,\ldots,z_n)$ which
belong to the tensor product $T^{j_1}\otimes \ldots \otimes T^{j_n}$, the scalar product
is given by
\begin{align}\label{sc2}
\vev{f_1,f_2}=
\left(\prod_{k=1}^n\int_{|z_k|<1} \!\!\!D_{j_k}\, z_k \right)\overline{f_1(z_1,\ldots,z_n)}\,
f_2(z_1,\ldots,z_n)\,.
\end{align}
It can be shown that the diagonal blocks (three-particle and four-particle) of the
evolution Hamiltonians are self-adjoint
with respect to the following scalar product:
\begin{align}\label{scqg}
\vev{\Psi_1|\Psi_2}=\sum_{i}^5 \vev{\Psi_1^{(i)}|\Omega_{ik}\Psi_2^{(k)}}\,,%%%%%+
%\sum_{i,k=4}^5 \vev{\Psi_1^{(k)}|\Omega_{ki}\Psi_2^{(i)
%}}
\end{align}
where the matrix
$\Omega$  is %the $2\times 2$ matrix
\begin{align}\label{scpsi}
\Omega=\begin{pmatrix}a_1 &   0 & 0  & 0 & 0\\
                      0   & a_2 & 0  & 0 & 0\\
                      0   & 0   & a_3& 0 & 0\\
                      0   &  0  & 0  & 2 & 1\\
                      0   &  0  & 0  & 1 & 2
\end{pmatrix}\,.
\end{align}
The coefficients $a_i=a$ for the chiral operators and $a_1=a_2=2a_3=a$
for the mixed ones, where $a>0$ is an arbitrary constant. We put $a=1$ in what
follows.
It is tacitly implied that $\vev{\Psi_1^{(i)}|\Omega_{ik}\Psi_2^{(k)}}$,  in (\ref{scpsi})
is given by
the $SU(1,1)$ scalar product (\ref{sc2}) with the spins $j_k$ equal to the conformal spins of
$\Psi_{1}^{(i)}$.

We would like to emphasize that the choice of the scalar product depends on the problem
under consideration, see, e.g. \cite{Kirch:2005tt} for a discussion.

%%%%%%%%%%%%%%%%%%%%%%%%%%%%%%%%%%%%%%%%%%%%%%%%%%%%%%%%%%%%%%%%%%%%%%%%%%%%%%%%%%%%%
%
\subsection{Chiral operators}
%
%%%%%%%%%%%%%%%%%%%%%%%%%%%%%%%%%%%%%%%%%%%%%%%%%%%%%%%%%%%%%%%%%%%%%%%%%%%%%%%%%%%%%
We are now in a position to write the explicit expression for the Hamiltonian that governs
the RG equation
for the chiral operators (\ref{Qoperators}), (\ref{Goperators}).
The quark block ${\mathbb H}_{q}^{\psi\psi\psi}$ can be written in terms of the two kernels defined in
(\ref{Hv}) and (\ref{He}):
\begin{align}\label{hup-1}
{\mathbb H}_{q}^{\psi\psi\psi}=\left(1+\frac1{N_c}\right)\, {\mathcal H}^{\psi\psi\psi}_q
\end{align}
with
\begin{align}\label{hup-2}
{\mathcal H}_q^{\psi\psi\psi} = \begin{pmatrix}
           \mathrm{H}&\mathcal{H}^e_{12}&\mathcal{H}^e_{13}\\
           \mathcal{H}^e_{21}&\mathrm{H}&\mathcal{H}^e_{23}\\
           \mathcal{H}^e_{31}&\mathcal{H}^e_{32}&\mathrm{H}\\
\end{pmatrix},
\end{align}
where
\begin{align}
 \mathrm{H} = \mathcal{H}_{12}^v+ \mathcal{H}_{23}^v+\mathcal{H}_{31}^v-\frac12\,.
\end{align}
As discussed in Sect.~3 the Hamiltonian ${\mathbb H}_q$ describes
renormalization of the three quark operators of both geometrical twist $T=3$ and $T=4$.
On the other hand, the Hamiltonian for the twist-3 operator $\psi_+\psi_+\psi_+$ is known
from Ref.~\cite{BDM}, where it is called $H_{3/2}$.
Thus one obtains a nontrivial consistency condition%
\footnote{$H_{3/2}$ on the r.h.s. of (\ref{consistency}) is given by the sum of three
``$v$''-type kernels for the conformal spins
pairs $\{j_i,j_k\}= \{1,1\}$ whereas $\mathrm{H}^0$ on the l.h.s. contains one kernel with
$\{j_i,j_k\}= \{1,1\}$ and two kernels
with $\{j_i,j_k\}= \{1,1/2\}$. The ``$e$''-type kernels appearing in the non-diagonal entries
$({\mathbb H}_{q})_{12}+({\mathbb H}_{q})_{13}$ correct for this difference.}
\begin{align}\label{consistency}
({\mathbb H}^{\psi\psi\psi}_{q})_{k1}+({\mathbb H}^{\psi\psi\psi}_{q})_{k2}
+({\mathbb H}^{\psi\psi\psi}_{q})_{k3}=H_{3/2}\,,\qquad k=1,2,3\,,
\end{align}
which, alternatively, can be used to restore $\mathbb{H}_{q}$ from the known
result~\cite{BDM} for $H_{3/2}$.

The ``gluon'' block ${\mathbb H}_g$ describes the renormalization of four-particle
quasi-partonic operators and can
be restored from the results existing in the 
literature~\cite{Bukhvostov:1984as,Bukhvostov:1985rn,Balitsky:1987bk}.
Though the baryon operators in question make sense for $N_c=3$  only, it is convenient
to separate the terms with the different color factor:
\begin{align}\label{gluonB}
{\mathbb H}_g^{\psi\psi\psi\bar f}=N_c \, {\mathbb H}_g^{(1)} + {\mathbb H}_g^{(0)} + \frac{1}{N_c}\,
{\mathbb H}_g^{(-1)} + \frac{21}{2}\,,
\end{align}
where the last term (a constant) stands for the self-energy type contributions.
{}For the diagonal elements we find
\begin{align}
[{\mathbb H}_g^{(1)}]_{kk}=&\mathcal{H}^v_{k4}-2\,\mathcal{H}^+_{k4}\,,\notag\\
[{\mathbb H}_g^{(0)}]_{kk}=&\mathcal{H}^v_{k+1,k-1}+\mathcal{H}^v_{k+1,4}+\mathcal{H}^v_{k-1,4}-
2 \left(
\mathcal{H}^+_{k+1,4}+\mathcal{H}^+_{k-1,4}+\mathcal{H}^-_{k+1,4}+\mathcal{H}^-_{k-1,4}
 \right)\,,\notag\\
[{\mathbb
H}_g^{(-1)}]_{kk}=&\mathcal{H}^v_{12}+\mathcal{H}^v_{23}+\mathcal{H}^v_{31}-2\,\mathcal{H}_{k4}^{-}\,.
\end{align}
In these expressions $k=1,2,3$ and  $k,k\pm 1$ appearing in the subscripts of the kernels on the r.h.s.
refer to arguments of the quark fields.  If  $k+1=4$ or $k-1=0$ the corresponding subscript
should be put to $1$ or $3$, respectively.

The off-diagonal matrix elements $i,k=1,2,3$, $i\not=k$, read
\begin{align}\label{offd}
[{\mathbb H}_g^{(0)}]_{ik}=&\mathcal{H}^v_{ik}-\mathcal{H}^v_{k4}+2\,\mathcal{H}^+_{k4}-\frac12\,,
\notag\\
[{\mathbb H}_g^{(-1)}]_{ik}=&-2\,\mathcal{H}^-_{k4}\,.
\end{align}
The identity in Eq.~(\ref{3gidentity}) implies the relations
\begin{align}\label{H-constraint}
[\mathbb{H}_g]_{11}-[\mathbb{H}_g]_{13}+[\mathbb{H}_g]_{21}-[\mathbb{H}_g]_{23}&=
[\mathbb{H}_g]_{33}-[\mathbb{H}_g]_{31}\,,\notag\\
[\mathbb{H}_g]_{22}-[\mathbb{H}_g]_{23}+[\mathbb{H}_g]_{12}-[\mathbb{H}_g]_{13}&=
[\mathbb{H}_g]_{33}-[\mathbb{H}_g]_{32}\,.
\end{align}

{}Finally, for the off-diagonal quark-gluon block we find
\begin{align}
{\mathbb H}_{qg}^{\rm chiral}=- \frac12 \mathcal{H}^{\rm chiral}_{qg}\,,
\end{align}
with
\begin{align}\label{H-off-1}
[\mathcal{H}^{\rm chiral}_{qg}]_{kk}=&\frac{1}{N_c}\left(\mathcal{V}^{(1)}_{k,k+1,(4)}+
\mathcal{V}^{(1)}_{k,k-1,(4)}\right)
-\mathcal{V}^{(2)}_{k,k+1,(4)}-\mathcal{V}^{(2)}_{k,k-1,(4)}\,,
\nonumber\\
[\mathcal{H}^{\rm chiral}_{qg}]_{ik}=&\mathcal{V}^{(1)}_{ik(4)}+\mathcal{V}^{(2)}_{ik(4)}\,,
\end{align}
where, as above,  the subscripts $k,k\pm 1$ take the values $1,2,3$.

%%%%%%%%%%%%%%%%%%%%%%%%%%%%%%%%%%%%%%%%%%%%%%%%%%%%%%%%%%%%%%%%%%%%%%%%%%%%%%%%%%%%%%%%%%%%%%
\subsection{Mixed chirality operators}
%%%%%%%%%%%%%%%%%%%%%%%%%%%%%%%%%%%%%%%%%%%%%%%%%%%%%%%%%%%%%%%%%%%%%%%%%%%%%%%%%%%%%%%%%%%%%%
{} For the  mixed chirality operators (\ref{qbx}), (\ref{qbg}) the quark block is
\begin{align}\label{mi-chi-o}
\mathbb{H}_{q}^{\psi\psi\chi}=\left(1+\frac1{N_c}\right) \mathcal{H}_q^{\psi\psi\chi}
\end{align}
where
\begin{align}\label{H:qmixed}
\mathcal{H}_q^{\psi\psi\chi}=\begin{pmatrix}
\mathrm{H}+\mathcal{H}^d_{13}-\mathcal{H}^+_{23} & & \mathcal{H}^e_{12} & &
z_{13}\mathcal{H}^{+}_{13}
\\[2mm]
\mathcal{H}^e_{21} & & \mathrm{H}+\mathcal{H}^d_{23}-\mathcal{H}^+_{13} &
&z_{23}\mathcal{H}^{+}_{23}
\\[2mm]
z^{-1}_{13}(\mathbb{I}-2\mathcal{H}^d_{13})&
&z^{-1}_{23}(\mathbb{I}-2\mathcal{H}^d_{23})& &
\mathrm{H}-2(\mathcal{H}^+_{13}+\mathcal{H}^+_{23})+3
\end{pmatrix}\,.
\end{align}
Similar to Eq.~(\ref{consistency}) one can derive the consistency relations
\begin{align}\label{H12cons}
[\mathbb{H}^{\psi\psi\chi}_{q}]_{11}+[\mathbb{H}^{\psi\psi\chi}_{q}]_{12}-
[\mathbb{H}^{\psi\psi\chi}_{q}]_{13}\, \partial_{z_3} &= H_{1/2}\,,
\nonumber\\
[\mathbb{H}^{\psi\psi\chi}_{q}]_{33}\,\partial_{z_3}-[\mathbb{H}^{\psi\psi\chi}_{q}]_{31}-
[\mathbb{H}^{\psi\psi\chi}_{q}]_{32} &= \partial_{z_3}\,H_{1/2}\,,
\end{align}
where the Hamiltonian $H_{1/2}$ describes the renormalization of the leading twist-three nucleon
distribution amplitude~\cite{BDKM}.

The gluon block $\mathbb{H}_g^{\psi\psi\chi\bar f}$ can be expanded in powers of $N_c$ in the same way
as in~(\ref{gluonB}),  with
\begin{align}
[{\mathbb H}^{\psi\psi\chi\bar
f,(1)}_g]_{kk}=&\mathcal{H}^v_{k4}-2\,(1-\delta_{k3})\mathcal{H}^+_{k4}\,,
\end{align}
\begin{align}
[\mathbb{H}^{\psi\psi\chi\bar f,(0)}_g]_{11}=&\mathcal{H}^v_{23}+\mathcal{H}^v_{24}+\mathcal{H}^v_{34}-
\mathcal{H}^+_{23}-2\mathcal{H}^+_{24}-2\mathcal{H}^-_{24}+P_{34}\mathcal{H}^e_{43}\,,
\notag\\
[\mathbb{H}^{\psi\psi\chi\bar f,(0)}_g]_{22}=&\mathcal{H}^v_{13}+\mathcal{H}^v_{14}+\mathcal{H}^v_{34}-
\mathcal{H}^+_{13}-2\mathcal{H}^+_{14}-2\mathcal{H}^-_{14}+P_{34}\mathcal{H}^e_{43}\,,
\notag\\
[\mathbb{H}^{\psi\psi\chi\bar f,(0)}_g]_{33}=&\mathcal{H}^v_{12}+\mathcal{H}^v_{14}+\mathcal{H}^v_{24}-
2(\mathcal{H}^+_{14}+\mathcal{H}^+_{24}+\mathcal{H}^-_{14}+\mathcal{H}^-_{24})\,,
\nonumber\\
[\mathbb{H}^{\psi\psi\chi\bar
f,(0)}_g]_{12}=&\mathcal{H}^v_{12}-\mathcal{H}^v_{24}+2\mathcal{H}^+_{24}-\frac12\,,
\qquad
[\mathbb{H}^{\psi\psi\chi\bar
f,(0)}_g]_{21}=\mathcal{H}^v_{21}-\mathcal{H}^v_{14}+2\mathcal{H}^+_{14}-\frac12\,,
\notag\\
[\mathbb{H}^{\psi\psi\chi\bar
f,(0)}_g]_{j3}=&\mathcal{H}^v_{j3}-\mathcal{H}^v_{34}-\phantom{2}\mathcal{H}^+_{j3}-\frac12\,,
\qquad
[\mathbb{H}^{\psi\psi\chi\bar f,(0)}_g]_{3j}=\mathcal{H}^v_{3j}-\mathcal{H}^v_{j4}-\mathcal{H}_{j3}^+ +
2\mathcal{H}^+_{j4}-\frac12\,,%&&\qquad k=1,2
%[\mathbb{H}_g^{(0)}]_{32}=\mathcal{H}^v_{23}-\mathcal{H}^v_{34}-\mathcal{H}^+_{23}-\frac12\,
\end{align}
\begin{align}\label{}
[\mathbb{H}^{\psi\psi\chi\bar f,(-1)}_g]_{kk}&=\mathcal{H}^v_{12}+\mathcal{H}^v_{13}+\mathcal{H}^v_{23}-
\mathcal{H}^+_{13}-\mathcal{H}^+_{23}
-2(1-\delta_{k,3})\mathcal{H}^-_{k4} +\delta_{k3} P_{34}\mathcal{H}^e_{43}\,,
\nonumber\\
[{\mathbb H}_g^{\psi\psi\chi\bar f,(-1)}]_{12}&=-2\mathcal{H}_{24}^-\,,
             \qquad
[{\mathbb H}_g^{\psi\psi\chi\bar f,(-1)}]_{21}=-2\mathcal{H}_{14}^-\,,
\notag\\
[{\mathbb H}_g^{\psi\psi\chi\bar f,(-1)}]_{j3}&=P_{34}\mathcal{H}_{43}^e \,,
               \qquad~~
[{\mathbb H}_g^{\psi\psi\chi\bar f,(-1)}]_{3j}=-2\mathcal{H}_{14}^-\,,
\end{align}
where $P_{34}$ is the permutation operator $P_{34}f(z_1,z_2,z_3,z_4) = f(z_1,z_2,z_4,z_3)$,
 $k=1,2,3$ and $j=1,2$.
The entries $[\mathbb{H}^{\psi\psi\chi\bar f}_g]_{ik}$ satisfy the same constraint~(\ref{H-constraint}).
Finally, for the off-diagonal  quark-gluon block we find
\begin{align}\label{H:qgmixed}
\mathbb{H}_{qg}^{\rm mixed}=-\frac12\mathcal{H}_{qg}^{\rm mixed}\,,
%+\Delta{\mathcal{H}}_{qg}\right)\,,
\end{align}
where
\begin{align}
 \left[\mathcal{H}_{qg}^{\rm mixed}\right]_{jk} =& \left[\mathcal{H}_{qg}^{\rm chiral}\right]_{jk} +
 \left[\Delta{\mathcal{H}}_{qg}\right]_{jk}\,, \qquad j,k=1,2
\nonumber\\
[\mathcal{H}^{\rm mixed}_{qg}]_{3k}=&\frac{2}{ z_{k3}}\left(
\mathcal{V}^{(b)}_{k3(4)}-\frac13\mathcal{V}^{(a)}_{k3(4)}-
\frac{1}{2}\mathcal{V}^{(3)}_{k3(4)}+
\frac{1}{2}\mathcal{V}^{(4)}_{k3(4)}\right)\,, \qquad k=1,2
\nonumber\\
[\mathcal{H}^{\rm mixed}_{qg}]_{33}=&- 2\sum_{j=1}^2\frac{1}{z_{j3}}
\left(\mathcal{V}^{(a)}_{j3(4)}-\frac13\mathcal{V}^{(b)}_{j3(4)}+ \frac43
\mathcal{V}^{(c)}_{j3(4)}+\frac{1}{6}\mathcal{V}^{(3)}_{j3(4)}+\frac{1}{2}\mathcal{V}^{(4)}_{j3(4)}
\right),
\end{align}
$\mathcal{H}^{\rm chiral }_{qg}$ is given in Eq.~(\ref{H-off-1}) and
%for $\Delta{\mathcal{H}}_{qg}$ we obtain
%
\begin{align}\label{}
[\Delta{\mathcal{H}}_{qg}]_{12}=&[\Delta{\mathcal{H}}_{qg}]_{21}=0\,,
\nonumber\\
[\Delta{\mathcal{H}}_{qg}]_{jj}=&
          \frac1{3}\mathcal{V}^{(a)}_{j3(4)}-\mathcal{V}^{(b)}_{j3(4)}\,,
\nonumber\\
[\Delta{\mathcal{H}}_{qg}]_{j3}=&
   \mathcal{V}^{(a)}_{j3(4)}-\frac13\mathcal{V}^{(b)}_{j3(4)}+\frac43\mathcal{V}_{j3(4)}^{(c)}  \,,
\end{align}
for $j=1,2$.

%%%%%%%%%%%%%%%%%%%%%%%%%%%%%%%%%%%%%%%%%%%%%%%%%%%%%%%%%%%%%%%%%%%%%%%%%%%%%%%%%%%%%%%%%%%%%%%%%
\section{Renormalization Group Equations II: Solutions}
%%%%%%%%%%%%%%%%%%%%%%%%%%%%%%%%%%%%%%%%%%%%%%%%%%%%%%%%%%%%%%%%%%%%%%%%%%%%%%%%%%%%%%%%%%%%%%%%%%
The  Hamiltonian~(\ref{mod-Hamiltonian}) has block-triangular structure.  In order to find
its eigenvalues it is, therefore, sufficient to consider
the diagonal quark and quark-gluon blocks separately.
%
%%%%%%%%%%%%%%%%%%%%%%%%%%%%%%
\subsection{Chiral quark operators}
%%%%%%%%%%%%%%%%%%%%%%%%%%%%%%
%
%
\subsubsection{Permutation symmetry}
The three-quark Hamiltonian $\mathcal{H}_q^{\psi\psi\psi}$  is given by  Eq.~(\ref{hup-2}) with
the kernels $\mathcal{H}^v$, $\mathcal{H}^{e}$ defined in (\ref{Hv}), (\ref{He}),
respectively.  It is easy to check that $\mathcal{H}_q^{\psi\psi\psi}$ commutes with the generator of
cyclic permutations
\begin{align}
 [\mathcal{P},\mathcal{H}_q^{\psi\psi\psi}]=0\,, \qquad \mathcal{P}=P_a\otimes
P_z\,,\qquad \mathcal{P}^3=1\,,
\label{cyclic}
\end{align}
 where $P_a$ permutes the quark quantum numbers and $P_z$ the quark coordinates, respectively:
\begin{align}\label{}
P_a\,\Psi^{(i)}_{N,q}(z_1,z_2,z_3)=\Psi^{(i+1)}_{N,q}(z_1,z_2,z_3)\,,\notag\\
P_z\,\Psi^{(i)}_{N,q}(z_1,z_2,z_3)=\Psi^{(i)}_{N,q}(z_3,z_1,z_2)\,.
\end{align}
The eigenfunctions of $\mathcal{H}_q^{\psi\psi\psi}$ can always be chosen to have
definite parity with respect to the cyclic permutations:
\begin{align}\label{permut}
 \mathcal{P}\,\Psi_{N,q}^{\varepsilon}=\varepsilon \,\Psi_{N,q}^{\varepsilon}\,, \qquad
\varepsilon\in \{1,e^{i2\pi/3},e^{-i2\pi/3}\}\,.
\end{align}
%where $\varepsilon$ is one of the eigenvalues of $\mathcal{P}$,
%$\varepsilon\in \{1,e^{2\pi/3},e^{-2\pi/3}\}$
The (vector) eigenfunction $\Psi^{(\epsilon)}_{N,q}$ can be written in terms of the single
function $\psi^{(\epsilon)}_{N,q}$ as
\begin{align}\label{psie}
\Psi_{N,q}^{\varepsilon}(z_1,z_2,z_3)=
\begin{pmatrix} \varepsilon^{0}&\psi_{N,q}^{\varepsilon}(z_1,z_2,z_3)\\
                \varepsilon^{1}&\psi_{N,q}^{\varepsilon}(z_2,z_3,z_1)\\
               \varepsilon^{2}&\psi_{N,q}^{\varepsilon}(z_3,z_1,z_2)
\end{pmatrix}.
\end{align}
The eigenfunctions of different parity are orthogonal with respect to the
quark part of the scalar  product~(\ref{scqg}),
$\vev{\Psi^\varepsilon|\Psi^{\varepsilon'}}\sim\delta_{\varepsilon\varepsilon'}$,
whereas for the eigenfunctions of the same parity one gets
\begin{align}\label{}
\vev{\Psi^\varepsilon_{N,q}|\Psi^{\varepsilon}_{N,q'}}=
3\,\vev{\psi_{N,q}^{\varepsilon}|\psi_{N,q'}^{\varepsilon}}\,.
\end{align}
The scalar product on the r.h.s. is given by Eq.~(\ref{sc2}) for the spins $j_1=1/2,\, j_2=j_3=1$.

In addition to the symmetry under cyclic permutations, the Hamiltonian $\mathcal{H}_q^{\psi\psi\psi}$
commutes with the permutation operator $\mathcal{P}_{12}=P_a^{(12)}\otimes P_z^{(12)}$ defined as
\begin{align}
P_a^{(12)}\Psi^{(1)}(\vec{z})=\Psi^{(2)}(\vec{z})\,,&&
P_a^{(12)}\Psi^{(2)}(\vec{z})=\Psi^{(1)}(\vec{z})\,,&&
P_z^{(12)}\Psi^{(i)}(z_1,z_2,z_3)=\Psi^{(i)}(z_2,z_1,z_3)\,.
\end{align}
Since $\mathcal{P}_{12}\,\mathcal{P}=\mathcal{P}^{-1}\,\mathcal{P}_{12}$ one concludes
that $\mathcal{P}_{12}\,\Psi^{\varepsilon}_{N,q}\sim \Psi^{\varepsilon^{-1}}_{N,q}$. Thus
the eigenvalues of the Hamiltonian $\mathcal{H}_q^{\psi\psi\psi}$ in the sectors with
$\varepsilon=e^{\pm i 2\pi/3}$ coincide.

Furthermore, if $\Psi_{N,q}^{\varepsilon}$ is the eigenfunction of the
Hamiltonian $\mathcal{H}_q^{\psi\psi\psi}$ with the eigenvalue
$\mathcal{E}_{N,q}^{\varepsilon}$\,,
then $\psi_{N,q}^{\varepsilon}$ is the eigenfunction with the same eigenvalue of the
effective Hamiltonian $\mathcal{H}(\varepsilon)$,
\begin{align}
\mathcal{H}(\varepsilon)\,\psi_{N,q}^{\varepsilon}(z_1,z_2,z_3)=
\mathcal{E}_{N,q}^{\varepsilon}\,\psi_{N,q}^{\varepsilon}(z_1,z_2,z_3)\,,
\end{align}
where
\begin{align}\label{Heff}
\mathcal{H}(\varepsilon)=\mathcal{H}_{3/2}-
\mathcal{H}_{12}^e(1-\varepsilon P_{z}^{-1})-\mathcal{H}_{13}^e(1-\varepsilon^{-1} P_{z})\,.
\end{align}
Here $\mathcal{H}_{3/2}$ is the evolution Hamiltonian for the twist-3 chiral quark operator~\cite{BDM},
\begin{align}\label{}
\mathcal{H}_{3/2}=\mathcal{H}_{12}^{v}+\mathcal{H}_{23}^{v}+\mathcal{H}_{31}^{v}+\frac32\,,
\end{align}
where the three kernels $\mathcal{H}^v_{ik}$ are all given by (\ref{Hv})
for the conformal spins $j_i=j_k=1$.

As was explained earlier the Hamiltonian $\mathcal{H}_q^{\psi\psi\psi}$, alias
$\mathcal{H}(\varepsilon)$,
describes the evolution of chiral operators of both geometrical  twist-4 and twist-3.
The twist-3 eigenfunctions $\Psi^{\rm tw-3}_{N,q}(z_1,z_2,z_3)$
which belong to the sector with $\mathcal{P}$-parity $\varepsilon$,
have  the same parity with respect to cyclic permutations of the coordinates $P_z$ alone,
 $P_z\, \Psi_{N,q}^{\rm tw-3}=\varepsilon \,\Psi_{N,q}^{\rm tw-3}$.
{}From the explicit expression in~(\ref{Heff}) it follows, obviously, that on such functions
$\mathcal{H}(\varepsilon)$ reduces to $\mathcal{H}_{3/2}$, as expected.

It follows that the eigenvalues (and eigenfunctions) of  $\mathcal{H}_{q}^{\psi\psi\psi}$
can be separated in three symmetry classes corresponding to the renormalization group
equation for:
\begin{itemize}
\item twist-3 operators;
\item twist-4 operators with $\mathcal{P}$-parity $\varepsilon=1$;
\item twist-4 operators with $\mathcal{P}$-parity $\varepsilon=e^{\pm i 2\pi/3}$.
\end{itemize}
In the application to the nucleon distribution amplitudes we are interested in
the eigenfunctions that satisfy the relations in~(\ref{q-symmetry})
(otherwise, the matrix elements of the corresponding local operators between the vacuum and
nucleon state vanish).  One can easily verify that the following
combination of the eigenfunctions with $\varepsilon=e^{+2i\pi/3}$ and $\varepsilon=e^{-2i\pi/3}$
has the required symmetry property:
\begin{align}
\Psi_{N,q}(\vec{z})\sim\left(1+\mathcal{P}_{12}\right)
\,\Psi_{N,q}^{\varepsilon}(\vec{z})\,.
\end{align}
%
%or, for the components,
%
%\begin{align}\label{}
%\Psi_{N,q}^{(i)}(\vec{z})=\Big(\varepsilon (\varepsilon P_z^{-1})^i+P_{23}\varepsilon^{-1}
%(\varepsilon^{-1} P_z)^i\Big) \psi_{N,q}^{\varepsilon}(\vec{z})\,.
%\end{align}
%
The twist-4 eigenfunctions with $\varepsilon=1$ do not satisfy the relation~(\ref{q-symmetry}).
They are relevant e.g. for the distribution amplitudes of baryons with isospin $I=3/2$.

%%%%%%%%%%%%%%%%%%%%%%%%%%%%%%%%%%%%%%%%%%%%%%%%%%%%%%%%%%%%%%%%%%%%%%%%%%%%%%%%%%%%%%%%%%%
\subsubsection{Complete Integrability}
%%%%%%%%%%%%%%%%%%%%%%%%%%%%%%%%%%%%%%%%%%%%%%%%%%%%%%%%%%%%%%%%%%%%%%%%%%%%%%%%%%%%%%%%%%%

The Hamiltonian~(\ref{hup-2}) possesses a hidden integral of motion.
Let
\begin{align}
 S_{ik}=\partial_k(z_k-z_i) \equiv (\partial/\partial z_k)(z_k-z_i)\,.
\end{align}
It is easy to see that $S_{ik}$ acts as the intertwining operator between the
representations  $T^{j_k=1/2}\otimes T^{j_i=1}$ and $T^{j_k=1}\otimes T^{j_i=1/2}$:
\begin{align}
 S_{ik}T^{j_k=1/2}\otimes T^{j_i=1}=T^{j_k=1}\otimes T^{j_i=1/2}S_{ik}\,.
\end{align}
We define two-particle ($3\times3$ matrix) operators $[Q_{ik}^\pm]^{i'k'}$, where $i<k$
and $i',k'=1,2,3$, by
\begin{align}
[Q^\pm_{ik}]^{ik}=S_{ik}\,,&&[Q^\pm_{ik}]^{ki}=S_{ki}
\end{align}
and all other off-diagonal matrix elements being zero. For the diagonal matrix elements
we put $[Q^\pm_{ik}]^{ii}=[Q^\pm_{ik}]^{kk}=\frac12$ and
\begin{align}
[Q^+_{ik}]^{jj}=\frac12+S_{ik} &&
[Q^-_{ik}]^{jj}=\frac12+S_{ki},
\end{align}
for $j$ different from $i$ and $k$. {}For example, explicit expressions for $Q_{12}^\pm$ are
\begin{align}
Q_{12}^{+}=\frac12\,\mathbb{I}+\begin{pmatrix}0&S_{12}&0\\
                                              S_{21}&0&0\\
                                              0&0& S_{12}
                                               \end{pmatrix},
&&
Q_{12}^{-}=\frac12\,\mathbb{I}+\begin{pmatrix}0&S_{12}&0\\
                                              S_{21}&0&0\\
                                              0&0& S_{21}
                                              \end{pmatrix}.
\end{align}
The two-particle Casimir operators $\widehat J^2_{ik}$ can be written in terms of $Q_{ik}^\pm$ as%
\footnote{It is easy to see that the
anticommutator $\widehat J_{ik}^2$ of the  $Q_{ik}^\pm$ operators is $SL(2,\mathbb{R})$
invariant whereas the $Q_{ik}^\pm$ themselves are not.}
\begin{align}\label{}
\widehat J_{ik}^2=\frac12\{Q_{ik}^+,Q_{ik}^{-}\}-\frac14
\end{align}
The  Hamiltonian $\mathcal{H}_q^{\psi\psi\psi}$ can be represented in the form
\begin{align}\label{H3q}
\mathcal{H}_q^{\psi\psi\psi}=\mathcal{H}_{12}+\mathcal{H}_{23}+\mathcal{H}_{31}+\frac32\,,
\end{align}
where
\begin{align}\label{pw}
\mathcal{H}_{12}=2\Big[\psi(\widehat J^2_{12})-\psi(2)\Big]=
\begin{pmatrix}\mathcal{H}_{12}^v-1&\mathcal{H}^{e}_{12}&0\\
\mathcal{H}^{e}_{12}&\mathcal{H}_{12}^v-1&0\\
0&0&\mathcal{H}_{12}^v
\end{pmatrix}
\end{align}
and similarly for $\mathcal{H}_{23},\mathcal{H}_{31}$.

{}Finally, let%
\footnote{
Note that $[\widehat J^2_{12},\widehat J^2_{23}]=[\widehat J^2_{13},\widehat
J^2_{12}]=[\widehat J^2_{23},\widehat J^2_{13}]$.
}
\begin{align}\label{Q3}
\widehat{Q}_3=\frac{i}{2}[\widehat J^2_{12},\widehat J^2_{23}]\,.
\end{align}
The operator $\widehat{Q}_3$  commutes
with the Hamiltonian $\mathcal{H}_q^{\psi\psi\psi}$:
\begin{align}
[\widehat{Q}_3,\mathcal{H}_q^{\psi\psi\psi}]=0
\label{Q311}
\end{align}
and defines, therefore, a nontrivial  integral of motion.
%One way to prove (\ref{Q311}) is making use of the general theory of integrable systems
%and observing that the system under consideration corresponds to
%a certain subsector of the $N=3$ integrable $SU(2,2)$ spin chain (for a discussion see
%e.g.~\cite{BFKS04}).
To prove~(\ref{Q311})  it is sufficient to  show that
\begin{align}\label{SKL}
[(\mathcal{H}_{ik}), \widehat Q_3]=i(\widehat J_{kj}^2-\widehat J_{ji}^2)\,,
\end{align}
where $j\neq i,k$, and the pair-wise Hamiltonians $(\mathcal{H}_{ik})$ are given in~Eq.~(\ref{pw}).
This can be done by calculating both sides of the relation~(\ref{SKL}) in the conformal basis,
see Refs.~\cite{BDKM,Braun:2001qx} for the details.

\begin{figure}
\centerline{\includegraphics[width=7cm]{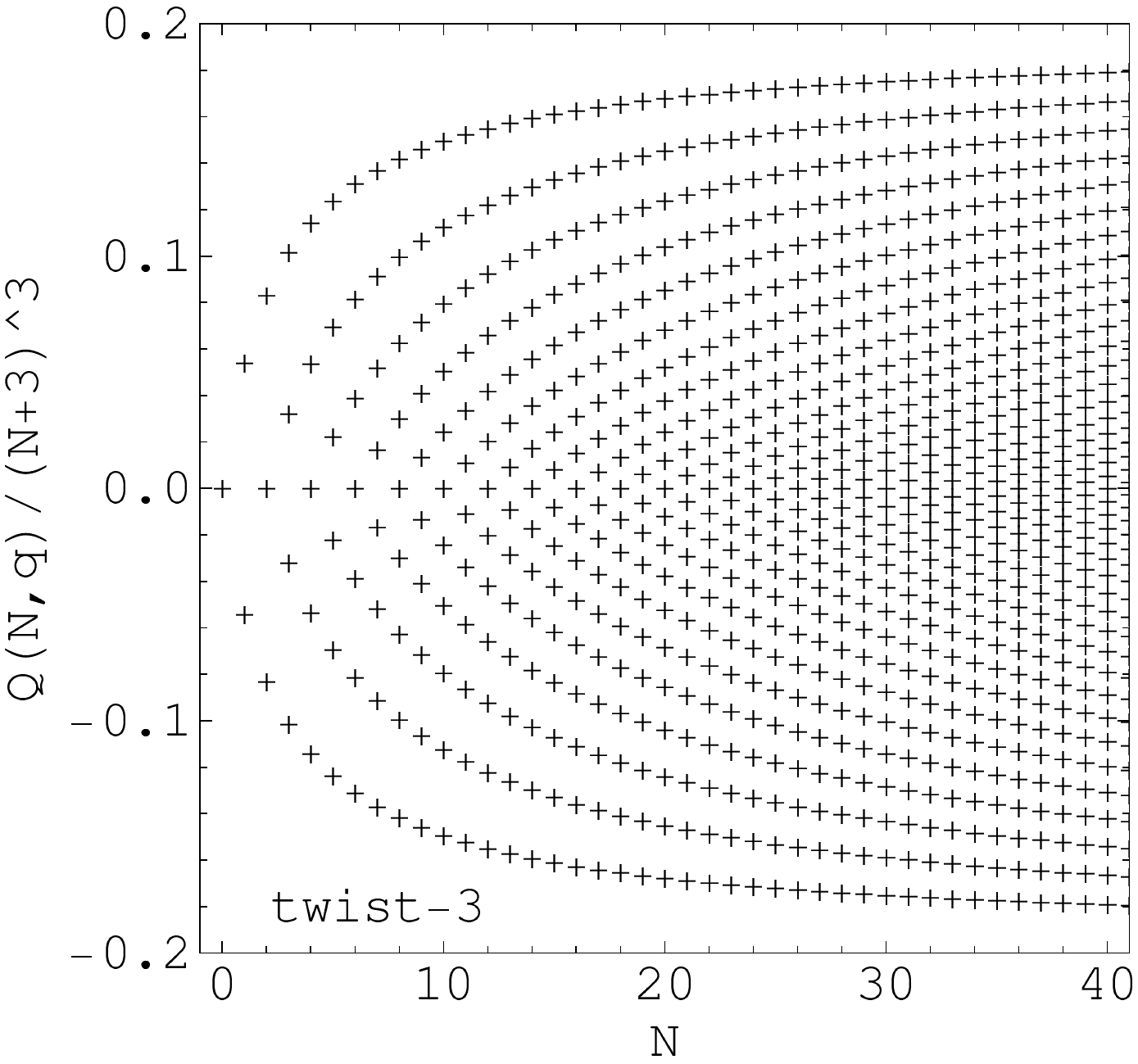}\hskip 0.5cm
\includegraphics[width=7cm]{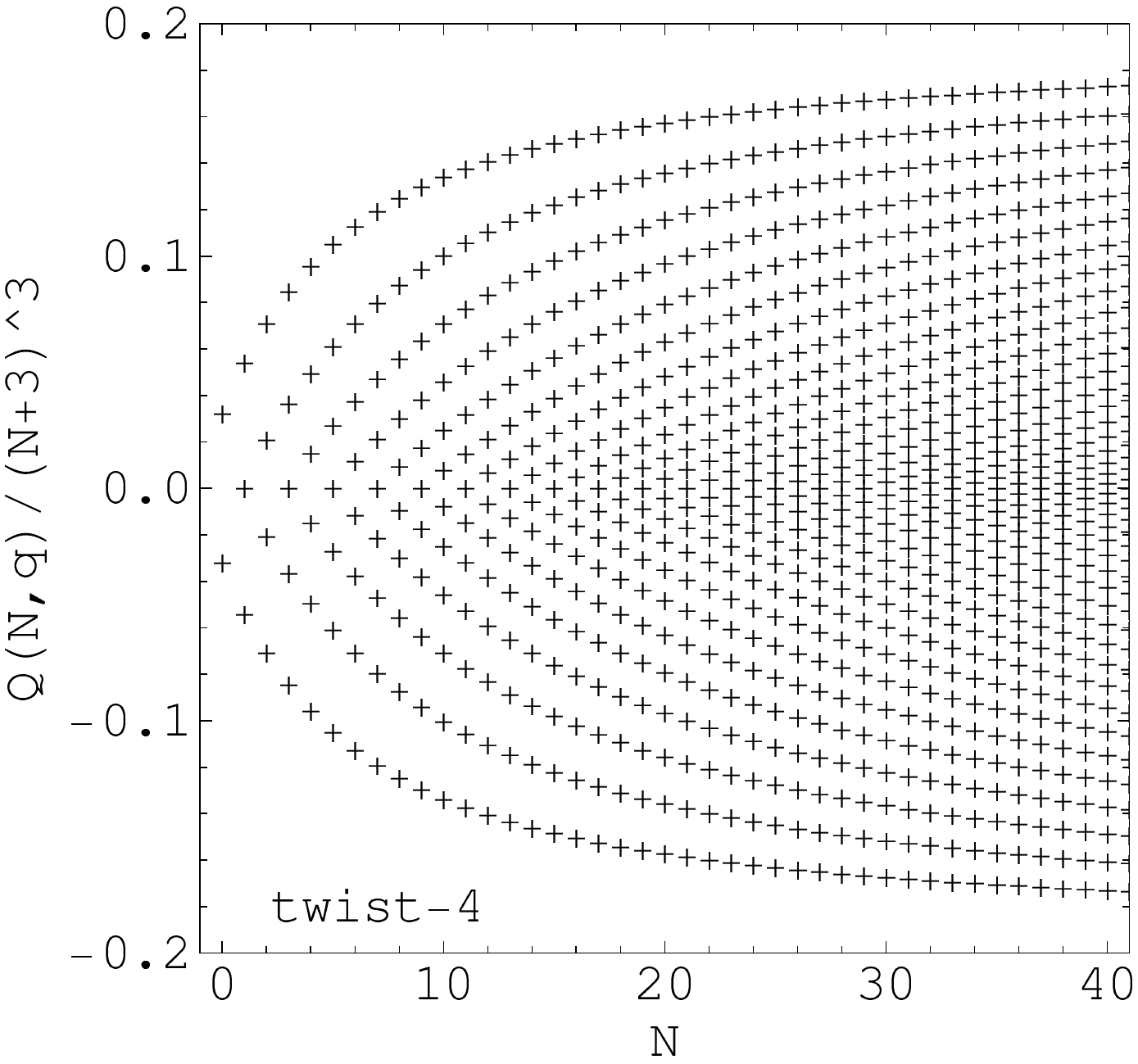}}
\caption{The spectrum of the conserved charge, $Q_3/(N+3)^3$, 
for twist-3 and twist-4  chiral quark operators.} 
\label{fig:Q3}
\end{figure}

It is straightforward to check that the operator $\widehat Q_3$ commutes with the operator
of cyclic permutations $\mathcal{P}$ and anticommutes with $\mathcal{P}_{12}$,
\begin{align}\label{QPP}
[\widehat Q_3,\mathcal{P}]=\{\widehat Q_3,\mathcal{P}_{12}\}=0\,.
\end{align}
Eq.~(\ref{QPP}) together with~(\ref{Q311}) imply that all eigenstates of the
Hamiltonian are double degenerate, except  for the states which  are annihilated by $\widehat Q_3$
i.e. $\widehat Q_3\Psi=0$, cf. \cite{BDKM}. The spectrum of $\widehat Q_3$ is shown and compared 
with the corresponding spectrum of the twist-3 conserved charge \cite{BDKM} in Fig.~\ref{fig:Q3}.

%
%%%%%%%%%%%%%%%%%%%%%%%%%%%%%%%%%%%%%%%%%%%%%%%%%%%%%%%%%%%%%%%%%
%%%%%%%%%%%%%%%%%%%%%%%%%%%%%%%%%%%%%%%%%%%%%%%%%%%%%%%%%%%%%%%%%
%

It was shown in Ref.~\cite{BFKS04} that the spectrum of one-loop anomalous dimensions of
(anti)chiral composite operators in QCD, i.e. operators constructed from the (anti)chiral
 fields and their derivatives, coincides with the spectrum of a certain (integrable) 
$SU(2,2)$-invariant spin chain. The Hamiltonian~(\ref{H3q}) can be viewed as the restriction of
the general $SU(2,2)$ spin chain Hamiltonian on the subspace with twist $E=4$ and helicity $H=1/2$.  
%The spectrum of the Hamiltonian $\mathcal{H}_q^{\psi\psi\psi}$ can be studied
%using powerful Quantum Inverse Scattering Methods (QISM) ~\cite{Faddeev}
%which is however beyond the scope of present paper.

%%%%%%%%%%%%%%%%%%%%%%%%%%%%%%%%%%%%%%%%%%%%%%%%%%%%%%%%%%%%%%%%%%%%%%%%%%%%%%%%%%%%%%%%
\subsubsection{The spectrum of anomalous dimensions}
%%%%%%%%%%%%%%%%%%%%%%%%%%%%%%%%%%%%%%%%%%%%%%%%%%%%%%%%%%%%%%%%%%%%%%%%%%%%%%%%%%%%%%%%
A short-distance expansion of the nonlocal
operator $\mathbb{O}(\vec{z})$,  see Eq.~(\ref{def:coef-fun}),
runs over a complete set of local operators $\mathbb{O}_{N,q}$
including operators with total derivatives. It is clear that in order to find  the anomalous dimensions
the operators with total derivatives can be omitted. The operators without total
derivatives can be singled out by their properties under conformal transformations: they
transform according to (\ref{Tg}) and are usually referred to as conformal operators.
The coefficient functions $\Psi_{N,q}(\vec{z})$ corresponding to the conformal operators
satisfy the following constraints
\begin{align}\label{shifts-inv}
&(\partial_1+\partial_2+\partial_3)\Psi_{N,q}(\vec{z})=0\,,\\
\label{Constraint}
&\partial_1 \Psi_{N,q}^{(1)}(\vec{z})+
\partial_2 \Psi_{N,q}^{(2)}(\vec{z})+\partial_3 \Psi_{N,q}^{(3)}(\vec{z})=0\,,
\end{align}
that follows from the requirement that such operators correspond to highest weights
of the corresponding  representation.
To get the first equation, we apply the operator
$P^{2\dot 2}$ which generates shifts along the ``plus'' light-cone direction to
Eq.~(\ref{def:coef-fun}). One gets
\begin{align}\label{}
\sum_{N,q}\Big[(\partial_1+\partial_2+\partial_3)\Psi_{N,q}(\vec{z})\Big]\,
\mathbb{O}_{N,q}=
\sum_{N,q}\Psi_{N,q}(\vec{z})\,
\partial_+\mathbb{O}_{N,q}\,.
\end{align}
The r.h.s. of this identity contains only operators with total derivatives, hence the
coefficients of the conformal operators on the l.h.s. must vanish.

To derive Eq.~(\ref{Constraint}) we
apply the transverse derivative $P^{1\dot 2}$ to the nonlocal operator of leading twist--3:
$\mathbb{O}^{tw-3}(\vec{z})=\epsilon^{ijk}\psi_+^{a,i}(z_1)\,\psi_+^{b,j}(z_2)\psi_+^{c,k}(z_3)$.
Taking into account that $i[\mathbf{P}^{1\dot 2}\psi_+](z)=-2\partial_z\psi_-(z)$, one obtains
\begin{align}\label{}
i[\mathbf{P}^{1\dot 2},\mathbb{O}^{tw-3}(\vec{z})]=-2\sum_{k=1}^3 \frac{\partial}{\partial z_k}
{Q}_k(\vec{z})=-2\sum_{N,q}\left(\sum_{k=1}^3 \frac{\partial}{\partial z_k}
\Psi^{(k)}_{N,q}(\vec{z})
\right)
\mathbb{Q}_{N,q}
\,,
\end{align}
where ${Q}_k(\vec{z})$ are defined in Eq.~(\ref{Qoperators}). Again, since
the l.h.s. only contains operators with the total derivatives, the coefficients of
conformal operators on the r.h.s. have to vanish.

%
%%%%%%%%%%%%%%%%%%%%%%%%%%%%%%%%%%%%%%%%%%%%%%%%%%%%%%%%%%%%%%%%%
\begin{figure}
\centerline{\includegraphics[width=7cm]{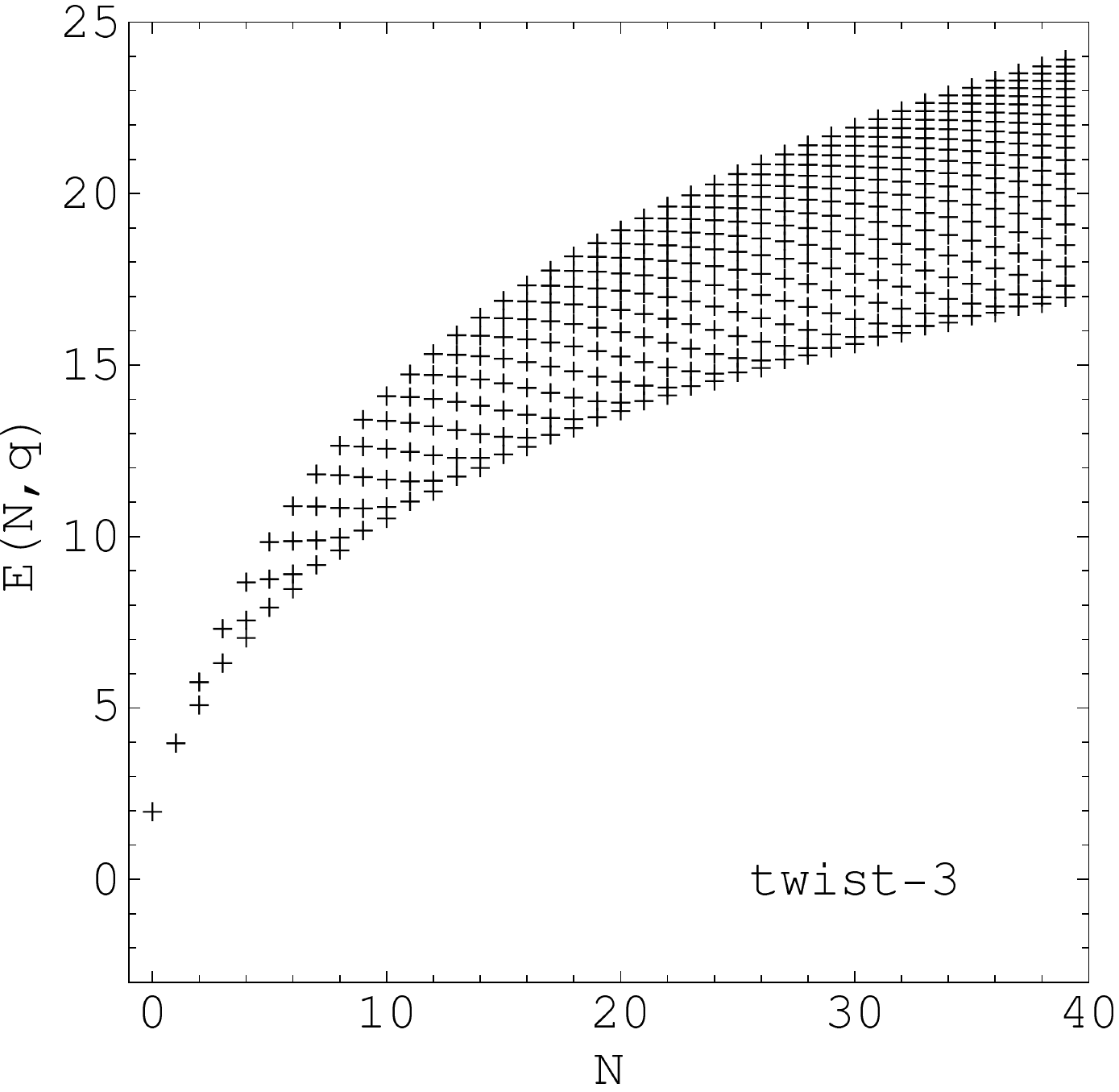}\hskip 0.5cm
\includegraphics[width=7cm]{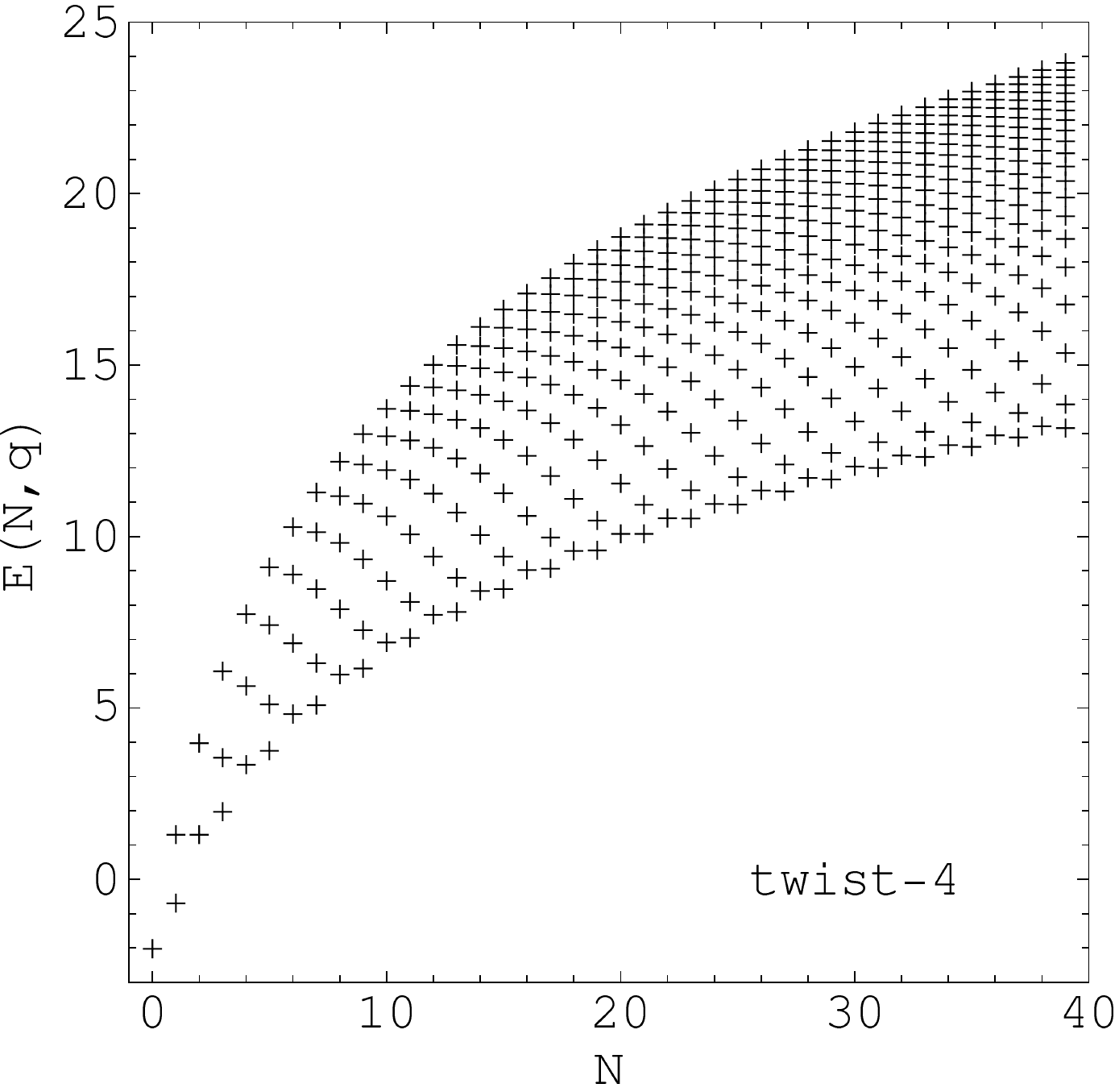}}
\vskip 5mm
\centerline{\includegraphics[width=7cm]{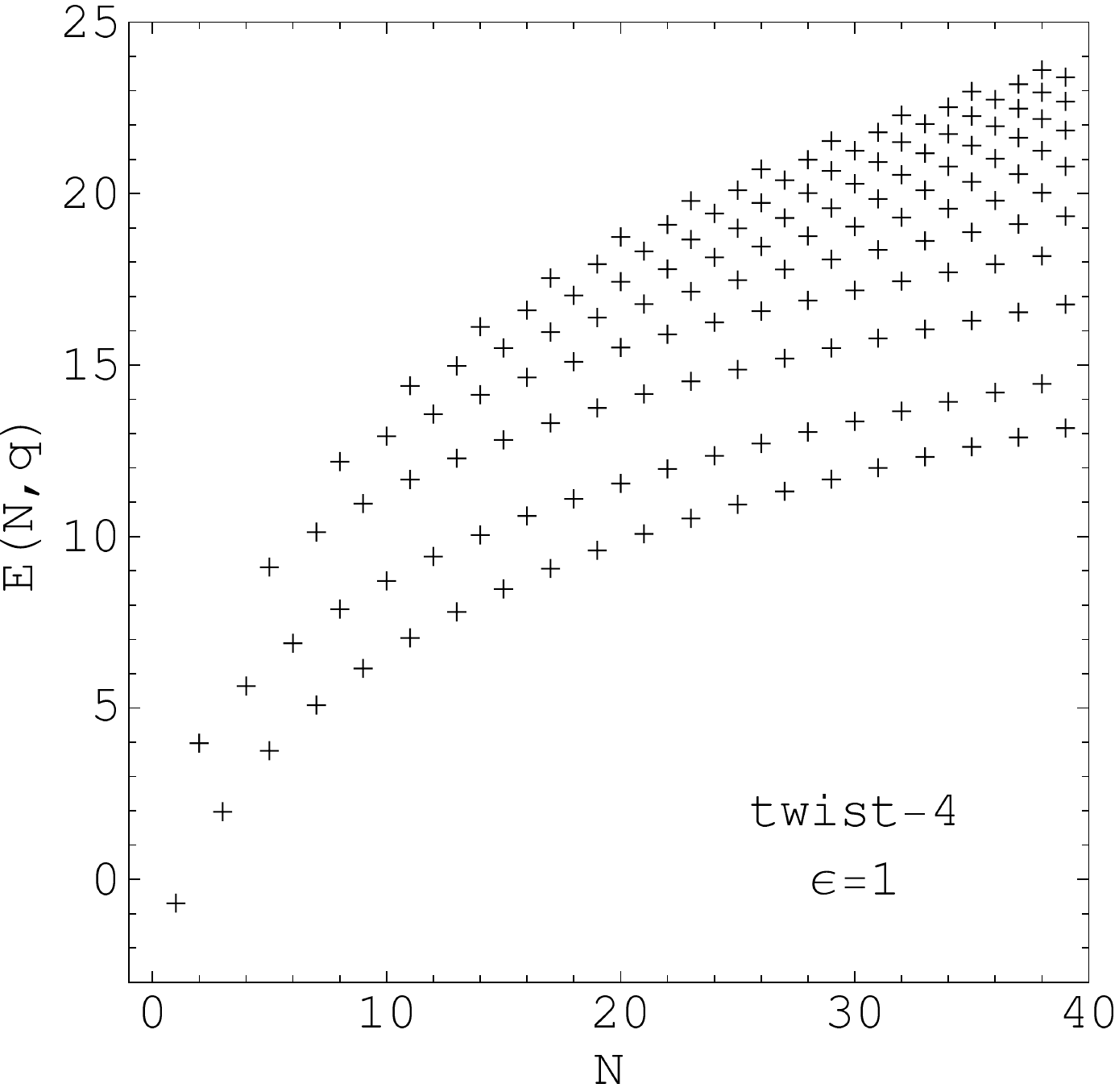}\hskip 0.5cm
\includegraphics[width=7cm]{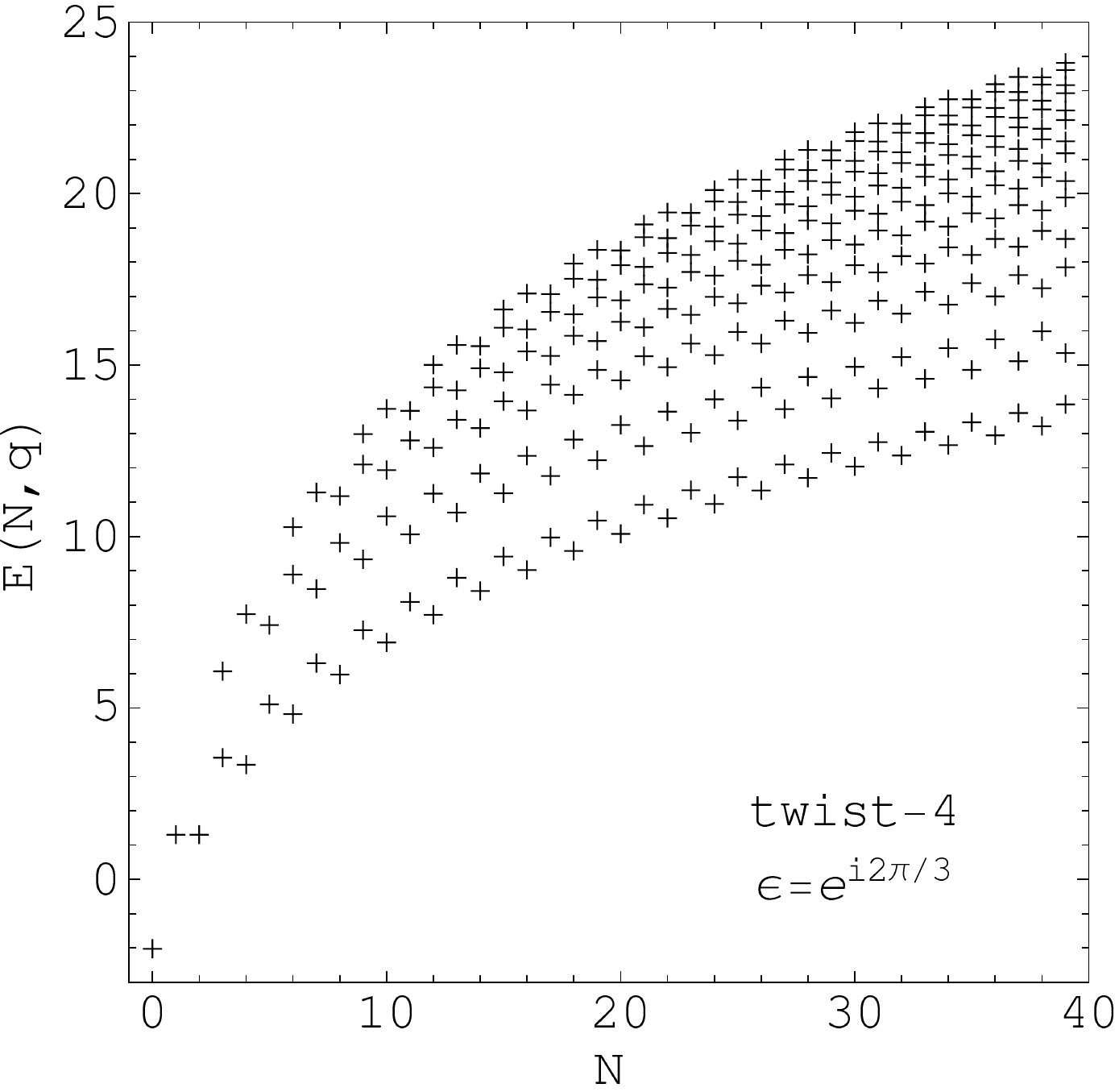}}
\caption{The spectrum of the Hamiltonian
$\mathbb{H}_q^{\psi\psi\psi}=(1+1/N_c)\mathcal{H}_q^{\psi\psi\psi}$.} 
\label{fig:chiral}
\end{figure}
%%%%%%%%%%%%%%%%%%%%%%%%%%%%%%%%%%%%%%%%%%%%%%%%%%%%%%%%%%%%%%%%%
%

The spectrum of the Hamiltonian $\mathcal{H}_q^{\psi\psi\psi}$ can be studied
using powerful Quantum Inverse Scattering Methods (QISM) ~\cite{Faddeev}
which is however beyond the scope of present paper. For this work we adopted
a ``brute-force'' method, calculating  $\mathcal{H}_q^{\psi\psi\psi}$ in the basis
of functions
\begin{align}
 e_{N,k}(z_1,z_2,z_3)=\frac{(z_1-z_2)^k\,(z_1-z_3)^{N-k}}{k!(N-k)!)}\,,
\label{basis1}
\end{align}
\begin{align}
   \mathcal{H}_q^{\psi\psi\psi} e_{N,k} = \sum_{k'=0}^N
(\mathcal{H}_q^{\psi\psi\psi})_{k'k}\, e_{N,k'} \,, 
\end{align}
and diagonalizing the resulting $(N+1)\times (N+1)$ matrix $(\mathcal{H}_q^{\psi\psi\psi})_{k'k}$
numerically.
The translation invariance of $e_{N,k}$ guarantees that the constraint  in
Eq.~(\ref{shifts-inv}) is satisfied identically.

The results are presented in Fig.~\ref{fig:chiral}. The spectra of twist-3 and twist-4
operators are shown in the upper left and upper right panels, respectively.
The two lower panels show the twist-4 spectra
in the sectors with $\varepsilon=1$ and $\varepsilon=e^{i2\pi/3}$ separately.
In addition, the numerical values of the eigenvalues for $N\le 6$ are collected
in Table~\ref{tab:Echiral}.

   %%%%%%%%%% Begin Table 1
    %%%%%%%%%%
\begin{table}[ht]
\renewcommand{\arraystretch}{1.3}
\begin{center}
\begin{tabular}{|c|c|c|c|c|}\hline
    $N$   & $E_{N,0}$    & $E_{N,1}$ & $E_{N,2}$ & $E_{N,3}$   \\ \hline
     0    &  $-2^\ast$   &    -               &  -   & -   \\
     1    &  $-\frac23$          &   $\frac43^\ast$  &  -  & -  \\
     2    &   $\frac43^\ast$     &   $4$        & -  & - \\
     3    &  $2$         &   $\frac{29-\sqrt{57}}6^\ast$   &  $\frac{29+\sqrt{57}}6^\ast$  &- \\
     4    &  $ \frac{167-3 \sqrt{481}}{30}^\ast$       &  $ \frac{17}{3}$ &
$ \frac{167+3 \sqrt{481}}{30}^\ast$  &-  \\
     5    &  $\frac{34}{9}$          &  $\frac{77}{15}^\ast$ &  $ \frac{67}9^\ast$  &
$\frac{137}{15}$
  \\
     6    &  $  3.633418^\ast$          &  $ 311/60$ &  $ 6.687457^\ast$  & $7.724361^\ast$ \\
\hline
\end{tabular}
\end{center}
\caption[]{\small  Anomalous dimensions of twist-4 chiral-quark operators
 in units of $\alpha_s/(2\pi)$; $N$ is the total number of covariant derivatives.
The entries marked with an asterisk correspond to the operators with
$\mathcal{P}$-parity $\varepsilon = e^{\pm i2\pi/3}$ and the remaining ones to
$\varepsilon=1$.
All anomalous dimensions except for the lowest ones, $E_{N,0}$,
for odd $N=2k+1$, are double degenerate.}
\label{tab:Echiral}
\vskip0.1cm
\renewcommand{\arraystretch}{1.0}
\end{table}

The general features of the twist-3 and twist-4 spectra are similar.
For both cases all  eigenvalues are double degenerate (see above), except for
those corresponding  to zero eigenvalue of the conserved charge
$\widehat Q_3\Psi_{N,q}= Q_3 \Psi_{N,q}$, $Q_3=0$.
These special eigenvalues turn out to be the lowest ones in the spectrum and can be found explicitly.
There are two series of such states, one in the twist-3 sector and one in the twist-4 sector.

The twist-3 eigenstates with $Q_3=0$ were studied in Ref.~\cite{BDKM}. They exist for even $N$,
are invariant under cyclic permutations and have the energy
\begin{align}\label{low-twist3}
E_{Q_3=0}^{tw-3}(N)=\left(1+\frac{1}{N_c}\right)
\left\{4\big[\psi(N+3)-\psi(2)\big]-\frac12\right\},\qquad N-\text{even}\,,
\end{align}
where $\psi(x)$ is the Euler $\psi$--function.
The twist-4 eigenstates with $Q_3=0$ exist for odd $N$. They are also invariant and cyclic permutations
($\varepsilon=1$) and have the  energy
\begin{align}\label{low-twist4}
E_{Q_3=0}^{tw-4}(N)=\left(1+\frac{1}{N_c}\right)
\left\{4\left[\psi\left(\frac{N+3}{2}\right)-\psi(2)\right]-\frac12\right\},\qquad N-\text{odd}.
\end{align}
The appearance of $(N+3)/2$ as argument of the Euler $\psi$--function is characteristic 
for systems involving half-integer conformal spins, cf.~\cite{Belitsky:1999bf,Beccaria:2008pp} 

As is seen from Fig.~\ref{fig:chiral} the upper parts of the twist-3 and twist-4 spectra
are very similar to
each other and are in fact interlacing. In particular the line of the largest eigenvalues
is the same for both twists and is given by
\cite{BDKM}
\begin{align}
  E_{\rm max}(N) = \left(1+\frac{1}{N_c}\right)
\left\{6 \ln N - 3 \ln 3 + 6\gamma_E +\frac32 +\mathcal{O}(1/N)\right\}.
\label{Emax}
\end{align}
The distance between the neighboring eigenvalues is in both cases $\mathcal{O}(1/N)$ in
the upper part and $\mathcal{O}(1/\ln^2 N)$ in the lower part of the spectra, respectively.

%%%%%%%%%%%%%%%%%%%%%%%%%%%%%%%%%%%%%%%%%%%%%%%%%%%%%%%%%%%%%%%%%%%%%%%%%%%%%%%%%%%%%%%%%%%%%%%
\subsection{Mixed chirality quark operators}
\begin{figure}[t]
\centerline{\includegraphics[width=7cm]{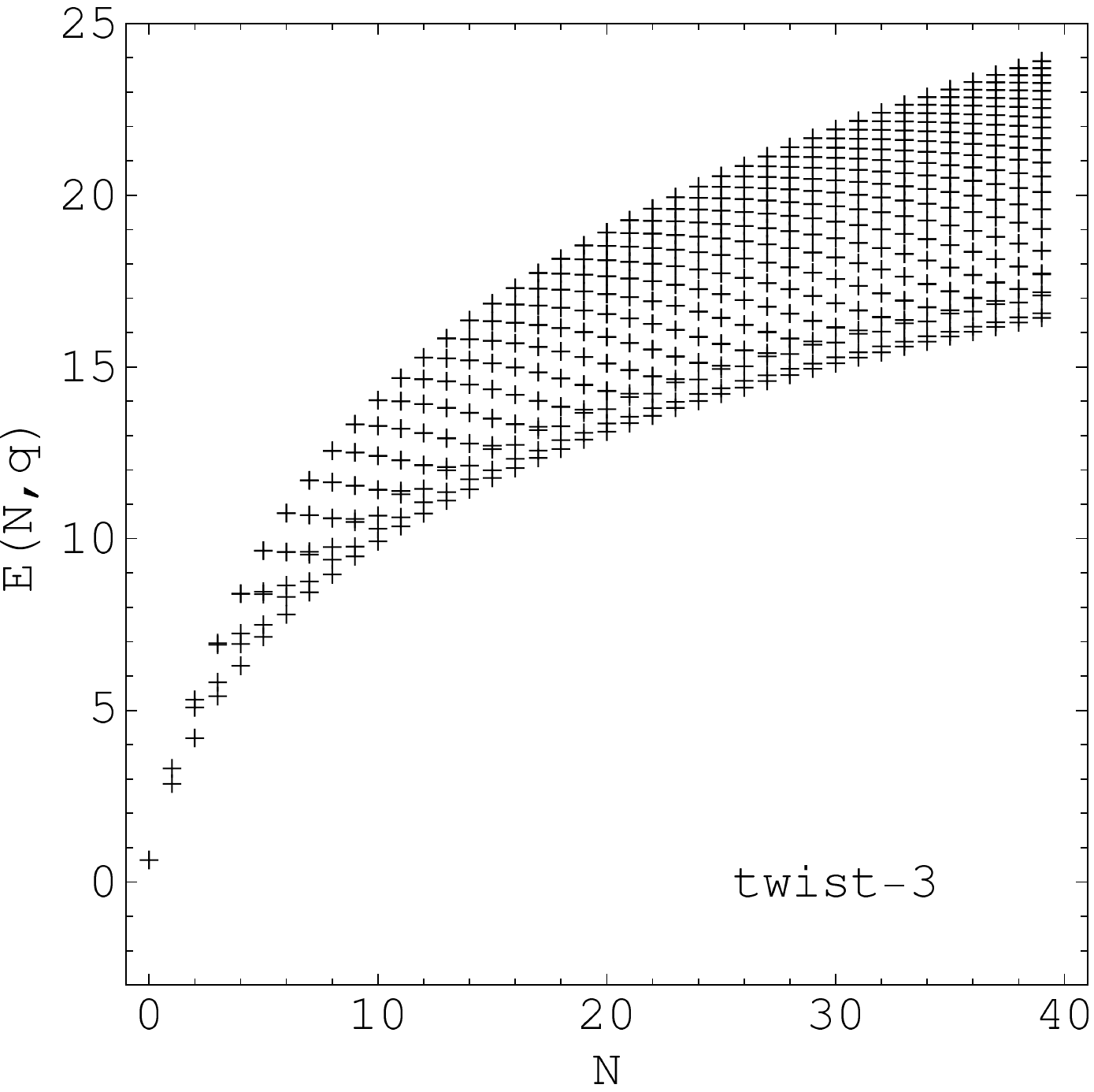}\hskip 0.5cm
\includegraphics[width=7cm]{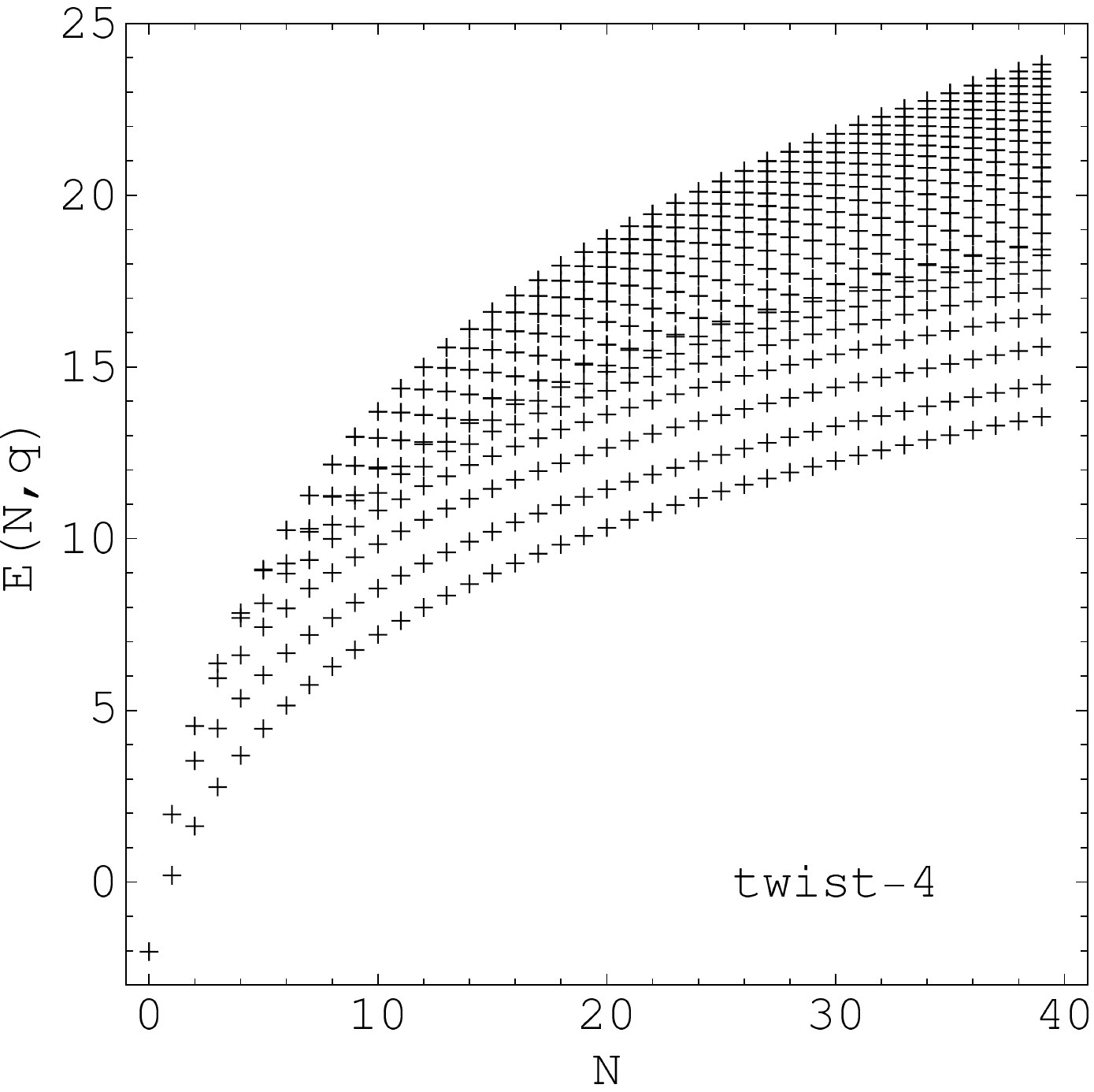}}
\caption{The spectrum of the Hamiltonian
$\mathbb{H}_q^{\psi\psi\bar\chi}=(1+1/N_c)\,\mathcal{H}_q^{\psi\psi\bar\chi}$.}
\label{fig:mixed}
\end{figure}

%%%%%%%%%%%%%%%%%%%%%%%%%%%%%%%%%%%%%%%%%%%%%%%%%%%%%%%%%%%%%%%%%%%%%%%%%%%%%%%%%%%%%%%%%%%%%%%
The analysis of operators of mixed chirality goes along the same lines.
The eigenfunctions of conformal twist-4 operators
have to satisfy the constraint in Eq.~(\ref{shifts-inv}) and in addition
\begin{align}\label{cons-2}
\frac{\partial}{\partial z_1}\Psi_{N,q}^{(1)}(\vec{z})+
\frac{\partial}{\partial z_2}\Psi_{N,q}^{(2)}(\vec{z})=\Psi^{(3)}_{N,q}(\vec{z})\,,
\end{align}
instead of (\ref{Constraint}).
The spectra of the operators of geometric twist-3 and twist-4 are shown in Fig.~\ref{fig:mixed}.
They are non-degenerate in both cases, in difference to the chiral operators.
The numerical values of the twist-4 eigenvalues for $N\le 4$ are collected in Table~\ref{tab:Emixed}.

%
%%%%%%%%%%%%%%%%%%%%%%%%%%%%%%%%%%%%%%%%%%%%%%%%%%%%%%%%%%%%%%%%%
%%%%%%%%%%%%%%%%%%%%%%%%%%%%%%%%%%%%%%%%%%%%%%%%%%%%%%%%%%%%%%%%%
%

   %%%%%%%%%% Begin Table 1
    %%%%%%%%%%
\begin{table}[ht]
\renewcommand{\arraystretch}{1.3}
\begin{center}
\begin{tabular}{|c|c|c|c|c|c|}\hline
$N$ & $E_{N,0}$ & $E_{N,1}$ & $E_{N,2}$ & $E_{N,3}$ & $E_{N,4}$   \\ \hline
     0    &  $-2$                 &    -               &  -   & -  &- \\
     1    &  $2/9$                    &   $2$ &  -  & - &- \\
     2    &  $\frac{2({14-\sqrt{43}})}{9}$     &   $32/9$   &  $\frac{2(14+\sqrt{43})}{9}$  &- &-\\
     3    &   $\frac{197-\sqrt{5089}}{45}$&   $\frac{49-\sqrt{73}}{9}$&
                                                $\frac{49+\sqrt{73}}{9}$  &
   $\frac{197+\sqrt{5089}}{45} $ &-  \\
     4    &  $3.706620 $ &
                           $ \frac{589-\sqrt{11161}}{90}$ &
                                                    $6.634936 $&
                                                      $ \frac{589+\sqrt{11161}}{90}$  &
$7.858442$
  \\
\hline
\end{tabular}
\end{center}
\caption[]{\small  Anomalous dimensions of twist-4 quark operators of mixed chirality
 in units of $\alpha_s/(2\pi)$; $N$ is the total number of covariant derivatives.}
\label{tab:Emixed}
\vskip0.1cm
\renewcommand{\arraystretch}{1.0}
\end{table}

The eigenvalues in the upper part of the spectra for twist-3 and twist-4 are very close to
each other and to the corresponding eigenvalues for chiral operators, cf. Fig.~\ref{fig:chiral}.
In particular, to the $\mathcal{O}(1/N)$ accuracy the line of the largest eigenvalues
does not depend on chirality. It is the same for both twists and is given by
Eq.~(\ref{Emax}).

%%%%%%%%%%%%%%%%%%%%%%%%%%%%%%%%%%%%%%%%%%%%%%%%%%%%%%%%%%%%%%%%%%%%%%%%%%%%%%%%%%%%%%%%%%%%%%%
\subsection{Quark-gluon operators}
%%%%%%%%%%%%%%%%%%%%%%%%%%%%%%%%%%%%%%%%%%%%%%%%%%%%%%%%%%%%%%%%%%%%%%%%%%%%%%%%%%%%%%%%%%%%%%%
Anomalous dimensions of four-particle twist-4 quark-gluon operators correspond to the
eigenvalues of the Hamiltonians $\mathbb{H}_g^{\psi\psi\psi \bar f}$ and
 $\mathbb{H}_g^{\psi\psi\chi \bar f}$
for the pure and mixed quark chirality cases, respectively.

The Hamiltonian  $\mathbb{H}^{\psi\psi\psi \bar f}$ is given explicitly in Eqs.~(\ref{gluonB})--(\ref{offd}).
It has the same symmetries as   $\mathbb{H}^{\psi\psi\psi}$ and its eigenfunctions can be classified by
parity  with respect to the cyclic permutation $\mathcal{P}$ of the three quarks  (\ref{cyclic}),
$\varepsilon=1, e^{\pm i2\pi/3}$
Due to the identity (\ref{3gidentity}) we are interested in eigenfunctions which satisfy the restriction
\begin{align}\label{}
\sum_{i=1}^3\Psi^{(i)}_{N,q}(z_1,z_2,z_3,z_4)=0\,.
\end{align}
The eigenfunctions belonging to the sectors with $\varepsilon=e^{\pm i2\pi/3}$ have the
same eigenvalues, $E_{N,q}(\varepsilon)=E_{N,q}(\varepsilon^{-1})$, and are related to each other as
$ \Psi^{\varepsilon^{-1}}_{N,q}(\vec{z})=\mathcal{P}_{12}\Psi^{\varepsilon}_{N,q}(\vec{z})$.
As in the case of quark operators, in the applications to nucleon distribution amplitudes
we are interested in the eigenfunctions of particular symmetry, cf.~(\ref{ggr-1}).
One can easily verify that the combination
%
%\begin{align}\label{}
$\Psi_{N,q}(\vec{z})=(1+\mathcal{P}_{12})\,\Psi^{\varepsilon}_{N,q}(\vec{z})$
%\end{align}
%
with $\varepsilon=e^{i2\pi/3}$ has all the necessary  properties.
In turn, the eigenvalues with $\varepsilon=1$ are not degenerate
and the corresponding operators (eigenfunctions) are relevant e.g. for the $\Delta$-baryon.

In order to calculate the spectrum we have used the basis of  functions
\begin{align}
 e_{N,k,m}(z_1,z_2,z_3,z_4)=\frac{(z_1-z_4)^k\,(z_2-z_4)^m(z_3-z_4)^{N-2-k-m}}{k!\,m!\,(N-2-k-m)!}
\label{basis2}
\end{align}
and diagonalized the resulting $N(N-1)\times N(N-1)$ matrix numerically.
The translation invariance of $e_{N,k.m}$ corresponds to the restriction to conformal operators,
cf. Eq.~(\ref{shifts-inv}).

%
%%%%%%%%%%%%%%%%%%%%%%%%%%%%%%%%%%%%%%%%%%%%%%%%%%%%%%%%%%%%%%%%%
\begin{figure}[t]
\centerline{\includegraphics[width=9cm]{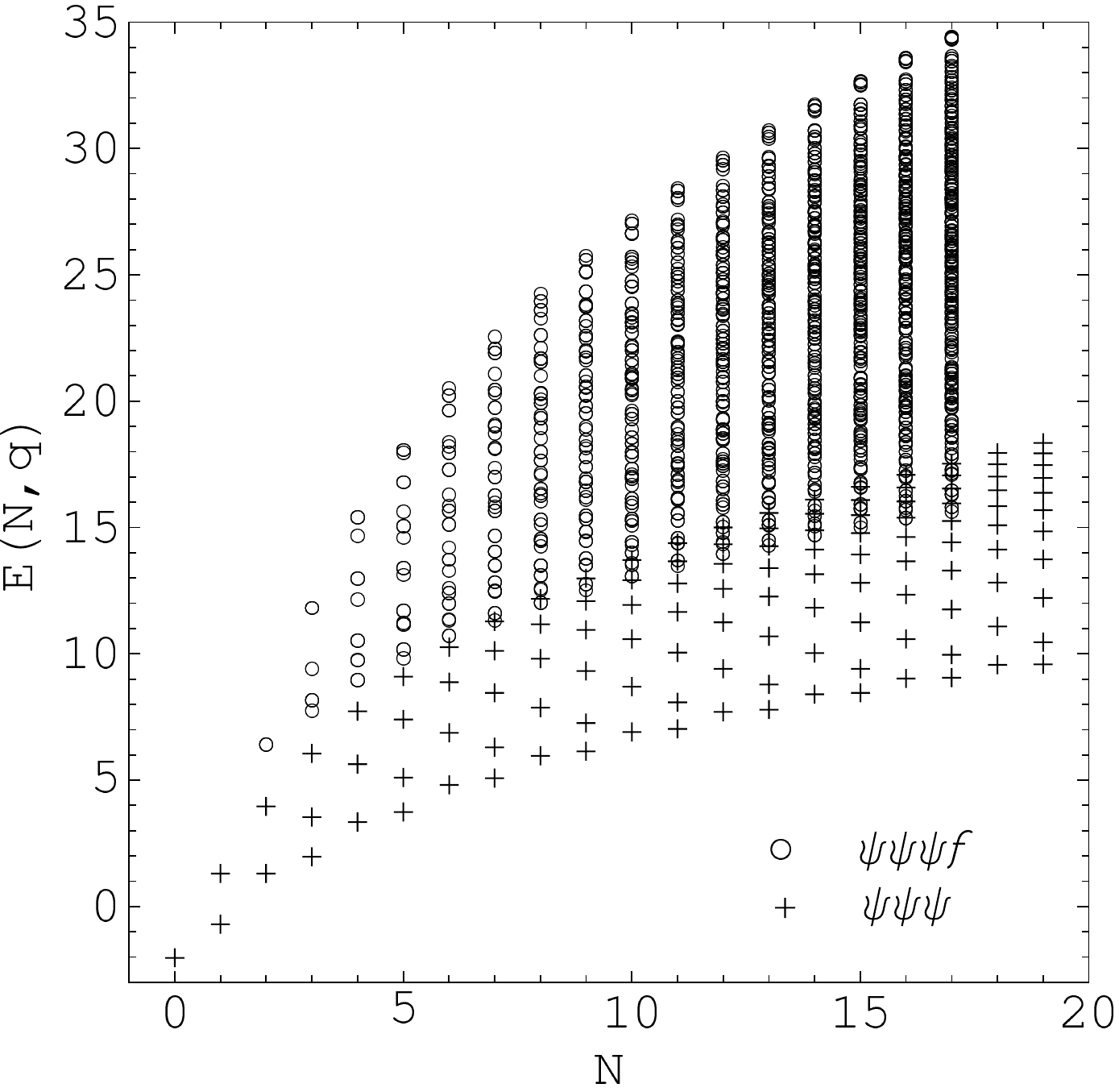}\hskip 0.5cm }
\caption{The combined spectrum of the Hamiltonians $\mathbb{H}_q^{\psi\psi\psi}$ (crosses) and
$\mathbb{H}_g^{\psi\psi\psi\bar f}$ (open circles), all parities.}
\label{fig:chiral-joint}
\end{figure}
%%%%%%%%%%%%%%%%%%%%%%%%%%%%%%%%%%%%%%%%%%%%%%%%%%%%%%%%%%%%%%%%%
%
%
%%%%%%%%%%%%%%%%%%%%%%%%%%%%%%%%%%%%%%%%%%%%%%%%%%%%%%%%%%%%%%%%%%
\begin{figure}[ht]
\centerline{\includegraphics[width=9cm]{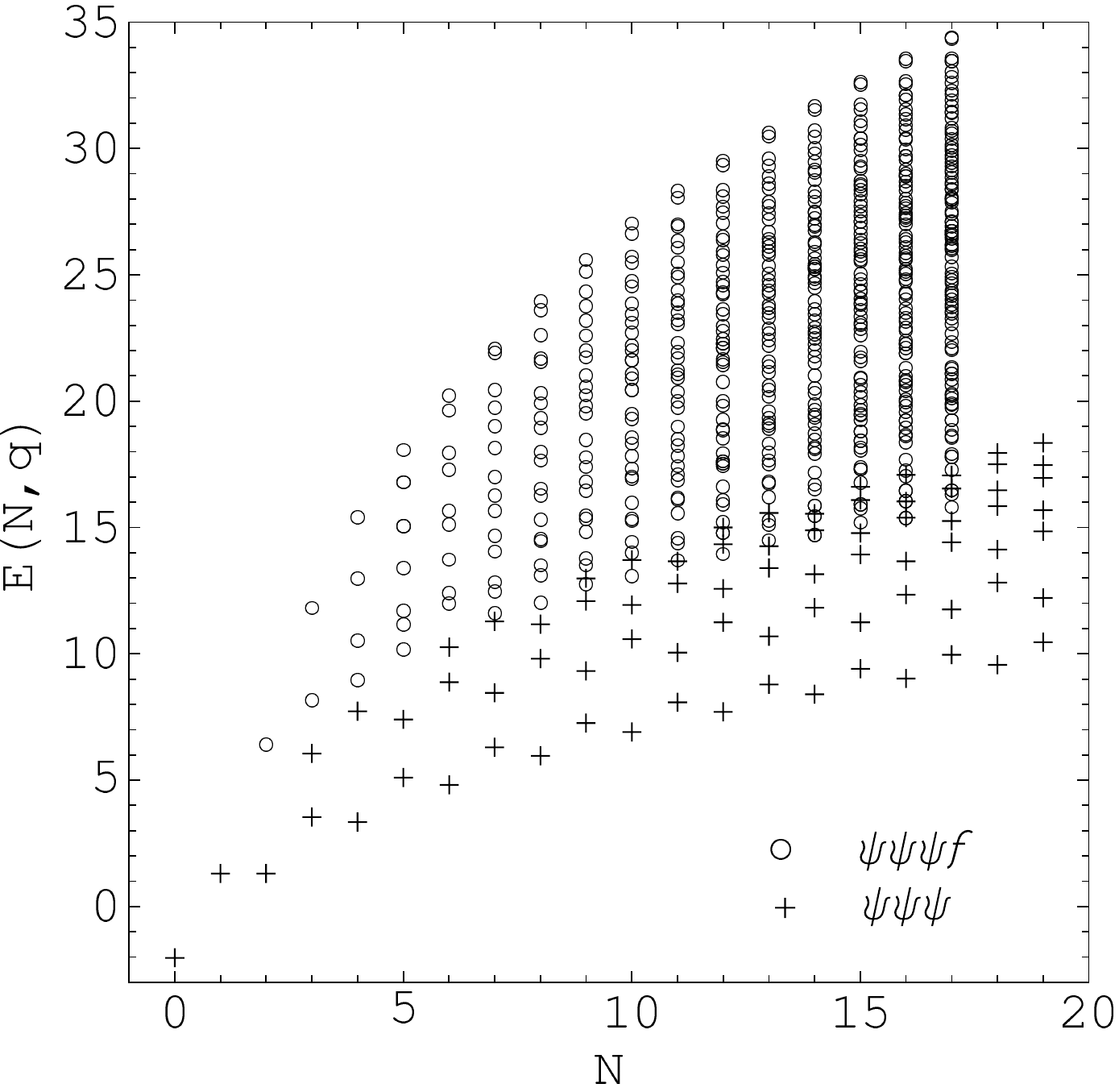}\hskip 0.5cm
}
\caption{The same as in Fig.~\ref{fig:chiral-joint}, 
but for $\varepsilon=e^{i2\pi/3}$ only. All eigenvalues
are double--degenerate.}
\label{fig:nucl-chiral}
\end{figure}
%%%%%%%%%%%%%%%%%%%%%%%%%%%%%%%%%%%%%%%%%%%%%%%%%%%%%%%%%%%%%%%%%
%

The combined spectrum of three-quark and three-quark-gluon chiral operators is shown
in Fig.~\ref{fig:chiral-joint}. (all parities) and in Fig.~\ref{fig:nucl-chiral} ( $\varepsilon=e^{i2\pi/3}$ only).
Note that the quark-gluon spectrum is much more dense as for given $N$ there  are $\mathcal{O}(N)$ quark
and $\mathcal{O}(N^2)$ quark-gluon eigenstates. Also, it is seen that for large $N$ the spectra overlap significantly.
It is easy to show that for $N\to\infty$ the eigenvalues lie within the bands
\begin{align}
\frac{16}{3} \ln N <  E_{N,q}^{\psi\psi\psi} < 8 \ln N\,,\qquad
\frac{16}{3} \ln N <  E_{N,q}^{\psi\psi\psi\bar f } < 14 \ln N\,.
\end{align}
Note that the lowest four-particle quark-gluon eigenvalue has the same logarithmic asymptotic as
the lowest three-quark one, so that they are separated at most by a constant. This suggests
that for most of the quark eigenstates there is strong mixing with the quark-gluon ones,
see the next Section.

\begin{figure}[ht]
\centerline{\includegraphics[width=9cm]{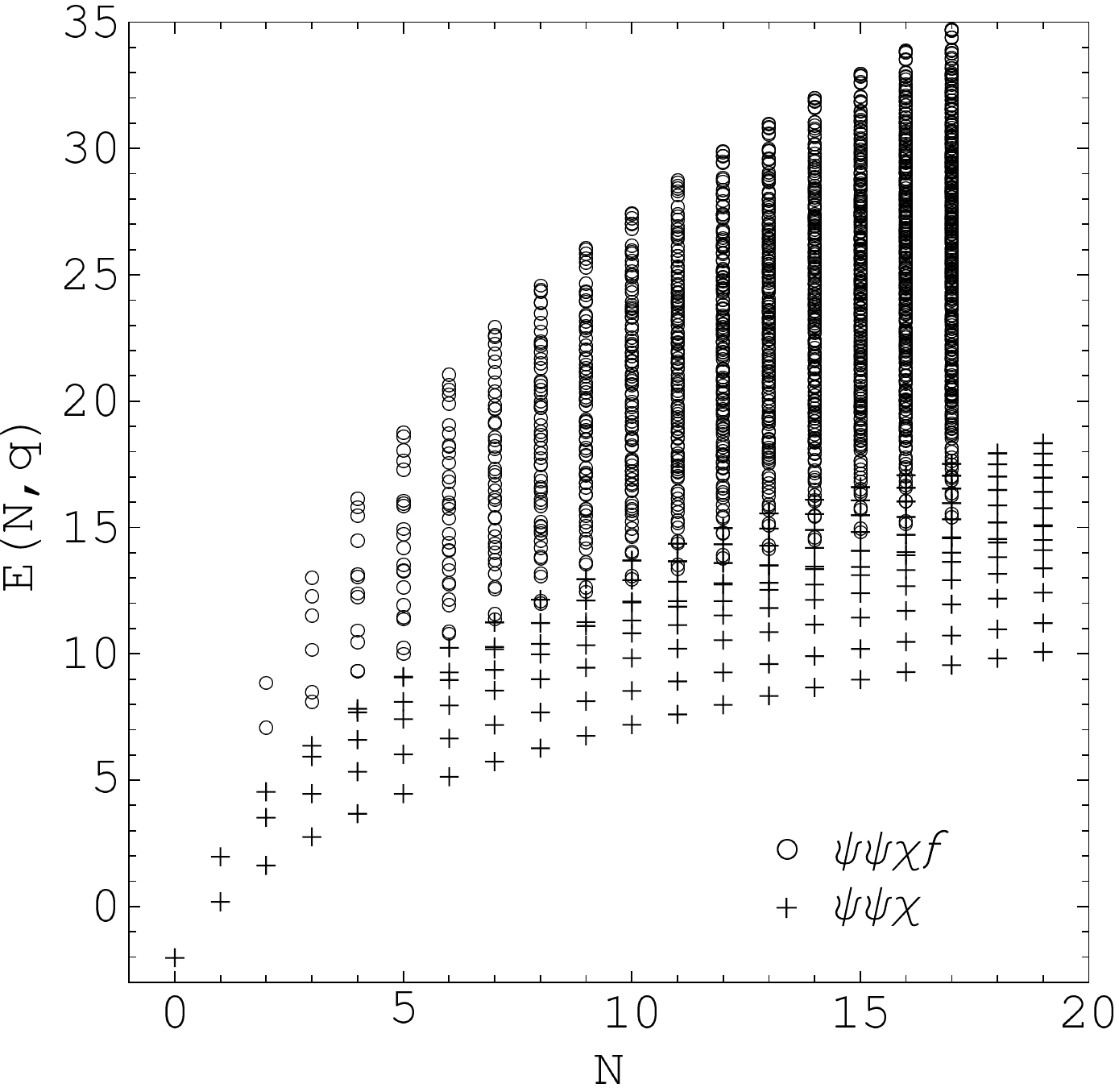}\hskip 0.5cm
}
\caption{The combined spectrum of the Hamiltonians $\mathbb{H}_q^{\psi\psi\chi}$ (crosses) and
$\mathbb{H}_g^{\psi\psi\chi\bar f}$ (open circles).}
\label{fig:joint-mixed}
\end{figure}

The combined spectrum of three-quark and three-quark-gluon operators with mixed chirality is shown
in Fig.~\ref{fig:joint-mixed}. It is similar to the chiral case, except that all
eigenvalues are nondegenerate.

%
%%%%%%%%%%%%%%%%%%%%%%%%%%%%%%%%%%%%%%%%%%%%%%%%%%%%%%%%%%%%%%%%%%
%%%%%%%%%%%%%%%%%%%%%%%%%%%%%%%%%%%%%%%%%%%%%%%%%%%%%%%%%%%%%%%%%
%

%%%%%%%%%%%%%%%%%%%%%%%%%%%%%%%%%%%%%%%%%%%%%%%%%%%%%%%%%%%%%%%%%%%%%%%%%%%%%%%%%%%%%%%%%%%%%%%%%%%
\subsection{Multiplicatively renormalizable operators}
%%%%%%%%%%%%%%%%%%%%%%%%%%%%%%%%%%%%%%%%%%%%%%%%%%%%%%%%%%%%%%%%%%%%%%%%%%%%%%%%%%%%%%%%%%%%%%%%%%%

For the construction of multiplicatively renormalizable operators one has to take into account the
off-diagonal quark-gluon blocks
$\mathbb{H}_{qg}^{\mathrm{chiral}}$ and  $\mathbb{H}_{qg}^{\mathrm{mixed}}$
for the chiral and mixed cases, respectively.
Note that the complete (reduced) evolution Hamiltonian $\widetilde{\mathbb{H}}$ is not Hermitian,
therefore its eigenfunctions are not mutually orthogonal.

In order to find the multiplicatively renormalizable operators we adopt the following procedure.
The nonlocal operator $\mathbb{O}(\vec{z})$ can be expanded over a complete  basis of
multiplicatively renormalizable local operators (\ref{def:coef-fun})
%
%\begin{align}
%\mathbb{O}(\vec{z})=\sum_{N,q}\Psi_{N,q}(\vec{z})\, \mathbb{O}_{N,q}\,,
%\label{def:coef-fun1}
%\end{align}
%
with the coefficient functions $ \Psi_{N,q}(\vec{z})$ that are  eigenfunctions of the Hamiltonian
$\widetilde{\mathbb{H}}$
\begin{align}\label{def:Schr-eq1}
[\widetilde{\mathbb{H}}\Psi_{N,q}](\vec{z})=E_{N,q}\Psi_{N,q}(\vec{z})\,,
\end{align}
cf.~(\ref{def:Schr-eq}).
Let $\Psi^\dagger_{N,q}$ be the eigenfunctions of the adjoint operator $\widetilde{\mathbb{H}}^\dagger$
\begin{align}\label{def:Schr-eq1-a}
[\widetilde{\mathbb{H}}^\dagger\Psi^\dagger_{N,q}](\vec{z})=E_{N,q}\Psi^\dagger_{N,q}(\vec{z})\,.
\end{align}
 The functions $\Psi_{N,q}(\vec{z})$ and $\Psi^\dagger_{N,q}(\vec{z})$ form a bi-orthogonal system:
\begin{align}\label{orth}
\langle\Psi^\dagger_{N',q'}|\Psi_{N,q}\rangle=\delta_{NN'}\delta_{qq'}\,.
\end{align}
Eigenstates corresponding to different eigenvalues are orthogonal with respect to the scalar product
(\ref{scqg}).
The multiplicatively renormalizable local operators $\mathbb{O}_{N,q}$ can be obtained as
the scalar product of the
nonlocal operator with the eigenfunction of the adjoint operator
\begin{align}
\mathbb{O}_{N,q}=\langle\Psi^\dagger_{N,q}|\mathbb{O}\rangle.
\end{align}
With the increasing $N$ the expressions rapidly become very cumbersome so that
in this work we present explicit results  for $N=0,1,2$ only.

To begin with, consider chiral operators. Let
\begin{align}\label{Qoperators_local}
Q_1^{(k_1,k_2,k_3)}=&\epsilon^{ijk} [(n\cdot D)^{k_1}\psi^{a}_-]^i\, [(n\cdot
D)^{k_2}\psi^{b}_+]^j \, [(n\cdot D)^{k_3}\psi^{c}_+]^k\,,
\nonumber\\
Q_2^{(k_1,k_2,k_3)}=&\epsilon^{ijk} [(n\cdot D)^{k_1}\psi^{a}_+]^i\, [(n\cdot
D)^{k_2}\psi^{b}_-]^j \, [(n\cdot D)^{k_3}\psi^{c}_+]^k\,,
\nonumber\\
Q_3^{(k_1,k_2,k_3)}=&\epsilon^{ijk} [(n\cdot D)^{k_1}\psi^{a}_+]^i\, [(n\cdot
D)^{k_2}\psi^{b}_+]^j \, [(n\cdot D)^{k_3}\psi^{c}_-]^k\,,
\end{align}
and
\begin{align}\label{Goperators_local}
{G}_1^{(k_1,k_2,k_3,k_4)}=&
        ig\epsilon^{ijk} \,(\mu\lambda)\,
[(n\cdot D)^{k_4}\bar f_{++}(n\cdot D)^{k_1}\psi^{a}_+]^i\, [(n\cdot D)^{k_2}\psi^{b}_+]^j\,
       [(n\cdot D)^{k_3}\psi^{c}_+]^k\,,
\notag\\
{G}_2^{(k_1,k_2,k_3,k_4)}=&
           ig\epsilon^{ijk} (\mu\lambda)\,
[(n\cdot D)^{k_1}\psi^{a}_+]^i\, [(n\cdot D)^{k_4}\bar f_{++} (n\cdot D)^{k_2}\psi^{b}_+]^{j}\,
[(n\cdot D)^{k_3}\psi^{c}_+]^k\,,
%\notag\\
%{G}_3(z_1,z_2,z_3,z_4)=&
%        ig\epsilon^{ijk} \psi^{a,i}_+(z_1)\, \psi^{b,j}_+(z_2)\,[\bar f_{++}(z_4) \psi^{c}_+(z_3)]^k\,,
\end{align}
As discussed above, the eigenfunctions alias multiplicatively renormalizable operators
can always be chosen in this case to have definite parity with respect to the cyclic
permutations~(\ref{permut}).
We remind that the spectrum for $\varepsilon = e^{\pm i2\pi/3}$ is double degenerate.
Any linear combination of the corresponding eigenfunctions satisfies the evolution equation.
We present the results corresponding to the operators of definite symmetry
under permutation of the first and the second quark
\begin{align}
\Psi^\pm_{N,q}=(1\pm \mathcal{P}_{12})\Psi_{N,q}^{\varepsilon}
\end{align}
which are more convenient for applications than those with definite $\varepsilon = e^{\pm i2\pi/3}$.
One obtains
\begin{align}\label{chiral+}
\mathbb{O}_{0,0}^{chiral,+}=&{{Q}}_1^{(000)}+{{Q}}_2^{(000)}-2{{Q}}_3^{(000)}\,,
\notag\\
%{x-3 y+2 z,-3 x+y+2 z,x+y-2 z}
\mathbb{O}_{1,1}^{chiral,+}=&
{{Q}}_1^{(100)}-\frac32{{Q}}_1^{(010)}+{{Q}}_1^{(001)}
%x-3 y+2 z
-\frac32 Q_2^{(100)}+{{Q}}_2^{(010)}+{{Q}}_2^{(001)}
%-3 x+y+2 z
+\frac12 {{Q}}_3^{(100)}
\notag\\&
+\frac12{{Q}}_3^{(010)}
-2{{Q}}_3^{(001)}\,,
%x+y-2 z
\notag\\
%{2 x^2-8 x y+12 y^2+4 x z-16 y z+6 z^2,12 x^2-8 x y+2 y^2-16 x z+4 y \
%z+6 z^2,-18 x^2+32 x y-18 y^2+4 x z+4 y z-4 z^2}
\mathbb{O}_{2,0}^{chiral,+}=&
{{Q}}_1^{(200)}+2{{Q}}_1^{(020)}+{{Q}}_1^{(002)}
-4 {{Q}}_1^{(110)} +2 {{Q}}_1^{(101)} - 4 {{Q}}_1^{(011)}
+2{{Q}}_2^{(200)}
\notag\\ &
+{{Q}}_2^{(020)}+{{Q}}_2^{(002)}
-4{{Q}}_2^{(110)} - 4{{Q}}_2^{(101)} +2 {{Q}}_2^{(011)}
-3{{Q}}_3^{(200)} -3 {{Q}}_3^{(020)}
\notag\\ &
- 2{{Q}}_3^{(002)}
+8{{Q}}_3^{(110)} +2 {{Q}}_3^{(101)} +2 {{Q}}_3^{(011)}
-\frac{7}{12}G_1^{(0000)}-\frac{7}{12} G_2^{(0000)}\,,
\notag\\
\mathbb{O}_{2,0}^{g,chiral,+}=& \frac32\left(G^{(0000)}_1+G^{(0000)}_2\right)\,,
\qquad E_{2,0}^{g,chiral}=19/3\,
\end{align}
and
\begin{align}\label{}
\mathbb{O}_{0,0}^{chiral,-}=&{{Q}}_1^{(000)}-{{Q}}_2^{(000)}\,,
\notag
\\
%
%3 x+y-4 z,-x-3 y+4 z,-5 x+5 y
\mathbb{O}_{1,1}^{chiral,-}=&3{{Q}}_1^{(100)}+\frac12{{Q}}_1^{(010)}-2\,{{Q}}_1^{(001)}
-\frac12 Q_2^{(100)}-3\,{{Q}}_2^{(010)}+2\,{{Q}}_2^{(001)}
\notag\\ &
-\frac52{{Q}}_3^{(100)}+\frac52{{Q}}_3^{(010)}\,,
\notag\\
%{2 x^2+8 y^2-4 x z-16 y z+10 z^2,-8 x^2-2 y^2+16 x z+4 y z-10 z^2,2 x^2-2 y^2-4 x z+4 y z}
\mathbb{O}_{2,0}^{chiral,-}=&{{Q}}_1^{(200)}+\frac43{{Q}}_1^{(020)}+\frac53{{Q}}_1^{(002)}
-2\,{{Q}}_1^{(101)}-4\,{{Q}}_1^{(011)}-\frac43{{Q}}_2^{(200)}-{{Q}}_2^{(020)}
-\frac53{{Q}}_2^{(002)}
\notag\\
&
+4Q_{2}^{(101)}+2Q_2^{(011)}
%\notag\\
%&
+\frac13{{Q}}_3^{(200)}-\frac13{{Q}}_3^{(020)}
-2{{Q}}_3^{(101)} +2 {{Q}}_3^{(011)}
\notag\\
&-\frac7{36}G_1^{(0000)}+\frac7{36}G_2^{(0000)} \,,
\notag\\
\mathbb{O}_{2,0}^{g,chiral,-}=& \frac12\left(G^{(0000)}_1-G^{(0000)}_2\right)\,,
\qquad E_{2,0}^{g,chiral}=19/3\,.
\end{align}
The multiplicatively renormalizable operators of the lowest dimension
in the $\varepsilon=1$ sector are
\begin{align}	
%\begin{eqnarray}\label{}
%{y-z,-x+z,x-y}
\mathbb{O}_{1,0}^{chiral,1}=&Q_1^{(010)}-Q_1^{(001)}-Q_2^{(100)}+Q_2^{(001)}+Q_3^{(100)}-Q_3^{(010)}\,,
\notag\\
%{-12 x y+6 y^2+12 x z-6 z^2,-6 x^2+12 x y-12 y z+6 z^2,6 x^2-6 y^2-12 x z+12 y z}
\mathbb{O}_{2,1}^{chiral,1a}=&Q_1^{(020)}-Q_1^{(002)}-6 Q_1^{(110)}+6 Q_1^{(101)}
-Q_2^{(200)}+Q_2^{(002)}+6Q_2^{(110)}-6 Q_2^{(011)}
\notag\\
&+Q_3^{(200)}-Q_3^{(020)}-6Q_3^{(101)}+6 Q_3^{(011)}\,,
\notag\\
%
%{2 x^2-2 x y-3 y^2-2 x z+8 y z-3 z^2,-3 x^2-2 x y+2 y^2+8 x z-2 y z-3 z^2,
%-3 x^2+8 x y-3 y^2-2 x z-2 y z+2 z^2}
\mathbb{O}_{2,1}^{chiral,1b}=&Q_1^{(200)}-\frac12 Q_1^{(020)}-\frac12 Q_1^{(002)}-
Q_1^{(110)}-Q_1^{(101)}+2\,Q_1^{(011)}
\notag\\
&-\frac12Q_2^{(200)}+Q_2^{(020)}-\frac12Q_2^{(002)}-Q_2^{(110)}+2\,Q_2^{(101)}-Q_2^{(011)}
\notag\\
&-\frac12Q_3^{(200)}-\frac12Q_3^{(020)}+Q_3^{(002)}+2\,Q_3^{(110)}-Q_3^{(101)}-Q_3^{(011)}\,.
%\end{eqnarray}
\end{align}
The operators $\mathbb{O}_{2,1}^{chiral,1a}$ and $\mathbb{O}_{2,1}^{chiral,1b}$ have the
same anomalous dimension $E^{\rm chiral}_{2,1}=4$, cf. Table~\ref{tab:Echiral}, so that any
linear combination of them is multiplicatively renormalizable as well. Note also that for $N=2$
there is no quark-gluon operator, the reason being that for
$\varepsilon=1$ the cyclic permutation symmetry picks up the combination
which is forbidden because of the condition in Eq.~(\ref{3gidentity}).

Next we consider the operators involving quark fields of mixed chirality. Let
\begin{align}\label{qbx_local}
{\mathcal{Q}}_1^{(k_1,k_2,k_3)}=&
\epsilon^{ijk}\,[(n\cdot D)^{k_1}\psi_-^{a}]^i\,
[(n\cdot D)^{k_2}\psi_+^{b}]^j\,[(n\cdot D)^{k_3}\bar\chi_+^{c}]^k\,,
\notag\\	
{\mathcal{Q}}_2^{(k_1,k_2,k_3)}=&
\epsilon^{ijk}\,[(n\cdot D)^{k_1}\psi_+^{a}]^i\,
[(n\cdot D)^{k_2}\psi_-^{b}]^j\,[(n\cdot D)^{k_3}\bar\chi_+^{c}]^k\,,
\notag\\
{\mathcal{Q}}_3^{(k_1,k_2,k_3)}=&\frac12
\epsilon^{ijk}\,[(n\cdot D)^{k_1}\psi_+^{a}]^i\, [(n\cdot D)^{k_2}\psi_+^{b}]^j\,[(n\cdot
D)^{k_3}\bar\chi_+^{3/2,c}]^k\,,
\end{align}
%
%({\it This $Q_3$=$2Q_3^{my\, notes}$}.)
where $\bar \chi_+^{3/2}\equiv \bar \chi_+^{(3/2,0)} = -(\mu D\bar\lambda)\bar\chi_+
\equiv  - D_{\mu\dot\lambda} \bar\chi_+$, cf. Eq.~(\ref{basis}), and
\begin{align}\label{qbg_local}
\mathcal{G}_1^{(k_1,k_2,k_3,k_4)}=&
ig\epsilon^{ijk}\,(\mu\lambda)\,
[(n\cdot D)^{k_4}\bar f_{++}(n\cdot D)^{k_1}\psi_+^a]^i\,  [(n\cdot D)^{k_2}\psi_+^{b}]^j\,
[(n\cdot D)^{k_3}\bar\chi_+^{c}]^k\,,
\notag\\
\mathcal{G}_2^{(k_1,k_2,k_3,k_4)}=&
ig\epsilon^{ijk}\,(\mu\lambda)\,
[(n\cdot D)^{k_1}\psi_+^{a}]^i\,[(n\cdot D)^{k_4}\bar f_{++}(n\cdot D)^{k_2}\psi_+^{b}]^j\,
[(n\cdot D)^{k_3}\bar\chi_+^{c}]^k\,.
\end{align}
In the mixed case there is no additional symmetry and all eigenvalues are non-degenerate.
The multiplicatively renormalizable operators of the lowest dimension read
\begin{align}\label{O-mixed}
\mathbb{O}_{0,0}^{mixed}=&{\mathcal{Q}}_1^{(000)}-{\mathcal{Q}}_2^{(000)}\,,
\notag\\
%\end{align}
%\begin{align}
\mathbb{O}_{1,0}^{mixed}=&\mathcal{Q}_1^{(100)}+\mathcal{Q}_1^{(010)}-\frac32
\mathcal{Q}_1^{(001)}
%z1/2+z2-(3 z3)/2
-\mathcal{Q}_2^{(100)}-\mathcal{Q}_2^{(010)}+\frac32
\mathcal{Q}_2^{(001)}\,,
%-z1-z2/2+(3 z3)/2
\notag\\
\mathbb{O}_{1,1}^{mixed}=&\mathcal{Q}_1^{(100)}-\mathcal{Q}_1^{(010)}+\frac12\mathcal{Q}_1^{(001)}
%z1-2z2+z3
-\mathcal{Q}_2^{(100)}+\mathcal{Q}_2^{(010)}+\frac12\mathcal{Q}_2^{(001)}+
%-2z1+z2+z3
\mathcal{Q}_3^{(000)}\,,
\notag\\
\mathbb{O}_{2,1}^{mixed}=&
\mathcal{Q}_1^{(200)}-\mathcal{Q}_1^{(020)}-\frac23\mathcal{Q}_1^{(002)}
-2\mathcal{Q}_1^{(101)}+3\mathcal{Q}_1^{(011)}%\notag\\
%z1^2/6-z2^2/2-(z1 z3)/3+z2 z3-z3^2/3
-\mathcal{Q}_2^{(200)}+\mathcal{Q}_2^{(020)}-\frac23\mathcal{Q}_2^{(002)}\notag\\
&-2\mathcal{Q}_2^{(011)}+3\mathcal{Q}_2^{(101)}
%-z1^2/2+z2^2/6+z1 z3-(z2 z3)/3-z3^2/3
%\notag\\
%&
+\mathcal{Q}_3^{(100)}+\mathcal{Q}_3^{(010)}-\frac43\mathcal{Q}_3^{(001)}
%z1/3+z2/3-(2 z3)/3
+\frac{91}{282} \mathcal{G}_1^{(0000)}+\frac{91}{282} \mathcal{G}_2^{(0000)}\,,
\notag\\
%\end{align}
%\begin{align}
\left(\begin{array}{c}
\mathbb{O}_{2,0}^{mixed} \\ \mathbb{O}_{2,2}^{mixed}
\end{array}\right)
%\mathbb{O}_{\scriptsize\begin{array}{c}2,0 \\[-1mm] 2,2\end{array}}^{mixed}
=&\mathcal{Q}_1^{(200)}+\mathcal{Q}_1^{(020)}+
\frac4{27}\left(4\pm\sqrt{43}\right)\mathcal{Q}_1^{(002)}+
\frac49\left(-5\pm\sqrt{43}\right)\mathcal{Q}_1^{(110)}\notag\\
&+\frac29\left(1\mp2\sqrt{43}\right)\mathcal{Q}_1^{(101)}-\frac19\left(17\pm
2\sqrt{43}\right)\mathcal{Q}_1^{(011)}
%2 z1^2-(40 z1 z2)/9+8/9 Sqrt[43] z1 z2+6 z2^2+(4 z1 z3)/9-8/9 \
%Sqrt[43] z1 z3-(68 z2 z3)/9-8/9 Sqrt[43] z2 z3+(32 z3^2)/9+(8 \
%Sqrt[43] z3^2)/9
%\notag\\&
-\mathcal{Q}_2^{(200)}-\mathcal{Q}_2^{(020)}
\notag\\
&
-\frac4{27}\left(4\pm\sqrt{43}\right)\mathcal{Q}_2^{(002)}
-\frac49\left(-5\pm\sqrt{43}\right)\mathcal{Q}_2^{(110)}
%\notag\\&
-\frac29\left(1\mp2\sqrt{43}\right)\mathcal{Q}_2^{(011)}
\notag\\
&
+\frac19\left(17\pm 2\sqrt{43}\right)\mathcal{Q}_2^{(101)}
%-6 z1^2+(40 z1 z2)/9-8/9 Sqrt[43] z1 z2-2 z2^2+(68 z1 z3)/9+8/9 \
%Sqrt[43] z1 z3-(4 z2 z3)/9+8/9 Sqrt[43] z2 z3-(32 z3^2)/9-(8 Sqrt[43] \
%z3^2)/9
%\notag\\
%&
+\frac19\left(19\mp
2\sqrt{43}\right)\left[\mathcal{Q}_3^{(100)}-\mathcal{Q}_3^{(010)}\right]
%(76 z1)/9-(8 Sqrt[43] z1)/9-(76 z2)/9+(8 Sqrt[43] z2)/9
\notag\\
&+\frac{1}{234}\left(33\mp
16\sqrt{43}\right)\,\left(\mathcal{G}_1^{(0000)}-\mathcal{G}_2^{(0000)}\right).
\notag\\
\mathbb{O}_{2,0}^{g,mixed}=&\mathcal{G}^{(0000)}_1-\mathcal{G}^{(0000)}_2\,,\qquad E^{g,mixed}_{2,0}=7\,,
\notag\\
\mathbb{O}_{2,1}^{g,mixed}=&3\left(\mathcal{G}^{(0000)}_1+\mathcal{G}^{(0000)}_2\right)\,,
\qquad E^{g,mixed}_{2,1}=79/9\,.
\end{align}
The corresponding eigenvalues (anomalous dimensions) are
listed in~Table~\ref{tab:Emixed}.

Closing this section we want to emphasize that all of the above results are written for generic quark flavors
and can be applied to the study of baryon states with arbitrary quantum numbers.

%%%%%%%%%%%%%%%%%%%%%%%%%%%%%%%%%%%%%%%%%%%%%%%%%%%%%%%%%%%%
%%%%%%%%%%%%%%%%%%%%%%%%%%%%%%%%%%%%%%%%%%%%%%%%%%%%%%%%%%%%
\section{Nucleon Distribution Amplitudes}
%%%%%%%%%%%%%%%%%%%%%%%%%%%%%%%%%%%%%%%%%%%%%%%%%%%%%%%%%%%%
%%%%%%%%%%%%%%%%%%%%%%%%%%%%%%%%%%%%%%%%%%%%%%%%%%%%%%%%%%%%

%%%%%%%%%%%%%%%%%%%%%%%%%%%%%%%%%%%%%%%%%%%%%%%%%%%%%%%%%%%%
%%%%%%%%%%%%%%%%%%%%%%%%%%%%%%%%%%%%%%%%%%%%%%%%%%%%%%%%%%%%
\subsection{The leading twist-3 distribution amplitude}
%%%%%%%%%%%%%%%%%%%%%%%%%%%%%%%%%%%%%%%%%%%%%%%%%%%%%%%%%%%%
%%%%%%%%%%%%%%%%%%%%%%%%%%%%%%%%%%%%%%%%%%%%%%%%%%%%%%%%%%%%
\label{sec:twist3DA}

{}For completeness and for further use in Sec.~\ref{sec:WW} we write down the expansion for the
leading twist-3 distribution amplitude $\Phi_3(x,\mu)$ (\ref{def:Phi3})
\begin{align}\label{phi3-expansion}
\Phi_3(x,\mu)=&x_1x_2x_3\sum_{N,q}c_{Nq}\,\phi_{Nq}(\mu)\, P_{Nq}(x)\,,
\end{align}
where $\mu$ stands for the scale dependence.
The polynomials $P_{N,q}(x)$ are related to the eigenfunctions
$\Psi_{N,q}$ of the Hamiltonian $H_{1/2}$ (\ref{H12cons}) and are given by
\begin{align}\label{def:P_rep}
P_{N,q}(x_1,x_2,x_3)=\vev{e^{\sum x_kz_k}|\Psi_{Nq}(\vec{z})}\,,
\end{align}
where $\langle\ldots\rangle$ is the $SU(1,1)$-invariant scalar product (\ref{sc2})
corresponding to the spins $j_1=j_2=j_3=1$.
The normalization constants $c_{N,q}$ are defined as
\begin{align}\label{NP}
c_{Nq}^{-1}=||\Psi_{Nq}(\vec{z})||^2/\Gamma(2N+6)=\int \mathcal{D}x\, x_1x_2x_3 |P_{Nq}|^2\,.
\end{align}
{}Finally, the scale-dependent coefficients $\phi_{N,q}(\mu)$ are defined as reduced matrix elements
of the multiplicatively renormalizable twist-3 operators
\begin{align}\label{def:t3oper}
\mathbb{O}^{tw-3}(\vec{z}) = &\epsilon^{ijk}\psi^{u,i}_+(z_1)\bar\chi_+^{u,j}(z_2) \psi_+^{d,k}(z_3)\,,
\notag\\
\mathbb{O}^{tw-3}_{Nq}(\mu)= &P_{Nq}(\partial_z)\,\mathbb{O}^{tw-3}({z},\mu)|_{\vec{z}=0},
\end{align}
\begin{align}\label{def:phi_Nq}
\vev{0|\mathbb{O}^{tw-3}_{Nq}(\mu)|N}=&+\frac12 (pn)  N^{\downarrow}_+\,(-ipn)^{N}\,\phi_{Nq}(\mu)\,.
\end{align}
Taking into account Eq.~(\ref{NP}) one can project the coefficients $\phi_{Nq}$ as
follows
\begin{align}\label{}
\phi_{Nq}(\mu)=\int \mathcal{D}x\, {P_{N,q}(x)}\,\Phi_3(x,\mu).
\end{align}
%
%The coefficients $\phi_{Nq}(\mu)$ evolve autonomously with $\mu$
The first few  terms in the expansion~(\ref{phi3-expansion}) read
\begin{align}\label{}
\Phi_3(x_1,x_2,x_3)=&120x_1x_2x_3\Big[\phi^{(2/3)}_{0}
+42\,\phi_{1,0}^{(26/9)}P_{1,0}(x)+14\,\phi_{1,1}^{(10/3)} P_{1,1}(x)+
\frac{63}{10}\,\phi^{(38/9)}_{2,0}P_{2,0}(x)
\notag\\&
+\frac{63}{2}\,\phi_{2,1}^{(46/9)}P_{2,1}(x)+
\frac9{5}\,\phi^{(16/3)}_{2,2}P_{2,2}(x)+\ldots
\Big],
\end{align}
where
\begin{align}\label{Ptwist3}
P_{1,0}(x)=&\frac12(x_1-x_3)\,,\notag\\
P_{1,1}(x)=& \frac12(x_1+x_3-2x_2)%=\frac12(1-3x_2)
\,,\notag\\
P_{2,0}(x)=&3 x_1^2-3 x_1 x_2+2 x_2^2-6 x_1 x_3-3 x_2 x_3+3 x_3^2\,,
\notag\\
P_{2,1}(x)=&(x_1-x_3)(x_1+x_3-3x_2)%=(x_1-x_3)\left(1-4 x_2\right)
\,,\notag\\
P_{2,2}(x)=&x_1^2 +x_3^2- 12 x_1 x_3  + 9 x_1 x_2 + 9 x_2 x_3 -
        6 x_2^2\,.
%\notag\\
%P^{E=46/9}_{2,-}(x)=&(x_1-x_3)(x_1+x_3-3x_2)%=(x_1-x_3)\left(1-4 x_2\right)
%\,,\notag\\
%P^{E=38/9}_{2,+}(x)=&3 x_1^2-3 x_1 x_2+2 x_2^2-6 x_1 x_3-3 x_2 x_3+3 x_3^2\,,
\end{align}
The superscript in $\phi_{N,q}^{(E_{Nq})}$ shows the corresponding anomalous dimension:
\begin{align}
\phi_{N,q}^{(E_{Nq})}(\mu)=\left(\frac{\alpha_s(\mu)}{\alpha_s(\mu_0)}\right)^{E_{Nq}/\beta_0}
\phi_{N,q}^{(E_{Nq})}(\mu_0)\,.
\end{align}
These expressions agree with the ones
existing in the literature, e.g.~\cite{Bergmann:1993az,BDKM,Stefanis:1999wy,Bergmann:1999ud}. 
In notations of Ref.~\cite{Braun:2000kw}  
\begin{align}\label{}
\phi_3^{0}=\phi_{0}^{(2/3)}\,,&&\phi_3^-=12\phi_{1,0}^{(26/9)}-\frac72\phi_{1,1}^{(10/3)}\,,&&
\phi_3^+=12\phi_{1,0}^{(26/9)}+\frac{21}2\phi_{1,1}^{(10/3)}\,.
\end{align}
%

%%%%%%%%%%%%%%%%%%%%%%%%%%%%%%%%%%%%%%%%%%%%%%%%%%%%%%%%%%%%
%%%%%%%%%%%%%%%%%%%%%%%%%%%%%%%%%%%%%%%%%%%%%%%%%%%%%%%%%%%%
\subsection{Twist-4 distribution amplitudes: General formalism}
%%%%%%%%%%%%%%%%%%%%%%%%%%%%%%%%%%%%%%%%%%%%%%%%%%%%%%%%%%%%
%%%%%%%%%%%%%%%%%%%%%%%%%%%%%%%%%%%%%%%%%%%%%%%%%%%%%%%%%%%%

The construction of twist-4 distribution amplitudes is somewhat more cumbersome because
$\mathbb{O}(\vec{z})$ (\ref{vector-operator}) involves five independent light-ray
operators (for both chiral 
and mixed cases), and also because triangular mixing between quark and quark gluon operators
makes the Hamiltonian non-hermitian.
Solving Eq.~(\ref{def:Schr-eq1}) one obtains the expansion of $\mathbb{O}(\vec{z})$ in the form
\begin{align}\label{exp-O}
\mathbb{O}(\vec{z})=\sum_{N,q}C_{Nq}\Psi_{Nq}(\vec{z})\, \mathbb{O}_{Nq}\,,
\end{align}
where\footnote{The coefficients $C_{Nq}$ are added (cf. Eq.~(\ref{def:coef-fun}))  to allow for
arbitrary normalization for the functions $\Psi_{Nq},\Psi^\dagger_{Nq}$.}
\begin{align}\label{aQ}
\mathbb{O}_{Nq}=\vev{\Psi^\dagger_{Nq}|\mathbb{O}}\,, &&
C_{Nq}^{-1}=\vev{\Psi^\dagger_{Nq}|\Psi_{Nq}}\,.
\end{align}
We remind that $\Psi^\dagger_{Nq}$ are the eigenfunctions of the adjoint Hamiltonian
 (with respect to the scalar product~(\ref{scqg})), cf.~ Eq.~(\ref{def:Schr-eq1-a}).
It is convenient to split the sum in (\ref{exp-O}) in two: one contains the ``three-quark
operators'', 
and the
other one  -- three-quark-gluon operators:
\begin{align}\label{exp-O2}
\mathbb{O}(\vec{z})=\sum_{N,q}A_{nq}\Psi_{Nq}(\vec{z})\, \mathbb{O}_{Nq}+
\sum_{N\geq 2,q}B_{nq}\Psi^g_{Nq}(\vec{z})\, \mathbb{O}^g_{Nq}\,.
\end{align}
Due to a block-triangular form of the Hamiltonian $\widetilde{\mathbb{H}}$, Eq.~(\ref{hamiltonian}),
only the first three (``quark'') components of
the coefficient functions $\Psi_{Nq}^a(\vec{z}), a=1,2,\ldots,5$
corresponding to the ``quark'' operators are nonzero, $\Psi^{a=4,5}_{N,q}=0$.
On the other hand, all five components of the coefficient functions $\Psi_{Nq}^g$ are
nonzero,  
in general.

For the eigenfunctions of the adjoint operator $\widetilde{\mathbb{H}}^\dagger$, (\ref{def:Schr-eq1-a}),
the situation is the opposite: the ``quark'' eigenfunctions have all components nonzero,
whereas ``quark'' components of the quark-gluon eigenfunctions  vanish, $\Psi^{g,\dagger,
a=1,2,3}_{Nq}=0$. 
Since the diagonal blocks of the Hamiltonian~(\ref{hamiltonian}) are  self-adjoint operators
one can choose the eigenfunctions $\Psi$ and $\Psi^\dagger$ as follows
\begin{align}
\Psi_{Nq}(\vec{z})=\begin{pmatrix}\overrightarrow{\Psi}_{Nq}(\vec{z})\\[2mm]
                                     \overrightarrow{0}\end{pmatrix},&&
\Psi_{Nq}^\dagger(\vec{z})=\begin{pmatrix}\overrightarrow{\Psi}_{Nq}(\vec{z})\\[2mm]
                                     \overrightarrow{\widetilde\Psi}^\dagger_{Nq}(\vec{z})\end{pmatrix},
\notag\\
\Psi^g_{Nq}(\vec{z})=\begin{pmatrix}\overrightarrow{\widetilde{\Psi}}^g_{Nq}(\vec{z})\\[2mm]
                                    \overrightarrow{\Psi}^{g}_{Nq}(\vec{z}) \end{pmatrix},&&
\Psi_{Nq}^{g,\dagger}(\vec{z})=\begin{pmatrix}\overrightarrow{0}
                                \\[2mm]
                                     \overrightarrow{\Psi}^g_{Nq}(\vec{z})\end{pmatrix}.
\end{align}
Here $\overrightarrow{\Psi}_{Nq}(\vec{z})\,,\,\overrightarrow{\widetilde{\Psi}}^g_{Nq}(\vec{z})$
are three--dimensional ``vectors'', whereas
$\overrightarrow{\Psi}^g_{Nq}(\vec{z})\,,\,\overrightarrow{\widetilde\Psi}_{Nq}(\vec{z})$
are two--dimensional.  It follows from~(\ref{exp-O2}) and~(\ref{aQ}) that the
only remaining ``daggered'' function $\overrightarrow{\widetilde\Psi}^\dagger_{Nq}(\vec{z})$
can be expressed as follows:
\begin{align}\label{}
\overrightarrow{\widetilde\Psi}^\dagger_{Nq}(\vec{z})=&-\sum_{q'} B_{Nq'}
\vev{\overrightarrow{\Psi}_{Nq}|\overrightarrow{\widetilde{\Psi}}^g_{Nq'}}\,
\overrightarrow{\Psi}^{g}_{Nq}(\vec{z}) \,,
\end{align}
i.e. explicit construction of the adjoint Hamiltonian is in fact not necessary.

Our task is to write the expansion of the nucleon distribution amplitudes in contributions
of multiplicatively renormalizable operators. To this end it is convenient to
introduce auxiliary amplitudes
\begin{align}\label{DA}
\vev{0|\mathbb{O}^a|P}=-\frac14(\mu\lambda)(-ipn)^{n_a} N^{\downarrow(\uparrow)}_+\int
\mathcal{D}x\, e^{-i(pn)\sum_k x_k z_k} \mathcal{O}^a({x})\,, \quad a=1,2,\ldots,5\,.
\end{align}
The r.h.s. of Eq.~(\ref{DA}) involves the nucleon spinor $N^\downarrow(N^\uparrow)$ for the chiral
(mixed)  operator, respectively. The factors $(-ipn)^{n_a}$ are introduced for later convenience:
For the chiral operators we choose $n=(0,0,0,2,2)$, and for the operators of mixed chirality
 $n=(0,0,1,2,2)$. It will always be clear from the context which operator, chiral or mixed,
is considered, so that we do not show it explicitly.

The ``standard'' nucleon distribution amplitudes introduced in Sec.~\ref{sect3} are related to
the amplitudes
$\mathcal{O}^{(a)}$ as follows:
\begin{align}\label{}
\Xi_4(x_1,x_2,x_3)=\mathcal{O}^{(1)}(x_1,x_2,x_3)\,,&&
\Xi_4^g(x_1,x_2,x_3,x_4)=\mathcal{O}^{(4)}(x_1,x_2,x_3,x_4)
\end{align}
for the chiral case and
\begin{align}\label{PPO}
\Psi_4(x_1,x_2,x_3)=&\mathcal{O}^{(1)}(x_2,x_3,x_1)\,,&
\Psi_4^g(x_1,x_2,x_3,x_4)=&\mathcal{O}^{(4)}(x_2,x_3,x_1,x_4),\notag\\
\Phi_4(x_1,x_2,x_3)=&\mathcal{O}^{(2)}(x_1,x_3,x_2)\,,&
\Phi_4^g(x_1,x_2,x_3,x_4)=&\mathcal{O}^{(5)}(x_1,x_3,x_2,x_4),\notag\\
D_4(x_1,x_2,x_3)=&\mathcal{O}^{(3)}(x_1,x_3,x_2)\,
\end{align}
for mixed chirality.

We define reduced matrix elements of the multiplicatively renormalizable local
operators as
\begin{align}\label{reduced}
\vev{0|\mathbb{O}^a_{Nq}|P}=&-\frac14(\mu\lambda)(-ipn)^{N} N^{\downarrow(\uparrow)}_+\,\phi_{Nq}\,.
\end{align}
It follows from Eqs.~(\ref{DA}) and (\ref{aQ}) that
\begin{align}\label{Qp}
\phi_{Nq}=\int\mathcal{D}x\, \sum_{ab}
\overline{P^{a,\dagger}_{Nq}(x)} \,\Omega_{ab}\, \mathcal{O}^b(x)\,.
\end{align}
Here
\begin{align}\label{}
P^{a,\dagger}_{Nq}(x)=\vev{e^{\sum_{k}x_kz_k}|\Psi^{\dagger,a}_{Nq}}_a\,,
\end{align}
where $\langle\ldots\rangle_a$ is the $SU(1,1)$ invariant scalar product~(\ref{sc2})
corresponding to the conformal spins of the function $\Psi^{\dagger,a}_{Nq}$.
We also define the functions
\begin{align}\label{Pepsi}
P^{a}_{Nq}(x)=\vev{e^{\sum_{k}x_kz_k}|\Psi^{a}_{Nq}}_a.
\end{align}
The scalar product for the functions $\Psi_{Nq}\,, \Psi^{\dagger}_{Nq}$ that corresponds
to conformal operators (i.e. $\Psi_{Nq}(z)$ and $\Psi^{\dagger}_{Nq}(z)$ are
shift-invariant polynomials) 
can be written as
\begin{align}\label{scP}
\vev{\Psi^\dagger_{Nq}|\Psi_{Nq'}}=c_N\int \mathcal{D}x \,\sum_{ab}
\mu^a(x)\,
\overline{P^{\dagger,a}_{Nq}(x)}\,\Omega_{ab}\, P^b_{Nq'}(x)\equiv
c_N(P^\dagger_{Nq}|P_{Nq'})\,,
\end{align}
where $c_N=\Gamma(2N+5)$
and the integration measure is defined as
\begin{align}\label{measure}
&\mu_1(x)=x_2x_3\,, \qquad \mu_2(x)=x_1x_3\,, \qquad  \mu_{4}(x)=\mu_{5}(x)=\frac12x_1x_2x_3x_4^2\,,
\notag\\
&\mu_3(x)=\begin{cases} x_1x_2 &\text{the chiral case}\\
\frac12x_1x_2x_3^2&\text{the mixed case}\,.
\end{cases}
\end{align}
Since the functions $ \Psi_{Nq}$ and $\Psi^\dagger_{Nq}$ form a bi-orthogonal system one obtains
\begin{align}\label{}
\mathcal{O}^a(x,\mu)=\mu_a(x)\sum_{Nq} c_{Nq} \,\phi_{Nq}(\mu)\, P_{Nq}^a(x)\,,
\end{align}
where
\begin{align}\label{}
c_{Nq}^{-1}=(P^\dagger_{Nq}|P_{Nq})=C_{Nq}^{-1}/c_N\,.
\end{align}
The scale dependence of the reduced matrix elements is given by
\begin{align}\label{}
\phi_{Nq}(\mu)=\left(\frac{\alpha_s(\mu)}{\alpha_s(\mu_0)}\right)^{E_{Nq}/\beta_0}\phi_{Nq}(\mu_0)\,.
\end{align}
We now specialize to the particular cases of interest.

%%%%%%%%%%%%%%%%%%%%%%%%%%%%%%%%%%%%%%%%%%%%%%%%%
%%%%%%%%%%%%%%%%%%%%%%%%%%%%%%%%%%%%%%%%%%%%%%
\subsection{Chiral amplitudes $\Xi_4$, $\Xi_4^g$}
%%%%%%%%%%%%%%%%%%%%%%%%%%%%%%%%%%%%%%%%%%%%%%%%%
%%%%%%%%%%%%%%%%%%%%%%%%%%%%%%%%%%%%%%%%%%%%%%
The expansion for the chiral  quark, $\Xi_4$, and gluon, $\Xi_4^g$ nucleon distribution
amplitudes reads
\begin{align}\label{}
\Xi_4(x,\mu)=&x_2x_3\,\left[\sum_{N,q} a_{Nq}\,\xi_{Nq}(\mu)\, \Pi_{Nq}(x)+
\sum_{N\geq 2,q} b_{Nq}\,\xi^g_{Nq}(\mu)\, {\widetilde\Pi}^g_{Nq}(x)
\right]\,,
\notag\\
\Xi^g_4(x,\mu)=&\frac12x_1x_2x_3x_4^2\sum_{N\geq 2, q}b_{Nq}\,\xi^g_{Nq}(\mu)\,\Pi^g_{N,q}(x)\,,
\end{align}
where $\Pi_{Nq}(x)=P_{Nq}^{a=1}(x)$,
${\widetilde\Pi}^g_{Nq}(x)={\widetilde{P}}^{g, a=1}_{Nq}(x)$,
$\Pi^g_{N,q}(x)=P^{g,a=1}_{Nq} $ and the expansion coefficients $\xi_{Nk}$,  $\xi^g_{Nk}$ are defined
as reduced matrix elements of the multiplicatively renormalizable operators
\begin{align}\label{reducedME}
\vev{0|\mathbb{O}^{chiral+}_{N,q}(\mu)|N}=&-\frac14 (\mu\lambda) m_N N^{\downarrow}_+\,(-ipn)^{N}\,
\xi_{Nq}(\mu)\,,
\notag\\
\vev{0|\mathbb{O}^{g,chiral+}_{N,q}(\mu)|N}=&-\frac14 (\mu\lambda) m_N N^{\downarrow}_+\,(-ipn)^{N}\,
\xi^g_{Nq}(\mu)\,.
\end{align}
Explicit expressions for the operators $\mathbb{O}^{chiral+}_{N,q}$, $\mathbb{O}^{g,chiral+}_{N,q}$
for $N\le 2$ are given in Eq.~(\ref{chiral+});
the corresponding anomalous dimensions can be found in Table~\ref{tab:Echiral}.

Taking into account the symmetry properties~(\ref{q-symmetry}),~(\ref{ggr-1}) one can write
the normalization constants as
\begin{align}\label{}
a_{Nq}^{-1}=&2\,\int \mathcal{D}x\,
\mu_1(x)\,
\overline{\Pi_{N,q}(x)}
\Big(2+ P_{23}\Big)\Pi_{N,q}(x)\,,\notag\\
b_{Nq}^{-1}=&2\,\int \mathcal{D}x\,\mu_4(x)\,
\overline{\Pi^g_{Nq}(x)}
\Big(2+ P_{23}\Big)\Pi^g_{Nq}(x)\,,
\end{align}
where $P_{23}$ is the permutation operator:  $P_{23}\Phi(x_1,x_2,x_3)=\Phi(x_1,x_3,x_2)$.
{}Projecting out the contribution of a particular operator one finds
\begin{align}\label{}
\frac12\xi^g_{Nq}=&\int \mathcal{D}x\,
\overline{\Pi^g_{Nq}(x)}
\Big(2+ P_{23}\Big)\,\Xi_4^g(x)\,,
\notag\\
\frac12\xi_{Nq}=&\int \mathcal{D}x\,
\overline{\Pi_{Nq}(x)}
\Big(2+ P_{23}\Big)\,\Xi_4(x)
-\sum_{q'} m^N_{qq'}\, \xi_{q'}^g\,,
\end{align}
where
\begin{align}\label{}
m^N_{qq'}=2\int\mathcal{D}x\,\mu_1(x)\,
{\Pi_{Nq}(x)}\Big(2+ P_{23}\Big)
\overline{\widetilde\Pi_{Nq'}^g(x)}\,.
\end{align}
{}Finally, taking into account the contributions with $N\leq 2$ we obtain the
distribution amplitudes
\begin{align}\label{result:Xi}
\Xi_4(x,\mu)=&4x_2x_3\Big[\xi_{0,0}(\mu)\,\Pi_0(x)+
9\,\xi_{1,1}(\mu)\,\Pi_1(x)
+12\,\xi_{2,0}(\mu)\,\Pi_2(x)+\frac{28}{3}\xi_{2,0}^g(\mu)\,{\widetilde \Pi}^g_{2}(x)
\Big],
\notag\\
\Xi_4^g(x,\mu)=& \frac13 8!\,x_1x_2x_3x_4^2\, \xi^g_2\,(\mu)\Pi_2^g(x)\,,
\end{align}
where
\begin{align}
\Pi_0(x)=&1\,,\notag\\
\Pi_1(x)=&x_1+x_3-\frac32 x_2\,,\notag\\
\Pi_2(x)=&x_1^2-4 x_1 x_2+2 x_2^2+2 x_1 x_3-4 x_2 x_3+x_3^2\,,
\notag\\
{\widetilde\Pi}^g_2(x)=&
\frac{43}2 x_1^2+4 x_1 x_2-2 x_2^2-47 x_1 x_3+4 x_2 x_3+\frac{13}{2} x_3^2\,.
\end{align}
and $\Pi^g_2(x)=1/2$.
In notation of Ref.~\cite{Braun:2000kw}
\begin{align}
\lambda_2=& \xi_{0,0}\,,
\notag\\
\lambda_1 f_2^d = &\frac{4}{15} \xi_{0,0} +\frac{2}{5} \xi_{1,0}\,.
\end{align}

%%%%%%%%%%%%%%%%%%%%%%%%%%%%%%%%%%%%%%%%%%%%%%%%
%%%%%%%%%%%%%%%%%%%%%%%%%%%%%%%%%%%%%%%%%%%%%%%
\subsubsection{Mixed chirality amplitudes $\Phi_4,\Psi_4, \Phi_4^g,\Psi_4^g$}
%%%%%%%%%%%%%%%%%%%%%%%%%%%%%%%%%%%%%%%%%%%%%%%%
%%%%%%%%%%%%%%%%%%%%%%%%%%%%%%%%%%%%%%%%%%%%%%%

The distribution amplitudes $\Phi_4$, $\Psi_4$, $D_4$ have collinear twist 4, but
receive contributions both from geometric twist-3 and twist-4 operators. 
Moreover, among  twist-4 operators there are descendants of twist-3 operators which matrix
elements do not involve new nonperturbative parameters, cf.~\cite{Ball:1998ff}.
In the discussion of the operator renormalization in Sect.~5 such operators were excluded 
by imposing appropriate symmetry conditions on the solutions of the renormalization 
group equations, Eqs.~(\ref{shifts-inv}), (\ref{cons-2}), and they are not listed in Sect.~5.4.
They do contribute to the distribution amplitudes, however.
The contributions to collinear twist-4 amplitudes that can be expressed in terms of the 
leading twist distribution are usually referred to as Wandzura-Wilczek contributions, with the
remaining parts being ``genuine'' twist-4:
\begin{align}
\Phi_4(x)=& \Phi_4^{WW}(x)~+~\Phi_4^{tw-4}(x)\,,\notag\\
\Psi_4(x)=& \Psi_4^{WW}(x)~+~\Psi_4^{tw-4}(x)\,,\notag\\
D_4(x)=& D_4^{WW}(x)~+~D_4^{tw-4}(x)\,.
\end{align}
In what follows we consider these two types of contributions separately.

%%%%%%%%%%%%%%%%%%%%%%%%%%%%%%%%%%%%%%%%%%%%%
%%%%%%%%%%%%%%%%%%%%%%%%%%%%%%%%%%%%%%%%%%%%
\subsubsection{Wandzura-Wilczek contributions $\Phi_4^{WW}, \Psi_4^{WW}$}
%%%%%%%%%%%%%%%%%%%%%%%%%%%%%%%%%%%%%%%%%%%%%
%%%%%%%%%%%%%%%%%%%%%%%%%%%%%%%%%%%%%%%%%%%%
\label{sec:WW}

Let
\begin{align}
{O}^{tw-3}(\vec{z},\lambda)\equiv {O}^{tw-3}(z_1,z_2,z_3;\lambda) =
\mathbb{O}^{tw-3}(z_1,z_3,z_2,\lambda) 
\end{align}
be the leading-twist light-ray operator that enters the definition of the leading twist-3
distribution amplitude: In comparison to (\ref{def:t3oper}) we have replaced $z_2\leftrightarrow z_3$
and added an argument $\lambda$ to remind that the ``plus'' projection is done with respect
to this particular spinor.

The short-distance expansion of ${O}^{tw-3}(\vec{z},\lambda) $ can be written as
\begin{align}\label{exp-tw-3}
{O}^{tw-3}(\vec{z},\lambda)=\sum_{N,q}C_{Nq}\,\left(\Phi_{Nq}(\vec{z})\,
\mathbb{O}^{tw-3}_{Nq}(\lambda)+\frac{1}{4(N+3)} S^+\Phi_{Nq}(\vec{z}) \,
i[\mathbf{P}_{\lambda\bar\lambda},\mathbb{O}^{tw-3}_{Nq}(\lambda)]\right)+ 
\ldots\,.
\end{align}
Here the first term in parenthesis corresponds to the contribution of the conformal twist-3 operators
$\mathbb{O}^{tw-3}_{Nq}(\lambda)$ (\ref{def:t3oper}) which are annihilated by the step-down 
operator of the $SL(2)$ algebra,  so that
$i[\mathbf{K}_{\mu\bar\mu},\mathbb{O}^{tw-3}_{Nq}(\lambda)]=0$.  
The  corresponding coefficient functions $\Phi_{Nq}(z_1,z_2,z_3)$
are shift-invariant homogeneous polynomials of degree $N$ and are related by a simple change 
of arguments $z_2\leftrightarrow z_3$ to  $\Psi_{Nq}(\vec{z})$ appearing in (\ref{def:P_rep}).
The normalization coefficients are chosen as 
$C_{Nq} = ||\Phi_{Nq}||^{-2} = c_{Nq}/\Gamma(2N+6)$, cf. (\ref{NP}).

The second term in (\ref{exp-tw-3}) corresponds to contribution of the operators that include
one total derivative (in ``plus'' direction).  Note that the corresponding coefficient functions 
can be obtained by application of the three-particle ``step-up'' operator
$S^+ = S^+_{j_1=1}+S^+_{j_2=1}+S^+_{j_3=1}= \sum_k (z_k^2\partial_{z_k}+2z_k)$ (\ref{diff-form}) 
so they are not independent and do not need to be calculated separately. 
The terms with two and more total derivatives have the similar structure. They are
indicated by ellipses. 

Omitting contributions of genuine geometric twist-4 operators the expansion of the 
light-ray collinear twist-4 vector-operator $\mathbb{O}^{mixed}(\vec{z})$
reads to the same accuracy 
\begin{align}\label{exp-lm}
[\mathbb{O}^a(\vec{z},\mu,\lambda)]_{tw-3}=&\sum_{N,q}C_{Nq}\Psi^a_{Nq}(\vec{z})\,
\mathbb{O}_{Nq}(\mu,\lambda)+\sum_{N\geq 1,q} C_{Nq}\,\widehat{\Psi}^a_{N+1,q}(\vec{z})
\widehat{\mathbb{O}}_{N+1,q}(\mu,\lambda)
\notag\\
&+\sum_{N,q}C_{Nq}\frac{1}{2(2N+5)}S^+\Psi^a_{Nq}(\vec{z})\,
i[\mathbf{P}_{\lambda\bar\lambda},\mathbb{O}_{Nq}(\mu,\lambda)]\,+
\ldots\,,
\,
\end{align}
where
\begin{align}\label{expWW}
\mathbb{O}_{Nq}(\mu,\lambda)=&\frac1{N+2}(\mu\partial_\lambda)\,\mathbb{O}^{tw-3}_{Nq}(\lambda)\,,
\notag\\
\widehat{\mathbb{O}}_{N+1,q}(\mu,\lambda)=&\frac{1}{4(N+3)^2}\left(i[\mathbf{P}_{\mu\bar\lambda}\,,
\mathbb{O}^{tw-3}_{Nq}(\lambda))]-\frac{N+2}{2N+5}i[\mathbf{P}_{\lambda\bar\lambda}\,,
\mathbb{O}_{Nq}^{tw-3}(\mu,\lambda)]\right)
\end{align}
are the light-cone conformal operators $i[\mathbf{K}_{\mu\bar\mu},\mathbb{O}_{Nq}(\mu,\lambda)]=
i[\mathbf{K}_{\mu\bar\mu},\widehat{\mathbb{O}}_{N+1,q}(\lambda,\mu)]=0$.
(It is tacitly assumed that the generator $S^+$  in the second line in Eq.~(\ref{exp-lm}) involves the 
conformal spins of the functions it acts on.) The operator $\mathbb{O}_{Nq}(\mu,\lambda)$
has geometric twist-3 
whereas $\widehat{\mathbb{O}}_{N+1,q}(\mu,\lambda)$ contains both twist-3 and twist-4 contributions.

The task is to find the corresponding (five-component) coefficient functions
$\Psi_{Nq},\widehat\Psi_{Nq}$.
This can be done by observing that $\mathbb{O}^a({z},\mu,\lambda)$ only depends linearly
on $\mu$ and essentially reduces to ${O}^{tw-3}(z,\lambda)$ after a formal substitution
$\mu \to\lambda$. 
Taking into account the identities
\begin{align}\label{}
\lambda\partial_\mu
\mathbb{O}^{(a=1,2)}(\vec{z},\mu,\lambda)=&{O}^{tw-3}(\vec{z},\lambda)\,, &
\lambda\partial_\mu
\mathbb{O}^{(a=3)}(\vec{z},\mu\lambda)=-\partial_{z_2}{O}^{tw-3}(\vec{z},\lambda)
\end{align}
and comparing the two representations in (\ref{exp-tw-3}) and (\ref{exp-lm})
one easily gets for $\Psi_{Nq}$
\begin{align}\label{}
\Psi_{Nq}^{a=1,2}(\vec{z})=&\Phi_{Nq}({z})\,,&
\Psi_{Nq}^{a=3}(\vec{z})=&-\partial_{3}\Phi_{Nq}(\vec{z})\,,&
\Psi_{Nq}^{a=4}(\vec{z}) =& \Psi_{Nq}^{a=5}(\vec{z}) = 0\,.
\end{align}
For the functions $\widehat\Psi^a_{Nq}$ 
one obtains, after some algebra
\begin{align}\label{widehatPsi}
\widehat\Psi^{a=1(2)}_{N+1,q}(\vec{z})=&-\Big[S^+-2(N+3)z_{1(2)}\Big]\,\Phi_{Nq}(\vec{z})\,,
\notag\\
\widehat\Psi^{a=3}_{N+1,q}(\vec{z})=&-\Big[2(2N+5)-\bigl[S^+-2(N+3)z_{3}\bigr]
\partial_3\Big]\,\Phi_{Nq}(\vec{z})\,,
\end{align}
where  $S^+=S_{j_1=1}^++S_{j_2=1}^++S_{j_3=1}^+$.
One can easily check that the functions $\widehat\Psi_{Nq}$ defined in Eq.~(\ref{widehatPsi}) are
shift-invariant polynomials, as they should be. They are normalized as
\begin{align}\label{}
||\Psi_{Nq}||^2=(N+2)||\Phi_{Nq}||^2\,, &&
||\widehat\Psi_{N+1,q}||^2=2(2N+5)(N+3)^2||\Phi_{Nq}||^2\,.
\end{align}
Going over to the ``P-representation'' as defined in Eq.~(\ref{Pepsi}) we obtain
\begin{align}\label{}
P^{(1)}_{Nq}(x)=&\partial_1 x_1\, P^{tw-3}_{Nq}(x)\,,&
P^{(2)}_{Nq}(x)=&\partial_2 x_2\, P^{tw-3}_{Nq}(x)\,,&
P^{(3)}_{Nq}(x)=&-2\partial_3\, P^{tw-3}_{Nq}(x)\,
\end{align}
and
\begin{align}\label{}
\widehat P^{(1)}_{N+1,q}(x)=&\Big(2N+5-X\partial_{x_1}\Big)\,x_1\,P^{tw-3}_{Nq}(x)\,,\notag\\
\widehat P^{(2)}_{N+1,q}(x)=&\Big(2N+5-X\partial_{x_2}\Big)\,x_2\,P^{tw-3}_{Nq}(x)\,,\notag\\
\widehat P^{(3)}_{N+1,q}(x)=&-2\Big(2N+5-X\partial_3\Big)\,P_{Nq}^{tw-3}(x)\,,
\end{align}
where $X=x_1+x_2+x_3$.

Taking appropriate matrix elements we end up with
the Wandzura-Wilczek contributions to the auxiliary distributions $\mathcal{O}^{(a)}$
(\ref{DA}) 
\begin{align}\label{}
[\mathcal{O}^{(a)}(x)]^{WW}=\mu_a(x)\sum_{Nq} \frac{c_{Nq}\,\phi_{Nq}}{(2N+5)}
\left(\frac1{N+2}P^{(a)}_{Nq}(x)-\frac{1}{N+3}\widehat P^{(a)}_{N+1,q}(x)\right),
\end{align}
where  $c_{Nq}, \phi_{Nq}$ are the twist-3 expansion coefficients,~(\ref{phi3-expansion}).
One can easily verify the following relation~(cf. Eq.~(\ref{DPPrel})):
\begin{align}\label{}
[\mathcal{O}^{(3)}(x)]^{WW}=x_1\, [\mathcal{O}^{(1)}(x)]^{WW}+x_2\, [\mathcal{O}^{(2)}(x)]^{WW}\,.
\end{align}

Finally, taking into account Eqs.~(\ref{PPO}) we obtain
\begin{align}\label{}
\Phi_4^{WW}(x)=&-\sum_{N,q}\frac{{c_{Nq}\,\phi_{Nq}}}{(N+2)(N+3)}
\left(N+2-\frac{\partial}{\partial x_3}\right)x_1x_2x_3\,
\,
P^{tw-3}_{Nq}(x_1,x_2,x_3)\,,\\
\Psi_4^{WW}(x)=&
-\sum_{N,q}\frac{{c_{Nq}\,\phi_{Nq}}}{(N+2)(N+3)}
\left(N+2-\frac{\partial}{\partial x_2}\right)x_1x_2x_3\,
\,
P^{tw-3}_{Nq}(x_2,x_1,x_3)\,,
\end{align}
where the polynomials $P_{Nq}^{tw-3}(x)$ are defined in
Eq.~(\ref{phi3-expansion}),~(\ref{Ptwist3}).
These expressions present one of the main results of this paper.

%%%%%%%%%%%%%%%%%%%%%%%%%%%%%%%%%%%%%%%%%%%%%%%%
%%%%%%%%%%%%%%%%%%%%%%%%%%%%%%%%%%%%%%%%%%%%%%%
\subsubsection{Genuine twist-4 contributions $\Psi^{tw-4}_4,\Phi_4^{tw-4}, \Psi_4^g,\Phi_4^g$}
%%%%%%%%%%%%%%%%%%%%%%%%%%%%%%%%%%%%%%%%%%%%%%%%
%%%%%%%%%%%%%%%%%%%%%%%%%%%%%%%%%%%%%%%%%%%%%%%

The expansion of the  twist-4 auxiliary amplitudes $\mathcal{O}^a(x)$  (\ref{DA}) reads
\begin{align}\label{expPP}
\mathcal{O}^{a}(x)=&\mu_a(x)\left[\sum_{Nq} A_{Nq}\,\eta_{Nq}\,
\mathcal{P}_{Nq}^{a}(x)+
\sum_{N\geq2,q}B_{Nq}\,\eta^g_{Nq}\, \widetilde {\mathcal{P}}_{Nq}^{a}(x)\right],\quad a=1,2,3\,,
\notag\\
\mathcal{O}^{a+3}(x)=&\frac12x_1x_2x_3x_4^2\sum_{N\geq 2,q} B_{Nq}\,\eta_{Nq}^g\,
\mathcal{P}^{g,a}_{Nq}(x)\,,\qquad a=1,2\,.
\end{align}
The coefficients $A_{N,q}$ and $B_{N,q}$ are given by
\begin{align}
A_{Nq}^{-1}=& \int \mathcal{D}x\,x_1x_2x_3\left(\frac1{x_1}\, |\mathcal{P}_{Nq}^{1}(x)|^2+
\frac1{x_2}\, |\mathcal{P}_{Nq}^{2}(x)|^2+\frac{x_3}{4} |\mathcal{P}_{Nq}^{3}(x)|^2\right),
\notag\\
B_{Nq}^{-1}=&\int \mathcal{D}x \,\mu_4(x)\,\sum_{ab}\mathcal{P}^{g,a}_{Nq}(x)\,
\omega_{ab}\,\mathcal{P}^{g,b}_{Nq}(x)\,,
\end{align}
where in the last line $\omega=\begin{pmatrix}2&1\\1&2\end{pmatrix}$.
Note a useful identity
\begin{align}\label{}
\mathcal{P}_{Nq}^{3}(x)=\frac2{x_3^2}\int_0^{x_3}d\tau\,\tau\,
\left(\partial_1 \mathcal{P}^{1}_{Nq}(x_1,x_2,\tau)+
\partial_2 \mathcal{P}^{2}_{Nq}(x_1,x_2,\tau)\right)
\end{align}
which is a consequence of Eq.~(\ref{cons-2}).

The reduced matrix elements $\eta_{Ng}(\mu)$ and $\eta_{Ng}^g(\mu)$ are
defined similar to (\ref{reducedME}), replacing the spin-down spinor $N^\downarrow$
by the spin-up one, $N^\uparrow$, cf. (\ref{DA}), (\ref{reduced}).
The expressions for the operators $\mathbb{O}^{mixed}_{N,q}$, $\mathbb{O}^{g,mixed}_{N,q}$
for $N\le 2$ and the corresponding anomalous dimensions are given in Eq.~(\ref{O-mixed})
and Table~\ref{tab:Emixed}, respectively.
They can be isolated by the projection
\begin{align}
\eta_{Nq}=&\sum_{a=1}^3\int
\mathcal{D}x\,
\mathcal{P}_{Nq}^{a}({x})\, \mathcal{O}^{a}({x})-\sum_{q'}M_{qq'}^{N}\,\eta_{Nq'}^g\,,
\notag\\
\eta_{Nq}^{g}=&\int \mathcal{D}x\, \sum_{ab}\mathcal{P}^{g,a}_{Nq}(x) \,\omega_{ab}\,
\mathcal{G}^{b}(x)\,,
\end{align}
where
\begin{align*}
M_{qq'}^{N}=B_{Nq'}\!\int\! \mathcal{D}x\, x_1x_2x_3\left(\frac{1}{x_1}
\mathcal{P}^{1}_{Nq}({x})\widetilde{\mathcal{P}}^{g,1}_{Nq'}({x})+
\frac{1}{x_2}
\mathcal{P}^{2}_{Nq}({x})\widetilde{\mathcal{P}}^{g,2}_{Nq'}({x})
+
\frac{x_3}{4}
\mathcal{P}^{3}_{Nq}({x})\widetilde{\mathcal{P}}^{g,3}_{Nq'}({x})
\right).
\end{align*}

Taking into account the contributions of local operators with $N\leq 2$ we obtain
\begin{align}\label{}
\Psi^{tw-4}_4(x,\mu)=&\phantom{-}12x_1x_3\Big[\eta_{0,0}(\mu)+
4\,\eta_{1,0}(\mu)\,\mathcal{P}_{1,0}(x_2,x_3,x_1)+
\frac{20}{3}\eta_{1,1}(\mu)\,\mathcal{P}_{1,1}(x_2,x_3,x_1)\notag\\
&
+\frac52\left(\frac{11}2+\frac{5}{\sqrt{43}}\right)\,\eta_{2,0}(\mu)\,\mathcal{P}_{2,0}(x_2,x_3,x_1)
+\frac{45}{2}\eta_{2,1}(\mu)\,\mathcal{P}_{2,1}(x_2,x_3,x_1)
\notag\\ &
+\frac52\left(\frac{11}2-\frac{5}{\sqrt{43}}\right)\eta_{2,2}(\mu)\,\mathcal{P}_{2,2}(x_2,x_3,x_1)
+\frac{140}{117}\eta^g_{2,0}(\mu)\,\widetilde{\mathcal{P}}^g_{2,0}(x_2,x_3,x_1)
\notag\\ &
+\frac{70}{47}\eta^g_{2,1}(\mu)\,\widetilde{\mathcal{P}}^g_{2,1}(x_2,x_3,x_1)\Big],
\notag\\
\Phi_4^{tw-4}(x,\mu)=&-12x_1x_2\Big[\eta_{0,0}(\mu)+4\,\eta_{1,0}(\mu)\,\mathcal{P}_{1,0}(x_3,x_1,x_2)-
\frac{20}{3}\eta_{1,1}(\mu)\,\mathcal{P}_{1,1}(x_3,x_1,x_2)
\notag\\ &
+\frac52\left(\frac{11}2+\frac{5}{\sqrt{43}}\right)\,\eta_{2,0}(\mu)\,\mathcal{P}_{2,0}(x_3,x_1,x_2)
-\frac{45}{2}\eta_{2,1}(\mu)\,\mathcal{P}_{2,1}(x_3,x_1,x_2)
\notag\\&
+\frac52\left(\frac{11}2-\frac{5}{\sqrt{43}}\right)\,\eta_{2,2}(\mu)\,\mathcal{P}_{2,2}(x_3,x_1,x_2)
+\frac{140}{117}\eta^g_{2,0}(\mu)\,\widetilde{\mathcal{P}}^g_{2,0}(x_3,x_1,x_2)
\notag\\ &
-\frac{70}{47}\eta^g_{2,1}(\mu)\,\widetilde{\mathcal{P}}^g_{2,1}(x_3,x_1,x_2)
\Big],
\notag\\
\Psi_4^g(x,\mu)=&~\phantom{-}\frac1{4}8!x_1x_2x_3x_4^2\left[\eta^g_{2,0}(\mu)+\frac13
\eta^g_{2,1}(\mu)\right],
\notag\\
\Phi_4^g(x,\mu)=&-\frac1{4}8!x_1x_2x_3x_4^2\left[\eta^g_{2,0}(\mu)-\frac13\eta_{2,1}^g(\mu)
\right],
\end{align}
where
\begin{align}\label{}
\mathcal{P}_{1,0}(x)=&x_1+x_2-\frac32 x_3\,,
\notag\\
\mathcal{P}_{1,1}(x)=&x_1-x_2+\frac12 x_3\,,
\notag\\
\mathcal{P}_{2,1}(x)=&x_1^2-x_2^2-2 x_1 x_3+3 x_2 x_3-\frac23 x_3^2\,,
\notag\\
\begin{pmatrix}\mathcal{P}_{2,0}(x)\\
\mathcal{P}_{2,2}(x)\end{pmatrix}
=&x_1^2+\frac49\left(-5\pm\sqrt{43}\right)x_1x_2
+x_2^2+\frac29\left(1\mp2\sqrt{43}\right)x_1x_3
\notag\\
&-\frac19\left(17\pm2\sqrt{43}\right)x_2x_3+
\frac{4}{27}\left(4\pm\sqrt{43}\right)x_3^2\,,
\notag\\
\widetilde{\mathcal{P}}_{2,0}^g(x)=&
64 x_1^2-55 x_1 x_2+\frac{11}2 x_2^2-73 x_1 x_3+11 x_2 x_3+\frac{17}{2} x_3^2\,,
\notag\\
\widetilde{\mathcal{P}}_{2,0}^g(x)=&16 x_1^2-\frac13x_2^2-32 x_1 x_3+x_2 x_3+5 x_3^2\,.
\end{align}
The expression for $D_4(x,\mu)$ (\ref{def:D4}) can be obtained using the integral representation in
Eq.~(\ref{DPPrel}).

The reduced matrix elements $\eta_{Nq}$ for $N=0,1$ are related to the
parameters introduced in Ref.~\cite{Braun:2000kw} as
\begin{align}\label{}
  \lambda_1=&-\eta_{00}\,,
\notag\\
 \lambda_1\,f_1^d=&-\frac16\phi_{00}-\frac3{10}\eta_{00}-\frac15 \eta_{10}+\frac13\eta_{11}\,,
\notag\\
 \lambda_1\,f_1^u=&-\frac16\phi_{00}-\frac1{10}\eta_{00}-\frac35 \eta_{10}+\frac13\eta_{11}\,.
\end{align}
The term in $\phi_{00}$ is the Wandzura-Wilczek contribution that can be traced to the twist-4 operator
containg a transverse derivative of a local twist-3 three-quark operator (and properly symmetrized).

%%%%%%%%%%%%%%%%%%%%%%%%%%%%%%%%%%%%%%%%%%%%%
%%%%%%%%%%%%%%%%%%%%%%%%%%%%%%%%%%%%%%%%%%%
\section{Conclusions}
%%%%%%%%%%%%%%%%%%%%%%%%%%%%%%%%%%%%%%%%%%%
%%%%%%%%%%%%%%%%%%%%%%%%%%%%%%%%%%%%%%%%%%%%%

The motivation for our study has been to work out  efficient techniques for a calculation
of anomalous dimensions of generic higher twist operators in QCD.
Apart from the applications to QCD phenomenology, this project was
fuelled by the recent progress in the understanding of the spectrum of the dilatation
operator in the maximally supersymmetric $N=4$ Yang-Mills theory~\cite{Beisert} and,
in particular, the  work \cite{BFKS04} where it was argued that
the diagonal part of one-loop QCD  RG equations (for arbitrary twist) can be written in a
Hamiltonian form in terms of  quadratic Casimir operators of the full conformal group $SO(4,2)$.
In simple words, the symmetry  under the conformal transformations in the directions
orthogonal to the light-cone plane implies  existence of relations between the renormalization group
equations of different (geometric) twist. This, in turn, suggests that the techniques
developed for the description of quasipartonic operators in QCD \cite{Bukhvostov:1985rn} 
can be generalized to include non-quasipartonic operators as well.
Our goal was to develop a consistent computational framework based on these ideas.

The first step was to construct the appropriate operator basis with ``good'' transformation properties.
We found that the complete basis of one-particle conformal operators for chiral quark and
self-dual gluon 
fields in QCD contains seven light-ray fields, and similar in the anti-chiral sector.  An interesting
feature of this basis is that it includes some, but not all, transverse derivatives: If the transverse
plane is parameterized in terms of a single complex variable as it is usually done in the studies
of high-energy scattering, then the basis fields only include holomorphic derivatives acting
on holomorphic components of the fields, and vice versa.

Although much of the formalism appears to be general, in this paper we concentrate on the
simplest  example of non-quasipartonic twist-four baryon operators that contain two
``plus''  and one ``minus'' quark field, schematically
$q_+ q_- q_+$,
and their mixing with (quasipartonic) four-particle operators involving a gluon field,  of the type
$q_+ q_+ q_+ F_{+\perp}$. For this setup we calculate all one-loop evolution
kernels and check that they are $SL(2)$ invariant, as expected.
The evolution equation for three-quark operators of the same chirality
turns out to be completely integrable. The spectrum of anomalous
dimensions coincides in this case with the energy spectrum of the twist-4 subsector of the
$SU(2,2)$ Heisenberg spin chain, confirming the prediction of \cite{BFKS04}.
We find the explicit expression of the corresponding conserved charge and
calculate its spectrum.
A simple analytic expression is found for the lowest anomalous dimension
for chiral quark twist-4 operators with odd number $N=2k+1$
of covariant derivatives. For other cases the spectra are studied numerically,
see Figs.~\ref{fig:chiral}--\ref{fig:joint-mixed} and Tables~\ref{tab:Echiral},\ref{tab:Emixed}.
It turns out that differences between twist-4 and twist-3 operators
and also between twist-4 operators of different chirality mostly affect a few lowest eigenstates
(for a given $N$); the upper part of the spectrum of anomalous dimensions is universal.
The spectrum of twist-4 quark operators overlaps strongly with that of
quark-gluon operators, apart from a few lowest states.

Finally, these results are applied to give a general classification and
calculate the scale dependence of subleading twist-4 nucleon distribution
amplitudes that are relevant for hard exclusive reactions involving a helicity flip.
In particular we introduce new four-particle distribution amplitudes
involving a gluon field, and derive explicit expressions for the expansion
of all distribution amplitudes in contributions of multiplicatively
renormalizable operators taking into account first three orders in the
conformal spin expansion. As a byproduct of our analysis, we
give an all-order expression (in conformal spin) for the contributions
of geometric twist-3 operators to the (light-cone) twist-4 nucleon
distribution amplitudes, which are usually referred to
as Wandzura--Wilczek contributions. 
The applications of these results to phenomenology of hard exclusive reactions will be 
considered elsewhere.

The techniques suggested in this paper can have a rather broad field of
applications, in particular to the calculation of twist-4 corrections to
the structure functions of deep-inelastic lepton-hadron scattering.
We plan to consider this problem in a separate publication.

\section*{Acknowledgements}

This work was supported by the German Research Foundation (DFG),
grants 92090175 and 9209282.

%%%%%%%%%%%%%%%%%%%%%%%%%%%%%%%%%%%%%%%%%%%%%%%%%%%%%%%%%%%%%%%%%%%%%%%%%%%%%%%%%%%%

\end{document}